%% file: info-as-maxcal.tex
\documentclass[12pt,stdletter,orderfromtodate,sigleft,dateno]{report}
\pdfoutput=1
\usepackage{arxiv}		
\usepackage{titlesec}	

\newcounter{chap}

\usepackage{amsmath}		
\usepackage{amssymb}		
\usepackage{amsfonts}		
\usepackage{amsthm}			
\usepackage{mathtools}		
\usepackage{tipa}			
\usepackage{bm}				
\usepackage{graphicx}		
\usepackage{url}			
\usepackage{algorithm,algpseudocode}
\usepackage{booktabs,multirow}		

\algnewcommand\algorithmicnotice{\textbf{Note:}}
\algnewcommand\Note{\item[\algorithmicnotice]}

\usepackage{tikz}			
\usepackage{tikz-network}	
\usepackage{caption}		
\usepackage{float}			
\usetikzlibrary{decorations.pathreplacing}

\usepackage{hyperref}

\newtheorem{thm}{Theorem}
\newtheorem{dfn}{Definition}

\newtheorem{crl}{Corollary}

\DeclareRobustCommand{\atez}{\text{\reflectbox{$\zeta$}}}		

\begin{document}
	
	\title{Information as Maximum-Caliber Deviation: A bridge between Integrated Information Theory and the Free Energy Principle}
	
	\author{Alexander Kearney\\
		University of Oxford\\
		\texttt{a.kearney.research@gmail.com}
	}
	
	\date{3rd May, 2026}
	
	\setcounter{secnumdepth}{3}
	\setcounter{tocdepth}{2}
	
	\maketitle
	
	\begin{abstract}
		The Free Energy Principle (FEP) is a leading framework for mathematically modeling self-organization and learning, while Integrated Information Theory (IIT) is a computational ontology of consciousness oriented around irreducible cause and effect.
		While conceptual unifications have been proposed and appear to be supported by empirical findings, the absence of a rigorous mathematical mapping places upper bounds on their precision and testability.
		This work proposes that information can be defined as the deviation $\psi$ of realized dynamics from a constrained maximum-caliber (MaxCal) path ensemble over a finite time horizon.
		Under this definition, each of the cause/effect repertoires central to IIT 3.0 emerge directly from MaxCal variational principles, allowing IIT's phenomenological calculus to be re-derived from constrained entropy-maximization (CMEP).
		This framework supplies a theoretical bridge to active inference, which is mathematically dual to CMEP under Langevin dynamics, and offers a principled route for extending IIT to new dynamical regimes.
		When the approach is applied under the Central Limit Theorem (CLT) for Markov chains and via large deviations theory (LDT) to Ising models, information $\psi$ is shown to be equivalent to prediction error under accompanying predictive coding models.
		This may hold relevance to the ``hill-shaped trajectory'' of $\Phi$ observed in neuronal cultures adapting to sensory inputs.
		Together, these results provide a physically and mathematically grounded rationale for studying the convergence of FEP, IIT, and thermodynamic frameworks of cognition such as recent work grounding consciousness in violations of the Fluctuation-Dissipation Theorem (FDT).
	\end{abstract}
	
	
	\clearpage
	\include{acknowledgements/ack.tex}
	
	\clearpage
	\tableofcontents
	\listoffigures
	\listoftables
	
	\include{1-intro/1_intro.tex}
	\include{2-IIT/2_IIT.tex}
	\include{3-bays/3_bays.tex}
	\include{4-info-dev/4_info_dev.tex}
	\include{5-IIT-FEP/5_IIT_FEP.tex}
	\include{6-complexity/6_complexity.tex}
	\include{7-conc/7_conc.tex}
	
	\addcontentsline{toc}{chapter}{Bibliography}
	
	\begingroup
	\scriptsize
	

	\endgroup
	
	\include{appendix/app.tex}

\end{document}

%% file: acknowledgements/ack.tex
\chapter*{Acknowledgments}\label{ack}

This work is a continuation and refinement of \textit{``Integrated Information Theory and Transitional Dynamics of Systems'',} conducted in 2022 at the Mathematical Institute, University of Oxford, under the supervision of Kobi Kremnizer.
I sincerely thank him for all of his support, including over the past year.
In addition to introducing me to the field, he has provided feedback, encouragement, and valuable insight.\\
\\
I would also like to thank my friend and collaborator Timothy Wroge for acquainting me with Free Energy Principle-based theories of cognition and for the feedback on earlier drafts.\\
\\
Finally, to all who have supported me in other capacities as I have conducted this work, I thank you too.

%% file: 1-intro/1_intro.tex
\chapter{Motivating a comparison of IIT and FEP}\label{1_intro}

\textsc{For the better part} of our history, humanity has questioned its existence.
Through philosophy, art, and science, we have grappled with two recurring questions: \textit{``why are we here?'',} and relatedly, \textit{``why are we conscious?''}\\
\\
1980s, California, engineers started tackling the latter question.
Owing to a decades-long collision of computer- and neuro-science, 2.9 billion square inches\footnote{
	According to research by Morder Intelligence: \url{https://www.mordorintelligence.com/industry-reports/ai-and-hpc-semiconductor-silicon-wafer-market}
} of silicon have become the world's testing ground for theories of neural computation.
Technology, however, is not the only field to enjoy the fruits of this era.
1994, it could be argued, marked a bifurcation point in brain research.\\
\\
From the Neurosciences Institute, San Diego, researchers demonstrated that sensory data can alter the value systems of synthetic neurons\cite{friston94} --- results which continue to influence work in dopaminergic signaling\cite{schultz98}, reinforcement learning\cite{sutton18}, and global workspace modeling\cite{dehaene01}.
Concurrently, complexity science was employed to frame the brain's segregation/integration trade-off, and information theory was leveraged to construct a mathematical solution\cite{tononi94}.
Rather than ripple --- or even shock --- the influence of these papers avalanched, finding expression in two sub-fields\cite{tononi04,friston03}.\\
\\
The \textit{Free Energy Principle (FEP)} is a framework for biological\cite{friston13} and physical\cite{sakthivadivel22,friston23} self-organization, proposing that homeostasis is maintained by minimization of a free energy functional.
Initially developed as a unifying framework for competing computational models of brain function\cite{Friston10}, it found prominence after using expectation maximization to derive the predictive coding algorithm\cite{rao99,friston03}.\\
\\
\textit{Integrated Information Theory (IIT)} addresses the modeling of consciousness\cite{tononi04,tononi08} by directly grappling with Chalmers' Hard Problem\cite{chalmers1995}.
Starting from the \textit{knowledge} that one is conscious and that this entails certain properties, IIT describes the ``mind's eye'' to then infer accompanying physical properties of conscious systems\cite{oizumi2014,albantakis2023}.\\
\\
There is no obvious connection between these two theories, and they can render substantively different predictions\cite{safron20,intrepid26}.
Yet, as we shall prove in chapter \ref{4_info_dev}, they are related by a shared basis in maximum-caliber (MaxCal) inference\cite{sakthivadivel23,maxwell23} --- which may address why an empirical relationship appears to exist\cite{albantakis14,olesen23,mayama25}.
While prior work suggests a conceptual basis for such a linkage\cite{safron20,safron22}, this thesis is (to the author's knowledge) the first which proposes a formal mathematical relationship.\\
\\
The shared origin in MaxCal inference draw questions regarding thermodynamic models of cognition\cite{kirkaldy65,vandervert95,buisan25}.
While not explored in its full depth here, it certainly motivates the methods.
In addition to offering a physically grounded basis of the mathematical methods\cite{zhou21}, thermodynamic perspectives may offer measurable physical predictions where computation of metrics proves intractable\cite{strelnikov19,buxton25}.

\section{Three ``disparate'' perspectives on cognition}\label{1.1}

\subsection{Integrated Information Theory (IIT)}\label{1.1.1}

Drawing inspiration from Descartes, IIT proposes an ontology in which the phenomenology of consciousness is epistemically prior to the physical realm\cite{oizumi2014,albantakis2023}.
The framework inverts the traditional direction of inference, beginning with the essential properties of experience to infer the features of a physical substrate.
While substantial support has been lended toward the framework\cite{consortium25,seth22}, rigorous critique has simultaneously been drawn\cite{tegmark2016,mediano18,doerig19,cea23}.\\
\\
IIT's fundamental method involves taking some system $(\bm{X}, \bm{P})$ (expressed as a random variable and transition probability matrix), observing its state $\bm{X}^0 = \bm{x}^0$, and comparing the predictions and retrodictions this yields against a counterfactual absence of information.
This absence is modeled by uniform perturbations, and the systems described in its third iteration (the one we focus on) are deterministic in nature.
Integration, meanwhile, occurs when information yielded by the whole exceeds information yielded by the sum of separate parts.
We model separation by random perturbation, which represent a disruption in transmission of information from one ``part'' to the other\cite{oizumi2014}.\\
\\
A substantial strength of IIT is its precision when compared with other theories of consciousness, though this also drives much critique.
Inconsistencies have been observed between the various $\Phi$ metrics proposed\cite{mediano18}, and the tractability of its methods has been questioned\cite{cerullo2015}.
In further addition, while IIT purports to apply at the fundamental level\cite{oizumi2014,tononi23,albantakis2023}, a point of contention has been its lack of alignment with fundamental physical models\cite{barrett14}.\\
\\
Field-theoretic\cite{barrett14} and quantum\cite{kremnizer15,zanardi18,prentner23} formulations have been proposed to solve the ``fundamental physics'' part of this problem, while category theoretic accounts have provided generalization\cite{tull21}.
However, falsifiability remains elusive (as in all theories of consciousness\cite{hoel21}) and scaling problems remain.
In addition, recent formulations of IIT have proposed a difference between fundamental causal events (which are modeled as Markovian) and physical manifestations (which may not be)\cite{albantakis2023}.
While the internal consistency of IIT has been preserved, it remains unclear whether this would sway skeptics.\\
\\
Nonetheless, these problems are common across theories of consciousness\cite{kleiner21} and IIT has been recognized by Chalmers as a leading candidate for resolving the Hard Problem\cite{chalmers2018}.
In further addition, the Perturbational Complexity Index (PCI) was developed in light of IIT as a potential marker of consciousness\cite{casali13}, and has since found clinical success in predicting recovery of non-communicative patients\cite{sarasso15,casarotto16}.
Thus, IIT continues to remain a well-researched, influential force within theoretical neuroscience\cite{consortium25,intrepid26}.

\subsection{Free Energy Principle (FEP)}\label{1.1.2}

The \textit{Bayesian Brain Hypothesis (BBH)} is a family of theories suggesting that (parts of) the brain performs Bayesian inference\cite{pratt1926,barlow61,marr82,friston03,herculano_houzel09}.
As these ideas surfaced, advances in computer science proposed methods for performing approximate Bayesian inference\cite{jordan99,blei17,ghahramani2000,beal03}.
\textit{Predictive Coding} is the hypotheses that the brain is composed of hierarchical layers, each exchanging predictions and errors\cite{rao99,millidge21}.
In 2003, this was formally connected to the BBH by showing that approximate inference over a \textit{free energy} functional yields the error terms from predictive coding\cite{friston03}.
This was part of a broader synthesis, the \textit{FEP,} which has since evolved into a full process theory of self-organization, termed \textit{active inference}\cite{friston13,buckley17,parr19,friston23}.\\
\\
While FEP makes no inherent claims about the character of consciousness, it has increasingly been explored in pursuit of answering such questions.
Markovian Monism has suggested that Markov blankets have properties relevant for explaining perceptions of mind-matter dualism\cite{friston20}, while neurorepresentationalism provides a principled account of sentience arising from hierarchical generative models\cite{pennartz15,pennartz22}.\\
\\
In contrast to IIT which is perceived as too precise, FEP's core critique centers around its generality.
In particular, some researchers have suggested that predictions made by active inference do not differ substantially enough from other theories to be deemed falsifiable\cite{biehl21,hodson24}, though it has also been noted by Friston himself that FEP is a mathematical construct and thus falsifiability is not a clear goal\cite{friston18}.
The process theory active inference is falsifiable, and recent attempts have been made to discern it from neurorepresentationalism and other theories of cognition\cite{intrepid26}.\\
\\
FEP and active inference has inspired an extremely broad body of literature, in areas including psychiatry\cite{adams16}, robotics\cite{pio-lopez16,pezzato20,kawahara22}, physics\cite{sakthivadivel22,maxwell23}, and even economics\cite{adra25,kuhn25}.
Within neuroscience itself, it has leveraged to explain proposed neuronal dynamics\cite{harrison05} as well as self-organized criticality\cite{bettinger23}.
The influence of FEP appears to lie in the way that its breadth combines with mathematical specificity and rigor.

\subsection{Thermodynamic models of cognition}\label{1.1.3}

Schr\"{o}dinger is well known for his work in theoretical physics, though comparatively less so for his characterizations of life and mind.
In his work \textit{``What is life?''}, he proposed that living organisms maintain non-equilibrium steady states (NESSs) via extraction of \textit{``negative entropy'' (negentropy)} from the environment.
With regards to cognition, his view was that this mining of negentropy imposes a requirement to learn and generates a state of consciousness, while ``established order'' constitutes the remit of the known and is unconscious\cite{schrodinger1944}.\\
\\
Kirkaldy echoed his perspective in 1965 with his conceptualization of the brain as an irrversible dynamical system.
Fuelled by continuous exchange of energy and information with the enrivonment, he proposed that ``unconsciousness'' corresponds to a stable saddle point with unchanging state variables, while consciousness is born from deviations away from this.
In addition, his view was that unconscious states should be resistant to perturbation while conscious ones are sensitive.
His framework also approached the question of learning, proposing that accumulation of free energy stored as stable (and thus un/less-conscious) saddle points drives learning and memory\cite{kirkaldy65}.\\
\\
Since then, several attempts have been made to ground consciousness (and broader cognition) in thermodynamic processes.
Vandervert proposed that consciousness can be understood as an embedding in space-time of continuously generated self-referential energy patterns (termed ``algorithms'')\cite{vandervert95}, while Collell and Fauquet suggested that existing energy-information couplings\cite{szilard29,landauer61} can be used to bridge thermodynamic and information theoretic accounts of the brain\cite{collell15}.
Ganesh, in 2020, advanced dissipation and breaking of detailed balance as necessary features of conscious processing\cite{ganesh20}.
Meanwhile, Strelnikov developed a full model of facial recognition which casts the brain as a hierarchically structured, fluctuation energy field.
From this, he was able to use a thermodynamic free energy function to yield predictive coding outcomes\cite{strelnikov19}.\\
\\
Further work has included utilizing a full thermodynamic model of the brain to accurately predict Alzheimer's disease\cite{zhou21}, while other research has used linear scaling of energy and entropy in the brain to account for Zipf law scaling in thermodynamic terms\cite{chialvo24}.
Task-dependent entropy flows have been observed in in vivo models\cite{ishihara25}, while ATP consumption have been linked to the neural effects of aging\cite{shichkova25}.
In the theoretical space, ``mental energy'', emotions, and decision-making has been modeled in thermodynamic terms\cite{deli21}.\\
\\
A related body of work has sought to associate critical dynamics with various facets of cognition.
Driven by observation of neuronal avalanches in circuits\cite{beggs03} and the wider resting brain\cite{shriki13}, the ``critical brain hypothesis''\cite{jerbi22} and related SOC-hypothesis\cite{cocchi17} suggest that criticality may be a driver of these behaviors.
Hierarchical network organization has even been proposed (and empirically tested) as a driver of these dynamics\cite{friedman13,signorelli21}, while theoretical work on artificial neural networks (ANNs) has shown that network depth matching or exceeding width can generate power-law behavior, large fluctuations, and critical dynamics\cite{roberts22}.
This work exists among a broader body of literature which observe dynamical complexity, such as metastability, within the brain\cite{hellyer15,hancock22}.
Further, SOC specifically has been advanced as a driver of consciousness\cite{walter22}.\\
\\
A particularly relevant point when considering unification of IIT with thermodynamic models would be that the PCI's success appears to validate Kirkaldy's hypothesis\cite{kirkaldy65}.
More precisely, the PCI-inspired Fluctuation-Dissipation Theorem (FDT) theories of consciousness suggest that consciousness occurs when the brain is in a far-from-equilibrium state.
The PCI acts as empirical support, as well as other data which show the FDT is violated in conscious states\cite{buisan25}.\\
\\
Overall, an increasing body of literature seeks to understand the brain in thermodynamic terms, and conceptual links are to be found both with predictive processing\cite{friedman13,signorelli21} and IIT\cite{buisan25}.
Relatedly, empirical and theoretical work both suggest that criticality --- which can be generated by hierarchy\cite{friedman13,signorelli21,roberts22} --- is relevant for understanding cognition.
Were IIT, FEP, or any other theory of consciousness to adopt a thermodynamic lens, new metrics such as ATP/ADP-ratios\cite{buxton25} in specific regions might be predicted, which could aide falsifiability of the framework.

\section{Convergence of IIT, FEP, and thermodynamic models of cognition}\label{1.2}

Before delving into the body of prior work, we shall briefly address the plausibility and desirability of a unified model --- which depends largely on what we mean by such a term.
At this stage of discovery, one would hope it is uncontroversial to suggest that reaching an entirely correct, complete account of human cognition is exceedingly unlikely, given the vast complexity of the brain.
This perspective is arguably reflected in IIT's commitment to continuous self-critique and revision\cite{tononi04,tononi08,oizumi2014,albantakis2023}, as well as (arguably) FEP's openness to adaptation\cite{Friston10}.\\
\\
However, the intractability that any single model might wholly account for cognition, arguably makes a ``soft-unification'' very plausible.
The breadth\cite{Friston10}, adaptability\cite{tononi04,albantakis2023}, or qualitative nature\cite{buisan25} of existing theories makes them as unlikely to be wholly incorrect as correct.
Thus, in addition to finding points of difference to assess their relative strengths\cite{consortium25,intrepid26}, one might also take interest in their points of convergence.
If multiple distinct theories point toward similar conclusions, this only strengthens our ability to account for the human brain.
Further, modeling how these frameworks interact with each other may help us to develop more sophisticated predictions, or to strengthen our mathematical models.

\subsection{Prior comparative work}\label{1.2.1}

While FEP and IIT differ in first principles, a growing body of literature seeks to assess how they relate to each other\cite{safron20,safron22,mayama25}, while adversarial collaborations test their differences\cite{intrepid26}.\\
\\
Empirical evidence suggests a mathematical relationship.
Studies on animats have demonstrated that interaction of an agent with a complex environment can generate $\Phi$-producing structures\cite{albantakis14}, while follow-up has observed that IIT's core metric $\Phi$ fluctuates alongside sensory surprisal\cite{olesen23}.
More recently, experiments on in vitro neurons demonstrated a strong correlation between $\Phi$ and Bayesian surprise (Spearman's $\rho = 0.879$), indicating that updates to a generative model may drive the development of $\Phi$\cite{mayama25}, echoing Schr\"{o}dinger's account of consciousness and learning.\\
\\
These experiments have built on conceptual unifications, such as Integrated World Modeling Theory (IMWT) which proposes that active inference drives internal resistance against entropy and subsequently generates integrated information.
Variational Autoencoders, Safron proposes, drive the hierarchical modeling while limiting complexity\cite{safron20,safron22}.
Further comparative work includes Markovian Monism, which suggests the interior-exterior separation imposed by Markov blankets underwrites the mind's perception of itself as separate to the physical world.
It is suggested from here that the axioms of IIT can be inferred from active inference processes\cite{friston20}.\\
\\
With regard to thermodynamics, dualisms between FEP and CMEP problems are already identified in the literature\cite{gottwald20,sakthivadivel23,maxwell23}, while the PCI provides a thermodynamic bridge to IIT\cite{buxton25}.
Empirical investigations have found that $\Phi$ (or its proxy measures\cite{tegmark2016}) acts as a signature of both information theoretic and dynamical complexity, including metastability, criticality, and distributed computation\cite{mediano19}.\\
\\
Thus, while the details of IIT, FEP, and thermodynamic accounts each differ\cite{safron20,intrepid26}, substantial overlap across each can be found.
Unifying theories often leverage principles of efficiency\cite{friston20} and entropic deviation\cite{safron20,safron22} to provide a functional bridge between the theories.

\section{The MaxCal framework and its key results}\label{1.3}

\subsection{Methodological overview}\label{1.3.1}

To formalize the mathematics of IIT, we introduce \textit{transition networks,} which are Dynamic Bayesian Networks (DBNs) consisting of causal inputs $\bm{X}^t$ and effective outputs $\bm{X}^{t+1}$, governed by a transition probability matrix $\bm{P}$.
We show in chapter \ref{4_info_dev} that IIT's assessment of integration is equivalent to applying a MaxCal ensemble over a transition network and then comparing it to some constrained version.
Isolating a subsystem corresponds to conditioning on exterior input nodes, while partitioning sets them to persistently occupy MaxCal ensembles.
In IIT 3.0, \textit{cause/effect repertoires} are compared against unconstrained variants via the Wasserstein metric\cite{oizumi2014}.
We generalize our description so that other measures of distance may be applied.\\
\\
In section \ref{5.2} of chapter \ref{5_IIT_FEP} we generalize this approach first to non-deterministic Markovian (termed \textit{single-step}) systems, before examining those over longer paths.
The longer paths entail a central limit (CLT) approximation as well as a large deviations (LDP) view.\\
\\
Our perspective on information impacts our mathematics.
The question of whether probability distributions are ontic or epistemic\cite{floridi10} impacts how we should handle them.
Further, we must also assess whether ``information'' lives inside a static state $\bm{X}^0 = \bm{x}^0$, or whether it is something which \textit{moves.}\\
\\
Ultimately, our single-step view is most compatible with an ontic, mobile interpretation of information, while our CLT and LDP views are more compatible with modeling probability distributions epistemically.
In the latter case, it is legitimate to posit whether we are calculating an intrinsic information measure or approximating it.\\
\\
With regard to FEP, we connect each of our models to its structure by understanding the system $\bm{X}$ as an interior of some system, being fed sensory data from some blanket $\bm{O}$.
In the single-step view, ``information'' is defined with respect to probability distributions which are marginal on blanket paths $\bm{o}^t \rightarrow \bm{o}^{t+1}$.
Under the CLT and LDP views, ``information'' constitutes an internal path $\bm{x}^0, \bm{x}^1, \ldots, \bm{x}^T$ which deviates from internal predictions due to partial dependence on blanket states.
In each case, ``information'' can be understood as a prediction error of some kind and is equivalent to Bayesian surprise under our CLT view.\\
\\
Following this, in chapter \ref{6_complexity} we examine single-step regimes which produce information, both theoretically and empirically.
In doing this, we link our metric to hallmarks of dynamical complexity.

\subsection{Mathematical results}\label{1.3.2}

The generalized unconstrained cause distribution which we derive is a Gibbs distribution in which conditional entropy $h( \bm{x}^t ) = \mathcal{H}( \bm{X}^{t+1} \vert \bm{X}^t = \bm{x}^t )$ acts in negated form as energy. When comparing the difference in path entropy between networks $\mathcal{G}$ which are initiated by $\mu$, and those which are initiated by some other input distribution $\rho$, we retrieve that ``information'' may be defined as KL divergence between the two.\\
\\
The MaxCal marginal $\mu$ may be interpreted as imposing a uniform distribution over the ``effective branches'' within the transition space, while ``information'' may be interpreted as the log-ratio of branching space lost due to initiating with $\rho$ instead of $\mu$.
For integration $\phi^{\mathcal{P}}( \rho )$, we assess whether path space has been compressed more across the ``in-tact'' mechanism or its constituent parts.
This leaves open the potential for something we shall term ``disintegration'', which simply means that fluctuations are ``felt'' more acutely by constituent parts than by the organized whole.\\
\\
When we apply these definitions to the IIT algorithm\cite{tull21}, we retrieve that a conscious system should have many interacting ``parts'' which combine to form greater multi-element ``parts'', by deviating further from their branching potential together than they each do separately.
Each mechanism should form currents which deviate maximally from baseline MaxCal paths, and the combined volume of fluctuations within the in-tact transition networks should be greater than that over any partitioned ``cut system''\cite{oizumi2014}.\\
\\
Over longer timescales (sections \ref{5.4} and \ref{5.5}), we apply our methods to systems representing generative models of an external world.
Predictions ($\pi$ at the central limit and $\pi \times \bm{P}$ under large deviations) are generated from internal parameters, while realized states $\rho$, $\Gamma$ are partially dependent on sensory signals $(\bm{o})_{t=0}^T$.
We quantify ``information'' via some sufficient statistic which contrasts realized values against predicted ones, and determine VFE by its standard formulations\cite{millidge21}.
Under CLT, we retrieve an equivalence between ``information'' and Bayesian surprise, while under our LDP view it constitutes an accuracy term.
Both of these results lend themselves to a perspective where integration increases model complexity, and can be induced by sensory surprisal.
These perspectives are complementary, rather than counteractive, because the LDP view sits one level ``below'' the CLT vision.\\
\\
From an FEP perspective, another result we retrieve is that the VFE functional our LDP perspective retrieves is an L2-regularized negative-log-likelihood loss function under LeCun's energy-based learning framework\cite{lecun06}.\\
\\
Turing our attention again toward single steps, we see that circulant matrices and generic random walks (GRWs) produce no information under our single-step framework.
Conversely, MERWs --- which exhibit traits such as metastability --- produce information under our framework.
Upon empirical investigation, we find that the sparsity of a transition matrix impacts the information it is likely to produce, with mildly sparse (a nontrivial minority of non-zero entries) TPMs producing the greatest quantities.

\section{Mathematical prerequisites}\label{1.4}

We assume familiarity with foundational probability theory, including the definition and structure of Markov chains.
We note that any irreducible, aperiodic, positive recurrent (including all finite) Markov chains will have a unique stationary distribution $\mu$ such that $\mu \bm{P} = \bm{P}$.
When denoting our transition probabilities $\mathbb{P}( \bm{X}^{t+1} = \bm{x}^{t+1} \vert \bm{X}^t = \bm{x}^t )$ we will use the notation $\bm{P}( \bm{x}^t, \bm{x}^{t+1} )$ to emphasize the matrix structure of $\bm{P}$.\\
\\
We will also assume familiarity with foundational information theory, including definitions of entropy $\mathcal{H}( \cdot )$, mutual information $\mathcal{I}( \cdot ; \cdot )$, conditional entropy $\mathcal{H}( \cdot \vert \cdot )$, and KL divergence $\mathcal{D}( \cdot \lvert \rvert \cdot )$.
The Wasserstein metric will be referred to in this paper as it is used in IIT\cite{oizumi2014}, however we will not use it mathematically.\\
\\
More advanced mathematics, such as that of Large Deviations, may be leveraged in specific applications of our techniques.
While full understanding of the methods is not essential for following our techniques, references are provided for curious readers\cite{linsker88,linsker90,ricci21,touchette12,adams23,roberts22,sutton18}.

%% file: 2-IIT/2_IIT.tex
\chapter{IIT in finite, deterministic, discrete-time systems}\label{2_IIT}

\textsc{IIT begins} with ``intrinsic existence'' --- the idea that consciousness is internally observed and therefore the physical substrate perceives itself.
Mathematically, this means the system exercises cause/effect power internally, and not just over external objects/systems.
This is formalised by identifying \textit{mechanisms}, which are subsystems that act as a holistic object.
The way these mechanisms combine with each other is studied to assess system integration ($\Phi$).\\
\\
To this end, we identify precisely one ``core cause" and ``core effect" per mechanism, which combine to produce \textit{concepts}.
A \textit{constellation of concepts} aggregates these into one \textit{conceptual structure}.
\textit{Conceptual information} can then be calculated to assess information contained in the whole system (of interacting parts).
This process is then repeated with partitions of the system to assess redundancy of the whole.
The minimal distance between the whole system's constellation and that of some partition quantifies the irreducibility of the whole, which constitutes \textit{integrated conceptual information}, $\Phi_{\bm{X}}$.
The rationale here is that conscious entities must receive and generate information ``as one'', while highly separable systems (from a consciousness perspective) disintegrate into ``many conscious subsystems''.\\
\\
For any system $\bm{X}$, $\Phi_{\bm{X}}$ is a quantifier of cause-effect power of  the whole over and above parts.
However, some subsystem $\bm{Y} \subsetneq \bm{X}$ or supersystem $\bm{X} \subsetneq \bm{X}'$ may also produce a positive $\Phi$ value.
If one is using $\Phi$ to predict consciousness, one must ask which of these systems is?
To this, we apply the axiom of \textit{exclusion,} stating that each element $X_i$ can only exist as part of one conscious system at a time.
Thus, we select the system with a global maximum of $\Phi$\cite{oizumi2014}, which can be justified when we view this calculation as a discovery process.
If we accept IIT's ontology\cite{tononi23} that consciousness is a fundamental property of existence, then we are simply identifying something which already exists by examining where its fundamental features are most applicable.
This is markedly different from suggesting that consciousness somehow ``chooses'' to occupy the maximal system.
IIT's reversal of the inference process which drives the Hard Problem\cite{chalmers1995} is a fundamental driver of its mathematical methods\cite{oizumi2014,albantakis2023}.

\section{Basic Definitions}\label{2.1}

\begin{dfn}[System; State Space]
	A \textup{system} refers to a finite, discrete-time random process $\mathbb{X} = \{ \bm{X}^t : 0 \leq t \leq T \}$, where $\bm{X}^t = (X_1^t, \ldots, X_n^t)$ and $T \in \mathbb{N}$.
	The \textup{state space} $(\Omega_i, d_i)$ of $X_i$ is a finite metric space associated with the variable.
	We write $(\Omega_{\bm{X}}, d)$, which is the product of the spaces $(\Omega_1, d_1), \dots, (\Omega_n, d_n)$.
\end{dfn}

Throughout Chapter 2, the systems we study will be deterministic in nature.

\begin{dfn}[Subsystem]
	A \textup{subsystem} $\mathbb{Y} = \{ \bm{Y}^t : 0 \leq t \leq T \}$ is obtained by choosing specific entries of $\bm{X}$ over all values of $t$, and observing the resulting chain. Formally, $\bm{Y} = ( Y_1, \dots, Y_r ) = ( X_{i_1}, \dots, X_{i_r} ) $, where $r \leq n$ and $i_k \neq i_m$ whenever $k \neq m$.
\end{dfn}

\subsection{Objective and subjective time}\label{2.1.1}

Above, we discussed time in our systems from an extrinsic vantage point: referring to changes of state over a defined period, with respect to an external clock ``ticking''.
However, IIT models consciousness \textit{intrinsically.}
To capture this, we use $t=0$ to refer to the \textit{present moment;} $t=-1$ to refer to the \textit{immediate past}; $t=1$ to label the \textit{immediate future}.
We will refer to this perspective as \textit{subjective time,} because $\bm{X}$ will always perceive itself to be at $t=0$, regardless of how long it has existed from an external point of view.\\
\\
The units used to model \textit{subjective time} are canonical, meaning that whichever unit of time maximises \textit{rate of information integration}\cite{oizumi2014}.
It is intuitive to grasp this when considering differences in processing across species.
While humans can process light flickers at a maximum of around 60 Hz\cite{hecht36}, the \textit{Lucilia sericata} does so at a maximum rate of 180 Hz\cite{rück61}, indicating that visual information is processed at different speeds.
Thus, the timescales one would use when considering integration of visual information would be different depending on the species.

\subsection{Notation}\label{2.1.2}

The blackboard bold $\mathbb{X}$ shall refer to the underlying stochastic process, while regular bold $\bm{X}$ is an instance of the random variable.
Lower case $\bm{x}$ is an observation of $\bm{X}$.
Superscripts of $t$ will be used to index time.\\
\\
Bold letters other than ``$\bm{X}$'' (i.e. $\bm{Y}$, $\bm{Z}$; $\bm{y}$, $\bm{z}$) will typically refer to subsystems and their observations.
Occasionally, they might refer to supersystems.
Non-bold letters $X$, $x$ will represent single elements and their observations.\\
\\
We'll make use of the symbol $\bot$ to denote complements.
When $\bm{X}$ is the underlying system being studied, $\bm{Y}_{\bot}$ shall refer to $\bm{X} \setminus \bm{Y}$, while $\bm{Z}_{\bot \vert \bm{Y}}$ will denote $\bm{Y} \setminus \bm{Z}$.
The expression $\mathcal{P}(\bm{X})$ will represent the power set of $\bm{X}$, while $\mathcal{P}^{*}(\bm{X})$ will be the power set excluding the empty set $\emptyset$.\\
\\
For any system $\bm{X}$ (including subsystems) let $\Omega_{\bm{X}}$ be its state space.
When our system is a collection of nodes $\bm{X} = (X_1, X_2, \ldots, X_n)$, we may use $\Omega_i$ to represent the state space of some $X_i$.
Finally, for any state space $\Omega$ we shall let $\mathbb{P}( \Omega )$ represent the set of probability distributions over it.

\section{Cause and Effect Functions}\label{2.2}

Cause/effect are fundamentally about impact, and in information theory ``impact" refers to transference of information.
An event $\mathcal{A}$ causes an outcome $\mathcal{B}$ when it \textit{gives} information which increases its likelihood.
An outcome $\mathcal{B}$ is affected when it \textit{receives} signals which change its state or properties somehow.\\
\\
We can consider a system $\bm{X}$ evolving over time --- for now we look at its immediate past, $t=-1$, and its present $t=0$.
There are \textit{many} ways we can describe $\bm{X}$ from an extrinsic point of view.
For example, we may think of it as one whole system evolving over time.
Or, we could think of it as many separate elements changing state individually.
In general, we can choose to think of it as any pair of subsystems $\{ \bm{V}^{-1}, \bm{V}_{\bot}^{-1} \}$ acting on any other pair of subsystems $\{ \bm{Y}^0, (\bm{Y})^{0} \}$ separately.
In other words $\bm{V}^{-1}$ influences the state of $\bm{Y}^0$ and $\bm{V}_{\bot}^{-1}$ independently influences $\bm{Y}_{\bot}^0$.
This idea is illustrated below in Figure 2.1.

\begin{figure}[h]
	\centering
	\begin{tikzpicture}
		\node at (-3, 2.5) {$t=-1$};
		\node at (0, 2.5) {$t=0$};
		
		\draw[thin, dashed, gray] (-1.5, 2.3) -- (-1.5, -3.2);
		
		\definecolor{past1}{RGB}{66, 133, 244}   
		\definecolor{past2}{RGB}{154, 190, 247}  
		\definecolor{future1}{RGB}{234, 67, 53}  
		\definecolor{future2}{RGB}{251, 188, 184} 
		
		\draw [fill=past1, fill opacity=0.35, draw=past1, draw opacity=0.8] (-3,0) ellipse (0.5cm and 1.85cm);
		\draw [fill=past2, fill opacity=0.5, draw=past2, draw opacity=0.8] (-3, -2.5) ellipse (0.5cm and 0.6cm);
		
		\Vertex[color=darkgray, x=-3, y=1.25]{}
		\Vertex[color=darkgray, x=-3, y=0]{}
		\Vertex[color=darkgray, x=-3, y=-1.25]{}
		\Vertex[color=darkgray, x=-3, y=-2.5]{}
		
		\draw [fill=future1, fill opacity=0.35, draw=future1, draw opacity=0.8] (0, 0.625) ellipse (0.5cm and 1.25cm);
		\draw [fill=future2, fill opacity=0.35, draw=future2, draw opacity=0.8] (0, -1.875) ellipse (0.5cm and 1.25cm);
		
		\Vertex[color=darkgray, x=0, y=1.25]{}
		\Vertex[color=darkgray, x=0, y=0]{}
		\Vertex[color=darkgray, x=0, y=-1.25]{}
		\Vertex[color=darkgray, x=0, y=-2.5]{}
		
		\draw [-stealth, thick, darkgray] (-2.5, 0.2) -- (-0.5, 0.725);
		\draw [-stealth, thick, darkgray] (-2.5, -0.2) -- (-0.5, -1.725);
		\draw [-stealth, thick, darkgray] (-2.5, -2.4) -- (-0.5, 0.525);
		\draw [-stealth, thick, darkgray] (-2.5, -2.6) -- (-0.5, -2.025);
		
		\node at (-3.6, 1.4) {$\bm{V}$};
		\node at (-3.8, -2.5) {$\bm{V}_{\bot}$};
		\node at (0.7, 1.35) {$\bm{Y}$};
		\node at (0.85, -1.875) {$\bm{Y}_{\bot}$};
	\end{tikzpicture}
	\caption[Partitioned system]{Our system $\bm{X}$ is partitioned into subsystems $(\bm{V}, \bm{V}_{\bot})$ at $t=-1$ and $(\bm{Y}, \bm{Y}_{\bot})$ at $t=0$.}
	\label{fig01}
\end{figure}

We can also think of any part of our system as a ``true insider'' to its evolution, while thinking of any other part as extrinsic to the conscious part (assuming there is one) of $\bm{X}$.	
Our task is to identify which of these perspectives matters most, and in what way, to the system $\bm{X}$.\\
\\
To formalise this perspective, we define one cause function per ordered pair of subsystems $(\bm{V}, \bm{Y})$.
Each cause function takes the pairs of their non-empty subsets $\mathcal{P}(\bm{X}) \times \mathcal{P}(\bm{X})$ to a probability distribution $\mathbb{P}(\bm{X}^{-1})$.
Specifically, by choosing a cause function (defined below) $\zeta_{\bm{V}, \bm{Y}}$ we fix the sub-state $\bm{V}_{\bot}^{-1} = \bm{v}^{-1}$ as background conditions, while excluding the nodes $\bm{Y}^0_{\bot}$ from consideration of the present.
Meanwhile, $\bm{V}^{-1}$ and $\bm{Y}^0$ continue to be modeled as random variables.\\
\\
The purpose of applying our cause function is to partition the remaining subsystems.
Specifically, $\bm{V}^{-1}$ becomes $\{ \bm{U}^{-1}, \bm{U}_{\bot \vert \bm{V}}^{-1}  \}$ and $\bm{Y}^0$ becomes $\{ \bm{Z}^0, \bm{Z}_{\bot \vert \bm{Y}}^0 \}$.
We seek to understand the relationship that $\bm{U}^{-1}$ has on $\bm{Z}^0$ (and $\bm{U}_{\bot \vert \bm{V}}^{-1}$ on $\bm{Z}_{\bot \vert \b,{Y}}^0$). See Figure 2.2 for more information.

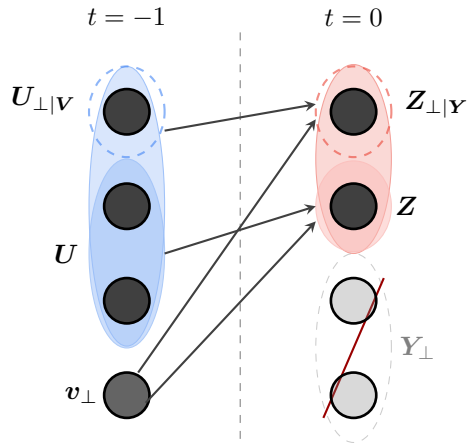
\begin{figure}[h]
	\centering
	\begin{tikzpicture}
		\node at (-3, 2.5) {$t=-1$};
		\node at (0, 2.5) {$t=0$};
		
		\draw[thin, dashed, gray] (-1.5, 2.3) -- (-1.5, -3.2);
		
		\definecolor{past1}{RGB}{66, 133, 244}
		\definecolor{past2}{RGB}{154, 190, 247}
		\definecolor{future1}{RGB}{234, 67, 53}
		\definecolor{future2}{RGB}{251, 188, 184}
		\definecolor{fixed}{RGB}{100, 100, 100}
		
		\draw [fill=past1, fill opacity=0.2, draw=past1, draw opacity=0.5] (-3,0) ellipse (0.5cm and 1.85cm);
		\draw [fill=past2, fill opacity=0.5, draw=past2, draw opacity=0.8] (-3, -0.625) ellipse (0.5cm and 1.25cm);
		\draw [draw=past1, draw opacity=0.6, dashed, thick] (-3, 1.25) ellipse (0.5cm and 0.6cm);
		
		\Vertex[color=darkgray, x=-3, y=1.25]{}
		\Vertex[color=darkgray, x=-3, y=0]{}
		\Vertex[color=darkgray, x=-3, y=-1.25]{}
		\Vertex[color=fixed, x=-3, y=-2.5]{}  
		
		\draw [fill=future1, fill opacity=0.2, draw=future1, draw opacity=0.5] (0, 0.625) ellipse (0.5cm and 1.25cm);
		\draw [fill=future2, fill opacity=0.5, draw=future2, draw opacity=0.8] (0, 0) ellipse (0.5cm and 0.6cm);
		\draw [draw=future1, draw opacity=0.6, dashed, thick] (0, 1.25) ellipse (0.5cm and 0.6cm);
		\draw [draw=gray, dashed, opacity=0.4] (0, -1.875) ellipse (0.5cm and 1.25cm);
		\draw [thick, red!60!black] (-0.4, -2.8) -- (0.4, -0.95);  
		
		\Vertex[color=darkgray, x=0, y=1.25]{}
		\Vertex[color=darkgray, x=0, y=0]{}
		\Vertex[color=gray, opacity=0.3, x=0, y=-1.25]{}
		\Vertex[color=gray, opacity=0.3, x=0, y=-2.5]{}
		
		\draw [-stealth, thick, darkgray] (-2.5, 1) -- (-0.5, 1.35);
		\draw [-stealth, thick, darkgray] (-2.5, -0.625) -- (-0.5, 0);
		\draw [-stealth, thick, darkgray] (-2.85, -2.2) -- (-0.5, 1.15);
		\draw [-stealth, thick, darkgray] (-2.75, -2.6) -- (-0.5, -0.2);
		
		\node at (-4.1, 1.4) {$\bm{U}_{\bot \vert \bm{V}}$};
		\node at (-3.8, -0.625) {$\bm{U}$};
		\node at (-3.6, -2.5) {$\bm{v}_{\bot}$};
		\node at (1.1, 1.35) {$\bm{Z}_{\bot \vert \bm{Y}}$};
		\node at (0.7, 0) {$\bm{Z}$};
		\node[opacity=0.5] at (0.85, -1.875) {$\bm{Y}_{\bot}$};
	\end{tikzpicture}
	\caption[Marginalized nodes (cause function)]{Applying the cause function $\zeta_{\bm{V}, \bm{Y}}$: $\bm{V}_{\bot}^{-1}=\bm{v}_{\bot}$ becomes a background condition, $\bm{Y}^0_{\bot}$ is discarded, and the remaining subsystems are partitioned.}
	\label{fig02}
\end{figure}

Before defining our cause function, we must understand the \textit{unconstrained cause repertoire}.
This represents the least knowledge a system can have of its past --- what it would perceive if all information exchange from past to present was disrupted.
Due to the Markovian nature of $\bm{X}$, this yields uniform distributions.

\begin{dfn}[Unconstrained Cause Repertoire]
	For a non-empty subsystem $\mathbb{V}$, the \textup{unconstrained cause repertoire} $\mathbb{P}^{\textup{uc}}( \bm{V}^{-1} )$ of a subsystem $\bm{V} \subseteq \bm{X}$ is simply a uniform distribution over $\Omega_{\bm{V}}$.
	For the empty set $\emptyset$, we have $\mathbb{P}^{\textup{uc}}(\emptyset) = 1$.
	For concision, we will $\mathbb{P}^{\textup{uc}}(\emptyset)$ not as the value of a standard $\mathbb{P}(\emptyset) = 0$, but as $\mathbb{P}^{\textup{uc}}(\emptyset)$.
\end{dfn}

Notice, we set $\mathbb{P}^{\textup{uc}}(\emptyset)$ not as the standard value in a probability distribution, $\mathbb{P}(\emptyset) = 0$, but as $\mathbb{P}^{\textup{uc}}(\emptyset) = 1$.
This is purely for notational purposes and holds no mathematical significance.

\begin{dfn}
	The \textup{cause function} $\zeta_{\bm{V}, \bm{Y}}$ of an ordered pair $(\bm{V}, \bm{Y})$ of subsystems, for a system $\mathbb{X}$ for which $\bm{X}^{-1} = \bm{x}^{-1}$ and $\bm{X}^0 = \bm{x}^0$, is a function from the domain $\mathcal{P}^{*}(\bm{V}) \times \mathcal{P}^{*}(\bm{Y})$ to the co-domain $\Pi(\Omega_{\bm{V}})$ defined by the following rules:
	\begin{enumerate}
		\item $\zeta_{\bm{V}, \bm{Y}} (\bm{V}, \bm{Z} \vert \bm{x}^{-1}, \bm{x}^0 ) = \zeta_{\bm{V}, \bm{Z}} (\bm{V}, \bm{Z} \vert \bm{x}^{-1}, \bm{x}^0 )$;
		\item $\zeta_{\bm{V}, \bm{Y}}(\bm{U}, \bm{Y} \vert \bm{x}^{-1}, \bm{x}^0 ) = \mathbb{P} (\bm{U}^{-1} \vert \bm{Y}^0 = \bm{y}^0, \bm{V}_{\bot}^{-1} = \bm{v}_{\bot}^{-1} ) \cdot \mathbb{P}^{\textup{uc}} (\bm{U}_{\bot \vert \bm{V}}^{-1})$;
		\item Otherwise, $\zeta_{\bm{V}, \bm{Y}}( \bm{U}, \bm{Z} \vert \bm{x}^{-1}, \bm{x}^0 ) =  \mathbb{P}( \bm{U}^{-1} \vert \bm{Z}^0 = \bm{z}^0, \bm{V}_{\bot}^{-1} = \bm{v}_{\bot}^{-1} ) \mathbb{P}( \bm{U}_{\bot \vert \bm{V}}^{-1} \vert \bm{Z}_{\bot \vert \bm{Y}}^0 = \bm{z}_{\bot \vert \bm{Y}}^0, \bm{V}_{\bot}^{-1} = \bm{v}_{\bot}^{-1} )$.
	\end{enumerate}
\end{dfn}

The effect functions and unconstrained effect repertoire are defined similarly.
We choose an ordered pair $(\bm{Y}, \bm{V})$ where is the subsystem we examine at $t=0$ and $\bm{V}$ is examined at $t=1$.
$\bm{Y}_{\bot}^0 = \bm{y}_{\bot}^0$ is treated as a background condition, and the nodes $\bm{V}_{\bot}^1$ are discarded because (due to Markovian properties) they do not materially impact the other nodes.
The unconstrained effect repertoire represents our knowledge a system's future state would have if all information sent about its present state was disrupted.
We note that in discrete-time, deterministic, finite-state systems it is not a uniform distribution due to forward-flow of information through time.

\begin{figure}[h]
	\centering
	\begin{tikzpicture}
		\node at (-3, 2.5) {$t=0$};
		\node at (0, 2.5) {$t=1$};
		
		\draw[thin, dashed, gray] (-1.5, 2.3) -- (-1.5, -3.2);
		
		\definecolor{present1}{RGB}{234, 67, 53}   
		\definecolor{present2}{RGB}{251, 188, 184} 
		\definecolor{future1}{RGB}{52, 168, 83}   
		\definecolor{future2}{RGB}{129, 201, 149}  
		\definecolor{fixed}{RGB}{100, 100, 100}
		
		\draw [fill=present1, fill opacity=0.2, draw=present1, draw opacity=0.5] (-3, 0.625) ellipse (0.5cm and 1.25cm);
		\draw [fill=present2, fill opacity=0.5, draw=present2, draw opacity=0.8] (-3, 0) ellipse (0.5cm and 0.6cm);
		\draw [draw=present1, draw opacity=0.6, dashed, thick] (-3, 1.25) ellipse (0.5cm and 0.6cm);
		\draw [draw=gray, dashed, opacity=0.6] (-3, -1.875) ellipse (0.5cm and 1.25cm);
		
		\Vertex[color=darkgray, x=-3, y=1.25]{}
		\Vertex[color=darkgray, x=-3, y=0]{}
		\Vertex[color=fixed, x=-3, y=-1.25]{}  
		\Vertex[color=fixed, x=-3, y=-2.5]{}   
		
		\draw [fill=future1, fill opacity=0.2, draw=future1, draw opacity=0.5] (0,0) ellipse (0.5cm and 1.85cm);
		\draw [fill=future2, fill opacity=0.5, draw=future2, draw opacity=0.8] (0, -0.625) ellipse (0.5cm and 1.25cm);
		\draw [draw=future1, draw opacity=0.6, dashed, thick] (0, 1.25) ellipse (0.5cm and 0.6cm);
		\draw [draw=gray, dashed, opacity=0.4] (0, -2.5) ellipse (0.5cm and 0.6cm);
		\draw [thick, red!60!black] (-0.4, -2.95) -- (0.4, -2.05);  
		
		\Vertex[color=darkgray, x=0, y=1.25]{}
		\Vertex[color=darkgray, x=0, y=0]{}
		\Vertex[color=darkgray, x=0, y=-1.25]{}
		\Vertex[color=gray, opacity=0.3, x=0, y=-2.5]{}
		
		\draw [-stealth, thick, darkgray] (-2.5, 1.25) -- (-0.5, 1.25);
		\draw [-stealth, thick, darkgray] (-2.5, 0) -- (-0.5, -0.625);
		\draw [-stealth, thick, darkgray] (-2.5, -1.45) -- (-0.5, 1);
		\draw [-stealth, thick, darkgray] (-2.5, -2.2) -- (-0.5, -1.25);
		
		\node at (-4.1, 1.35) {$\bm{Z}_{\bot \vert \bm{Y}}$};
		\node at (-3.7, 0) {$\bm{Z}$};
		\node at (-3.7, -1.875) {$\bm{y}_{\bot}$};
		\node at (1.1, 1.4) {$\bm{U}_{\bot \vert \bm{V}}$};
		\node at (0.7, -0.625) {$\bm{U}$};
		\node at (0.85, -2.5) {$\bm{V}_{\bot}$};
	\end{tikzpicture}
	\caption[Marginalized nodes (effect function)]{Applying the effect function $\atez_{\bm{Y}, \bm{V}}$: $\bm{Y}^0_{\bot}=\bm{y}_{\bot}$ becomes a background condition, $\bm{V}^1_{\bot}$ is discarded, and the remaining subsystems are partitioned.}
	\label{fig03}
\end{figure}

\begin{dfn}[Unconstrained Effect Repertoire]
	The \textup{unconstrained effect repertoire} of a subsystem $\bm{V}$ with respect to subsystem $\bm{Y}$ (for a system $\mathbb{X}$ which has been observed as $\bm{X}^0 = \bm{x}^0$), is the probability distribution of $\bm{V}^1$ given that $\bm{Y}^0$ is uniformly distributed and $\bm{Y}_{\bot}^0 = \bm{y}_{\bot}^0$.
	We write $\mathbb{P}^{\textup{uc}}(\bm{V}^1 \vert \bm{Y}_{\bot}^0 = \bm{y}_{\bot}^0)$ to denote the unconstrained effect repertoire for $\bm{V}$ with respect to $\bm{Y}$.
	We often use a ``pure" unconstrained effect repertoire $\mathbb{P}^{\textup{uc}}(\bm{V}^1)$ of a subsystem $\bm{V}$, which is the probability distribution of $\bm{V}^1$ conditioned on $\bm{X}^0$ being uniform.
\end{dfn}

\begin{dfn}[Effect Function]
	We define the \textup{effect function} $\atez_{\bm{Y}, \bm{V}}$ of an ordered pair $(\bm{V}, \bm{Y})$, for a system $\mathbb{X}$ for which $\bm{X}^0 = \bm{x}^0$, as a function with domain $\mathcal{P}^* (\bm{Y}) \times \mathcal{P}^* (\bm{V})$ and codomain $\Pi ( \Omega_{\bm{V}} )$ such that:
	\begin{enumerate}
		\item $\atez_{\bm{Y}, \bm{V}}( \bm{Y}, \bm{U} \vert \bm{x}^0 ) = \mathbb{P}( \bm{U}^1 \vert \bm{Y}^0 = \bm{y}^0, \bm{Y}_{\bot}^0 = \bm{y}_{\bot}^0 ) \cdot \mathbb{P}^{\textup{uc}} ( \bm{U}_{\bot \vert \bm{V}}^1 \vert \bm{Y}_{\bot}^0 = \bm{y}_{\bot}^0 )$;
		\item $\atez_{\bm{Y}, \bm{V}}( \bm{Z}, \bm{V} \vert \bm{x}^0 ) = \mathbb{P}( \bm{V}^1 \vert \bm{Z}^0 = \bm{z}^0, \bm{Y}_{\bot}^0 = \bm{y}_{\bot}^0 )$;
		\item $\atez_{\bm{Y}, \bm{V}}( \bm{Z}, \bm{U} \vert \bm{x}^0 ) = \mathbb{P}( \bm{U}^1 \vert \bm{Z}^0 = \bm{y}^0, \bm{Y}_{\bot}^0 = \bm{y}_{\bot}^0 ) \cdot \mathbb{P}( \bm{U}_{\bot \vert \bm{V}}^1 \vert \bm{Z}_{\bot \vert \bm{Y}}^0 = \bm{z}_{\bot \vert \bm{Y}}^0, \bm{Y}_{\bot}^0 = \bm{y}_{\bot}^0 )$ otherwise.
	\end{enumerate}
\end{dfn}

\subsection{Interpretation}\label{2.2.1}

The unconstrained cause and effect repertoires are designed to represent the system's knowledge about its causes/effects were it to have access to minimal information, which in this case is its causal structure and uniform perturbations\cite{oizumi2014}.
Another is to imagine the system $\bm{X}$ is in a state of maximal uncertainty at $t=-1$, which we explore in later sections.
In deterministic, conditionally independent systems, these are both modeled by assigning each element a uniform distribution at $t=-1$.\\
\\
The cause/effect functions, introduced here, focus our attention on a particular trajectory of the system's evolution across time, and then allow us to investigate the independence of sub-transitions.\\
\\
The states $\bm{x}^{-1}$, $\bm{x}^0$, $\bm{x}^1$ represent the state of the system's past, present, and future.
Mathematically, they (as applicable) can be modeled as inputs to the cause and effect functions, or as constraints which the cause and effect functions are dependent on.
Here, we generally frame them as the latter.

\section{Using Cause and Effect Functions to Identify Mechanisms}\label{2.3}

We use the cause and effect functions to identify mechanisms within the system --- that is, subsystems which generate cause and effect information.
First order mechanisms are elements which are meaningfully affected by the past-state of the system, while meaningfully affecting its future-state.
Higher order mechanisms are multi-element subsets which integrate information within themselves (and thus can be claimed to act as single entities), while contributing both cause and effect information to the system as a whole\cite{oizumi2014}.

\subsection{First Order Mechanisms}\label{2.3.1}

Suppose $X_k$ is a single-element subsystem.
To ascertain how much cause-information it generates within the present-state $\bm{X}^0 = \bm{x}^0$, we compare our knowledge of the system's past-state generated by the present-state of $X_k$ (i.e. the conditional distribution of $\bm{X}^{-1}$ given that $X_k^0 = x_k^0$) against the na\"{i}ve perspective of the past ( i.e. the uniform distribution $\mathbb{P}^{\text{uc}}(\bm{X}^{-1})$).\\
\\
To state $\mathbb{P}^{\text{uc}}(\bm{X}^{-1} = \bm{x}^{-1} \vert X_k^0 = x_k^0)$ in terms of our cause function, we note that we are examining the whole system in the past-state but restricting our attention to $X_k$ in the present, meaning we choose the function corresponding to $(\bm{X}, X_k)$.
We examine the impact $X_k^0 = z_k^0$ has on \textit{all} of $\bm{X}^{-1}$, so we choose $\bm{X}, X_k$ as our inputs.
Formally, we can prove this by noting
\begin{equation}\label{eq2.3.1}
	\begin{split}
		\zeta_{\bm{X}, X_k} &= \mathbb{P} (\bm{X}^{-1} \vert X_k^0 = x_k^0, \emptyset^{-1} = \emptyset^{-1}) \cdot \mathbb{P}^{\textup{uc}} (\emptyset) \\
		&= \mathbb{P} (\bm{X}^{-1} \vert X_k^0 = x_k^0).
	\end{split}
\end{equation}
We will write $\zeta_k(X_k \vert \bm{x}^0)$ as a shorthand because the value does not depend on $\bm{x}^{-1}$, as we have eliminated background conditions by choosing $\bm{X}$ as our past-subsystem.\\
\\
Recall, the cause information is ascertained by comparing the two distributions:

\begin{dfn}[Cause Information (single-element subsystem)]
	The \textup{cause information} of $X^0_k = x_k^0$ is defined as,
	\begin{equation*}\label{dfn2.3.1}
		\textup{ci}(X_k \vert \bm{x}^0) = d( \zeta_k(X_k \vert \bm{x}^0) \lvert \rvert \mathbb{P}^{\text{uc}}(\bm{X}^{-1}) ),
	\end{equation*}
	where $d( \cdot \lvert \rvert \cdot )$ is the distance metric of our system $\mathbb{X}$.
\end{dfn}

Analogously, to ascertain whether $X_k^0 = x_k^0$ generates effect information, we need to compare the information $\bm{X}^1$ receives from $X_k^0 = x_k^0$ against the na\"{i}ve case.\\
\\
The na\"{i}ve distribution is the ``pure" unconstrained effect repertoire $\mathbb{P}^{\text{uc}}(\bm{X}^1)$.
To represent the distribution $\bm{X}^1 \vert X_k^0 = x_k^0$ as an effect function, we note that we are considering $X_k$ in the present while eliminating any background conditions, and examining $\bm{X}$ in the future.\\
\\
To eliminate background conditions, we choose $\bm{X}$ as our left-tuple.
$\bm{X}$ is clearly our right-tuple.
Next, we are considering the impact $X_k^0 = x_k^0$ has on $\bm{X}^1$, so we choose $(X_k, \bm{X})$ as our entries.
We can prove that this works by calculating,
\begin{equation}
	\begin{split}
		\atez_{\bm{X}, \bm{X}} (X_k, \bm{X} \vert \bm{x}^0) &= \mathbb{P} ( \bm{X}^1 \vert X_k^0 = x_k^0, \emptyset^0 = \emptyset^0 ) \\
		&= \mathbb{P} ( \bm{X}^1 \vert X_k^0 = x_k^0 ).
	\end{split}
\end{equation}

Similarly to before, we shall use $\atez_k ( X_k \vert \bm{x}^0 ) = \atez_{\bm{X}, \bm{X}} (X_k, \bm{X} \vert \bm{x}^0)$ as a shorthand.
Now, we can compare to define the effect information.

\begin{dfn}[Effect Information (single-element subsystem)]
	The \textup{effect information} of $X^0_k = x_k$ is defined as,
	\begin{equation}\label{2.6}
		\textup{ei}(X_k \vert \bm{x}^0) = d( \atez_k(X_k \vert \bm{x}^0) \lvert \rvert \mathbb{P}^{\textup{uc}}( \bm{X}^1 ) ),
	\end{equation}
	where $d( \cdot \lvert \rvert \cdot )$ is the distance metric used for distributions.
\end{dfn}

Consciousness is fundamentally a form of self-observation.
Thus, IIT states a subsystem must be both affected-by and affecting-of the system in order to contribute to consciousness.
To this end, we measure the extent to which a subsystem does both.

\begin{dfn}[Cause-Effect Information]
	The \textup{cause-effect information} generated by $X_k^0 = x_k^0$ when the system $\bm{X}^0$ is in state $\bm{x}^0$ is defined by,
	\begin{equation}\label{2.7}
		\textup{cei}(X_k \vert \bm{x}^0) = \min \{ \textup{ci}(X_k \vert \bm{x}^0), \textup{ei}(X_k \vert \bm{x}^0) \}.
	\end{equation}
\end{dfn}

Finally, we can define first order mechanisms\cite{oizumi2014}.

\begin{dfn}[First Order Mechanism]
	A single-element subsystem $X_k$ is a \textup{first order mechanism} of the system $\mathbb{X}$ in the state $\bm{X}^0 = \bm{x}^0$ if and only if $\textup{cei}(X_k \vert \bm{x}^0) > 0$.
\end{dfn}

\subsection{Higher Order Mechanisms}\label{2.3.2}

Groups of first order mechanisms \textit{can} combine to form higher order mechanisms, but how should we differentiate between when they're acting as one whole or when they're acting as a ``group of individuals"?\\
\\
``Whole'' subsystems will generate more information as a single object than as separate, individuated parts.
We capture this concept when defining integrated (cause/effect) information for higher order mechanisms.
If the subsystem generates positive integrated information, we consider it to be a higher order mechanism\cite{oizumi2014}.

\subsubsection{Integrated Cause Information}\label{2.3.2.1}

Here we are examining the impact $\bm{Y}$ has on itself, and so we choose our cause function to be $\zeta_{\bm{Y}, \bm{Y}}$.
Next, we note that when $\bm{A}, \bm{Z}$ are non-empty, proper subsystems of $\bm{Y}$ then the outputs of $\zeta_{\bm{Y}, \bm{Y}}$ are as follows:
\begin{eqnarray*}
	&\zeta_{\bm{Y}, \bm{Y}}(\bm{A}, \bm{Z} \vert \bm{x}^0) = \mathbb{P}(\bm{A}^{-1} \vert \bm{Z}^0 = \bm{z}^0) \mathbb{P}(\bm{A}_{\bot \vert \bm{Y}}^{-1} \vert \bm{Z}_{\bot \vert \bm{Y}}^0 = \bm{z}_{\bot \vert \bm{Y}}^0) \\
	&\zeta_{\bm{Y}, \bm{Y}}(\bm{Y}, \bm{Z} \vert \bm{x}^0) = \mathbb{P}(\bm{Y}^{-1} \vert \bm{Z}^0 = \bm{z}^0) \\
	&\zeta_{\bm{Y}, \bm{Y}}(\bm{A}, \bm{Y} \vert \bm{x}^0) = \mathbb{P}(\bm{A}^{-1} \vert \bm{Y}^0 = \bm{y}^0) \mathbb{P}^{\textup{uc}}(\bm{A}_{\bot \vert \bm{Y}}^{-1}).
\end{eqnarray*}

The distribution $\zeta_{\bm{Y}, \bm{Y}}(\bm{Y}, \bm{Y} \vert \bm{x}^0) = \mathbb{P}(\bm{Y}^{-1} \vert \bm{Y}^0 = \bm{y}^0)$ represents the knowledge mechanism $\bm{Y}$ has about its past given its current state.
Meanwhile, let $(\bm{A}^{\textup{min}}_{\textup{cause}}, \bm{Z}^{\textup{min}}_{\textup{cause}})$ refer to the element of $\mathcal{P}^*(\bm{Y}) \times \mathcal{P}^*(\bm{Y}) \setminus \{ (\bm{Y}, \bm{Y}) \}$ which minimizes the distance between $\zeta_{\bm{Y}, \bm{Y}}(\bm{A}, \bm{Z} \vert \bm{x}^0)$ and $\zeta_{\bm{Y}, \bm{Y}}(\bm{Y}, \bm{Y} \vert \bm{x}^0)$.
Now, we can define the cause-information of $\bm{Y}$ as a mechanism\cite{oizumi2014}:

\begin{dfn}[Integrated Cause-Information (Mechanism)]
	The \textup{integrated cause information of subsystem $\bm{Y}$ given its present-state $\bm{y}^0$} is defined to be:
	\begin{equation}\label{2.8}
		\phi^{\textup{cause}}_{\bm{Y}}(\bm{y}^0) = d( \zeta_{\bm{Y}, \bm{Y}}(\bm{Y}, \bm{Y} \vert \bm{x}^0) \lvert \rvert \zeta_{\bm{Y}, \bm{Y}}( \bm{A}^{\textup{min}}_{\textup{cause}}, \bm{Z}^{\textup{min}}_{\textup{cause}} \vert \bm{x}^0) )
	\end{equation}
\end{dfn}

\subsubsection{Integrated Effect Information}\label{2.3.2.2}

We, similarly, want to find $(\bm{Z}^{\textup{min}}_{\textup{eff}}, \bm{U}^{\textup{min}}_{\textup{eff}})$ which minimizes the distance between $\atez_{\bm{Y}, \bm{Y}}(\bm{Y}, \bm{Y} \vert \bm{x}^0)$ and $\atez_{\bm{Y}, \bm{Y}}(\bm{Z}, \bm{U} \vert \bm{x}^0)$.
We use this to define define\cite{oizumi2014}:

\begin{dfn}[Integrated Effect-Information (Mechanism)]
	The \textup{integrated effect information of subsystem $\bm{Y}$ given the present-state $\bm{y}^0$} is defined to be:
	\begin{equation}\label{2.9}
		\phi^{\textup{eff}}_{\bm{Y}}(\bm{y}^0) = d( \atez_{\bm{Y}, \bm{Y}}(\bm{Y}, \bm{Y} \vert \bm{x}^0) \lvert \rvert \atez_{\bm{Y}, \bm{Y}} (\bm{Z}^{\textup{min}}_{\textup{eff}}, \bm{U}^{\textup{min}}_{\textup{eff}} \vert \bm{x}^0 ) ).
	\end{equation}
\end{dfn}

\subsubsection{Combining Cause and Effect}\label{2.3.2.3}

As before, we calculate the integrated information \textit{within} a mechanism by combining cause and and effect:

\begin{dfn}[Integrated Information (Mechanism)]
	The \textup{integrated information} of a \textup{mechanism} $\bm{Y}$ in state $\bm{y}$ is defined to be,
	\begin{equation}\label{2.10}
		\phi_{\bm{Y}}(\bm{y}^0) = \min \{ \phi^{\textup{cause}}_{\bm{Y}}(\bm{y}^0), \phi^{\textup{eff}}_{\bm{Y}}(\bm{y}^0) \}.
	\end{equation}
\end{dfn}

If a subsystem has positive integrated information, then we discuss it as a single mechanism which impacts the entire system.
From here, we can define its cause information $\textup{ci}_{\bm{Y}}(\bm{y}^0)$ and effect information $\textup{ei}_{\bm{Y}}(\bm{y}^0)$ as we did in the case of first order mechanisms.\\
\\
However, we can go further.
An integrated \textit{system} is not merely a collection of integrated mechanisms.
We need to understand how these mechanisms relate so that the system functions ``as one".
For a system to observe itself as a single whole, the mechanisms integrate together (producing information larger than the sum of their parts), \textit{and} the integration of $\bm{X}$ needs to be stronger than that of competing systems\cite{oizumi2014}.

\section{Integration of Systems}\label{2.4}

Now that we can identify our mechanisms $\mathbb{M} = \{ \bm{M}_1, \dots, \bm{M}_N \}$ within our system $\bm{X}$, we need to assess how they combine together.\\
\\
In IIT, we attach a \textit{concept} to each mechanism.
This can be thought of as a vector containing data about the cause, effect, and overall volume of irreducible information contained across these $\bm{M}_i^0 = \bm{m}_i^0$.\\
\\
We collate all of the concepts to form a ``big picture'', which we refer to as a \textit{constellation}.
If we then create \textit{cut systems} for each partition of $\bm{X}$, each along with their own constellations, we can compare the information observed by our overall system to the information observed by some sum of its parts.\\
\\
If there is a partition $\bm{Y} \cup \bm{Y}_{\bot}$ which produces the same constellation (collation of concepts) as $\bm{X}$, then we can rule out the possibility that $\bm{X}$ is one single conscious system, since all of the consciousness-producing information is contained in two smaller parts.
However, if $\bm{X}$ does integrate system-level information this is still not necessarily sufficient for treating it as conscious, because separate conscious agents may integrate information with each other, or small parts of a larger, conscious system may integrate information internally.
So, we need a method of identifying ``where the mind sits.''\\
\\
We resolve this by stating a stating the \textit{conscious complex} (mind) is a local maximum of integrated information, meaning it integrates more system level information than any of its subsystems or \textit{supersystems}\cite{oizumi2014}.

\subsection{Identifying Core Causes and Effects}\label{2.4.1}

As a predecessor to identifying concepts, we identify \textit{core causes} and \textit{core effects}.
For $\bm{M} \in \mathbb{M}$, we can typically model a number of subsystems of $\bm{X}^{-1}$ as having ``caused'' $\bm{M}^0 = \bm{m}^0$, but for the purposes of IIT we must identify a ``main'' one.
The exclusion postulate states that it's the maximally informative subsystem.
In other words, the subsystem of $\bm{X}^{-1}$ which is measured to have the largest impact on $\bm{M}^0 = \bm{m}^0$ is identified as its \textit{core cause}.
Analogously, the subsystem of $\bm{X}^1$ which is measured to be most impacted by $\bm{M}^0 = \bm{m}^0$ is identified as its \textit{core effect}.\\
\\
Our first step is to extend integrated information to be defined \textit{between} mechanisms.
For each pair of subsystems $(\bm{V}^{-1}, \bm{M}^0)$, let $\phi^{\textup{cause}}_{\bm{V}, \bm{M}}$ be a function from $\mathcal{P}^*(\bm{V}) \times \mathcal{P}^*(\bm{M})$ to $\mathbb{R}^{\geq 0}$, conditional on background conditions $(\bm{x}^{-1}, \bm{x}^0) \in \Omega_{\bm{X}} \times \Omega_{\bm{X}}$, defined by
\begin{equation}
	\phi^{\textup{cause}}_{\bm{V}, \bm{M}} ( \bm{U}, \bm{N} \vert \bm{x}^{-1}, \bm{x}^0 ) = d ( \zeta_{\bm{V}, \bm{M}} ( \bm{V}, \bm{M} \vert \bm{x}^{-1}, \bm{x}^0 ) \lvert \rvert \zeta_{\bm{V}, \bm{M}} ( \bm{U}, \bm{N} \vert \bm{x}^{-1}, \bm{x}^0 ) )
\end{equation}
Our function, $\phi^{\textup{cause}}_{\bm{V}, \bm{M}}$ takes $(\bm{U}, \bm{N})$ to the difference between the cause function's output of the transition from $\bm{V}^{-1}$ to $\bm{M}^0$, and the cause function's output of this transition when each set is partitioned by $\bm{U}^{-1} \cup \bm{U}^{-1}_{\bot \vert \bm{V}}$ and $\bm{N}^0 \cup \bm{N}^0_{\bot \vert \bm{M}}$.
Like before, we can quantify how much of the transition from $\bm{V}^{-1}$ to $\bm{M}^0$ is truly unique to these mechanisms as-separate-from-their-constituent-parts by taking the minimum, resulting in a function $\phi^{\textup{cause}} \ : \ \mathcal{P}^*(\bm{X}^{-1}) \times \mathbb{M} \ \rightarrow \mathbb{R} $ taking the pair $( \bm{V}^{-1}, \bm{M}^0 )$ to the cause information $\min \{ \phi^{\textup{cause}}_{\bm{V}, \bm{M}} ( \bm{U}, \bm{N} ) : \bm{U} \subset \bm{V}, \bm{N} \subset \bm{M}, \ \text{and} \ (\bm{U}, \bm{N}) \neq (\bm{V}, \bm{M})  \}$.\\
\\
(This is, of course, all conditional on background conditions $\bm{x}^{-1}, \bm{x}^0$).\\
\\
Now, we are finally in a position to define the ``core cause'' of $\bm{M}^0 = \bm{m}^0$.

\begin{dfn}[Core Cause]
	Given that the system $\bm{X}^0$ is in the state $\bm{x}^0$, the \textup{core cause} of $\bm{M}^0 = \bm{m}^0$ (for some mechanism $\bm{M}$) is the subsystem $\bm{V}_{\bm{M}}^{\textup{cause}, -1}$ of $\bm{X}^{-1}$ which produces maximal integrated cause information. I.e,
	\begin{equation}
		\bm{V}_{\bm{M}}^{\textup{cause}} = \textup{argmax}_{\bm{V} \in \mathcal{P}^{*} (\bm{X})} \phi^{\textup{cause}} ( \bm{V}^{-1}, \bm{M}^0 \vert \bm{x}^0, \bm{x}^{-1} ).
	\end{equation}
\end{dfn}

The ``core effect'' of a mechanism is defined analogously.
For each ordered pair $(\bm{M}, \bm{V})$ belonging to $\mathcal{P}^* (\bm{X}^0 \times \mathcal{P}^* (\bm{X}^1))$, we define a function $\phi^{\textup{eff}}_{\bm{M}, \bm{V}}$ from $\mathcal{P}^* (\bm{M}) \times \mathcal{P}^* (\bm{V})$ (conditional on $\bm{x}^0 \in \Omega_{\bm{X}}$) to $\mathbb{R}^{\geq 0}$ representing the integrated effect information $\bm{V}^1$ receives from $\bm{M}^0$, relative to some partition:
\begin{equation}
	\phi^{\textup{eff}}_{\bm{M}, \bm{V}} ( \bm{N}, \bm{U} \vert \bm{x}^0 ) = d ( \atez_{\bm{M}, \bm{V}} ( \bm{M}, \bm{V} \vert \bm{x}^0 ) \lvert \rvert \atez_{\bm{M}, \bm{V}} ( \bm{N}, \bm{U} \vert \bm{x}^0 ) ).
\end{equation}
The overall integrated effect information $\bm{V}^1$ receives from $\bm{M}^0$ is the minimum across all of these partitions:
\begin{equation}
	\phi^{\textup{eff}} ( \bm{M}^0, \bm{V}^1 \vert \bm{x}^0 ) = \min \{ \phi^{\textup{eff}}_{\bm{M}, \bm{V}} ( \bm{N}, \bm{U} ) : \bm{N} \in \mathcal{P}^* (\bm{M}), \bm{U}  \in \mathcal{P}^* (\bm{V}), \ \textup{and} \ ( \bm{N}, \bm{U} ) \neq ( \bm{M}, \bm{V} )  \}.
\end{equation}
Thus, the core effect can be defined\cite{oizumi2014}.
\begin{dfn}[Core Effect]
	Given that the system $\bm{X}^0$ is in the state $\bm{x}^0$, the core effect of $\bm{M}^0 = \bm{m}^0$ for some mechanism $\bm{M}$ is the subsystem $\bm{W}_{\bm{M}}^{\textup{eff}, 1}$ of $\bm{X}^1$ which produces the most integrated effect information. I.e.
	\begin{equation}
		\bm{W}_{\bm{M}}^{\textup{eff}} = \textup{argmax}_{\bm{} \in \mathcal{P}^*(\bm{X})} \phi^{\textup{eff}} ( \bm{M}^0, \bm{V}^0 \vert \bm{x}^0 ).
	\end{equation}
\end{dfn}

\subsection{Concepts, Constellations, and Conceptual Information}\label{2.4.2}

\begin{dfn}[Concept]
	For a system $\bm{X}$ in state $\bm{x}^0$, the concept of mechanism $\bm{M}$ is an ordered triple $( \bm{V}_{\bm{M}}^{-1} , \bm{M}^0 , \bm{W}_{\bm{M}}^{1} )$ where $\bm{V}_{\bm{M}}^{-1} = \bm{V}_{\bm{M}}^{\textup{cause}}$ and $\bm{W}_{\bm{M}}^{1} = \bm{W}_{\bm{M}}^{\textup{eff}}$ are the core cause and core effect of $\bm{M}^0 = \bm{m}^0$ respectively.
\end{dfn}

\begin{dfn}[Integrated Information (Concept)]
	The \textup{integrated information} produced by a concept $c = ( \bm{V}_{\bm{M}}^{-1} , \bm{M}^0 , \bm{W}_{\bm{M}}^{1} )$ is the minimum of its integrated cause information and integrated effect information:
	\begin{equation}
		\phi(c \vert \bm{x}^0) = \min \{ \phi^{\textup{cause}} ( \bm{V}_{\bm{M}}^{-1}, \bm{M}^0 \vert \bm{x}^{-1}, \bm{x}^0 ) , \phi^{\textup{eff}} ( \bm{W}_{\bm{M}}^{1}, \bm{M}^0 \vert \bm{x}^0 ) \}
	\end{equation}
\end{dfn}

\begin{dfn}[Concept Repertoire]
	The \textup{concept repertoire} associated with a concept $c = ( \bm{V}_{\bm{M}}^{-1} , \bm{M}^0 , \bm{W}_{\bm{M}}^{1} )$ is the combined probability distribution of $\bm{X}^{-1}$ and $\bm{X}^1$ conditional on $\bm{M}^0 = \bm{m}^0$, when we consider $(\bm{V}_{\bm{M}}^{-1}, \bm{M}^0)$ and $(\bm{M}^0, \bm{W}_{\bm{M}}^1)$ to have been acting independently of their respective complements. I.e,
	\begin{equation}
		\textup{rep} (c) = \zeta_{\bm{X}, \bm{M}} ( \bm{V}_{\bm{M}}, \bm{M} \vert \bm{x}^0 ) \atez_{\bm{M}, \bm{X}} ( \bm{M}, \bm{W}_{\bm{M}} \vert \bm{x}^0 ).
	\end{equation}
\end{dfn}

\begin{dfn}[Constellation]
	The \textup{constellation} of a system $\bm{X}$ in a state $\bm{x}^0$ is a set
	\begin{equation}
		C(\bm{x}^0) = \{ (c, \textup{rep}(c)) : \text{$c$ is a concept of some mechanism} \}
	\end{equation}
	of each concept and its respective repertoire.
\end{dfn}
We can measure the distance between two concepts using an extended version of the Wasserstein metric --- of, for that matter, any metric we are choosing to model our system.
Measuring the distance between our constellation and the ``na{\"i}ve'' one will tell us how much information our system possesses.
However, to measure integrated information we must examine its partitions.\\
\\
For any partition $\bm{z} = ( \bm{Y}, \bm{Y}_{\bot} )$ recall that we can create a cut system $\bm{X}(z)$, which has the same nodes as $\bm{X}$ but no dependency relations across the partition.
Let $C^z (\bm{x}^0)$ represent the constellation of our cut system $\bm{X}(z)$ in the initial state $\bm{X}^0 (z) = \bm{x}^0$.
Let $D ( C(\bm{x}^0) \lvert \rvert C^z (\bm{x}^0) )$ represent the distance between $C(\bm{x}^0)$ and $C^z (\bm{x}^0)$\cite{oizumi2014}.
\begin{dfn}[Integrated Conceptual Information]
	The \textup{integrated conceptual information} of a system $\bm{X}$ in state $\bm{x}^0$ is the minimal difference between $C(\bm{x}^0)$ and $C^z (\bm{x}^0)$ across all partitions $z$ and their associated cut systems $\bm{X}(z)$. I.e,
	\begin{equation}
		\Phi ( \bm{X} \vert \bm{x}^0 ) = \min_{z \in \textup{Partitions}(\bm{X})} D ( C(\bm{x}^0) \lvert \rvert C^z (\bm{x}^0) ).
	\end{equation}
\end{dfn}

\subsection{Identifying the Subject in the System}\label{2.4.3}

Positive conceptual information tells us there is a subject within $\bm{X}$ which integrates information, but that doesn't mean $\bm{X}$ experiences itself as a single whole.
For example, our brains almost certainly integrate information with our whole bodies, but this does not automatically mean it is conscious.
\textit{Exclusion} postulates that this happens because (some region of) your brain integrates more information with it-with-your-body does.
We formalize this with the language of a \textit{core complex}\cite{oizumi2014}.

\begin{dfn}[Core Complex]
	A \textup{core complex} of a system $\bm{X}$ in the state $\bm{x}^0$ is a subsystem $\bm{Y}$ which generates a local maximum of integrated conceptual information.
	I.e. $\Phi ( \bm{Y}^0 \vert \bm{y}^0 )$ is greater than $\Phi ( \bm{Z}^0 \vert \bm{z}^0 )$ for any subsystem $\bm{Z}$ of $\bm{X}$ which shares elements with $\bm{Y}$.
\end{dfn}

\subsection{Interpretation}\label{2.4.4}

The way concepts combine to create system level information appears to be designed to embody how our minds combine information.
As you (the reader) glaze this dissertation, your brain combines light, shape, color and abstract concepts to confer meaning upon these written symbols.
You also hear, feel, and perhaps smell or taste things, and these amalgamate to form a cohesive experience.
This combinatorial process is what we model with \textit{integration.}\\
\\
Whether or not integration of system information (and being a local maximum of it) is enough to ``be conscious'' is a difficult question.
IIT states that yes, it is, because there are no other aspects of experience which aren't reducible to IIT's axioms, and thus the two can be identified\cite{oizumi2014,albantakis2023}.
However, IIT itself is a work in progress and the scientific approach for studying consciousness remains elusive\cite{kleiner21,hoel21}.

%% file: 3-bays/3_bays.tex
\chapter{Bayesian Methods in Theoretical Neurobiology}\label{3_bays}

\textsc{Helmholtz proposed} in 1867 that perception arises from unconscious inference\cite{pratt1926}.
Almost a century later, Barlow identified elimination of redundant signals as a motivating factor for this\cite{barlow61}.
In 1982, Marr proposed that perception entails recovering causes from sensory statistical noise\cite{marr82}.
Thereafter, the Helmholtz Machine was created, aiming to model human learning by inferring patterns from data through a stochastic generative model\cite{dayan95}.
Theoretical neuroscience then formally adopted Bayesian inference\cite{rao99}, while computer scientists advanced methods to perform it more efficiently\cite{jordan99,ghahramani2000,beal03}.
These synthesized, which resulted in a family of theories applying approximate Bayesian inference to the brain\cite{friston03,friston05,harrison05}.
A subset of these incorporated action, in addition to perception, as a core driver of the process\cite{friston11,friston16,pezzulo24}.\\
\\
Though related operationally and historically, these advancements can be split into specific threads.
\textit{Bayesian Brain Hypothesis} is a very general conjecture proposing that the brain performs Bayesian inference\cite{pratt1926,barlow61,marr82,friston03,friston05}.
\textit{Variational inference} names a family of algorithms which iterate on an approximate posterior to estimate Bayesian inference\cite{jordan99,ghahramani2000,beal03}.
\textit{Predictive coding} is best summarized as the suggestion that hierarchical gradient descent on prediction errors drives dynamics in the brain\cite{rao99,millidge21}.
In 2003, Friston proved that variational inference on Gaussian neural models could drive this\cite{friston03,friston05}.\\
\\ 
Friston's proof was part of a broader framework termed the \textit{Free Energy Principle (FEP),} which proposed that several competing frameworks in theoretical neurobiology could be unified by a hypothesis that the brain minimizes a functional termed free energy\cite{Friston10}.
Since then, FEP has taken a life of its own and inspired a process theory called \textit{Active Inference}, modeling biological\cite{friston13,pezzulo24} and physical\cite{parr19,maxwell23,friston23} self-organization\cite{friston11,pezzulo24}.
While employable as a framework to motivate predictive coding\cite{buckley17}, it is compatible with many other regimes\cite{parr19,daCosta22}.\\
\\
In the following chapter, we aim to provide the na\"{i}ve reader with a generalized understanding of these regimes, including commonalities and distinctions.
We start in section \ref{3.1} by introducing variational inference, then in section \ref{3.2} describe predictive coding in Rao and Ballard's original terms\cite{rao99}.
In section \ref{3.3} we describe the FEP framework before detailing its application to the process theory of active inference.

\section{Introduction to Variational Inference}\label{3.1}

In variational inference we begin with observables $O$ and some latent state $X$ which is hidden.
They are modeled to arise from the same generative model: $O, X \sim p$ with $p(o, x) = p(o \vert x) p(x)$ for any $(o, x)$ in state space $\Omega_O \times \Omega_X$.
Given the observation $O = o$, Bayes' Theorem provides us a formula to reverse-engineer the distribution of the latent variable:

\begin{equation*}
	p( x \vert o )
	=
	\dfrac{
		p( o \vert x ) p(x)
	}{
		\sum_{\hat{x} \in \Omega_{X}} p( o \vert \hat{x} )p( \hat{x} )
	}.
\end{equation*}

In line with our variational analysis and upcoming MaxCal framing, we note this is equivalent to calculating a Gibbs distribution with temperature $\beta = 1$ and energy levels $\bm{E}( x \vert o ) = - \log p(o \vert x) - \log p(x)$\cite{jaynes57,gottwald20}.
Irrespective, this method is very computationally expensive because it increases exponentially with the number of latent variables\cite{jordan99}.\\
\\
Thus, it becomes more computationally feasible to perform numerical approximation on some function $q(x; o)$ which approximates $p(x \vert o)$.
Mathematically, $q(x ; o)$ may be thought of as a distribution conditional on $O=o$, or equivalently as a series of distributions over $X$ parameterized by $o$.
Note though that the parameters $\theta$ trained will not depend on $o$.\\
\\
If $p$ belongs to an exponential family then approximating $p(x \vert o)$ is equivalent to approximating the sufficient statistics contained in the exponential.
From this frame, it is operationally coherent to equate approximating $p(x \vert o)$ with approximating $\ell(x \vert o) = \log p( x \vert o )$.
Depending on the properties of our observable $O$, there are two possible ways we might Taylor-expand this:

\begin{gather}
	\ell(x \vert o)
	=
	\ell( \hat{x} \vert o )
	+
	\ell'( \hat{x} \vert o )
	( x - \hat{x} )
	+
	\ell''( \hat{x} \vert o )
	( x - \hat{x} )^2
	+
	\ldots
	\\
	\ell(x \vert o)
	=
	\ell( \hat{x} \vert \hat{o} )
	+
	\nabla \ell( \hat{x} \vert \hat{o} )
	( (x, o) - (\hat{x}, \hat{o}) )^{\top}
	+
	( (x, o) - (\hat{x}, \hat{o}) )
	\nabla^2 \ell( \hat{x} \vert \hat{o} )
	( (x, o) - (\hat{x}, \hat{o}) )^{\top}
	+
	\ldots
\end{gather}

Thus, dirac-delta and normal approximations $q( x ; o )$ are commonly used\cite{millidge21}.
However, there are notable exceptions.\\
\\
To measure our approximation, we maximize an objective function referred to as the \textit{Evidence Lower Bound (ELBO):}

\begin{equation}
	\textup{ELBO}(q)
	=
	\mathbb{E} \left[
	\log p( o, x )
	\right]
	-
	\mathbb{E} \left[
	\log q( x \vert o )
	\right],
\end{equation}

In a machine learning context this provides a lower bound for the expected model evidence $\log p(O)$, while simultaneously minimizing expected KL divergence between $q(x; o)$ and $p(x \vert o)$\cite{blei17}.
While originally developed under the assumption the generative model is known\cite{jordan99,ghahramani2000,beal03}, techniques such as energy-based learning allow us to conduct it on generative models which are unknown\cite{lecun06}.
Further, methods such as Stochastic Gradient Variational Bayes (SGVB) allow one to parameterize and learn the generative model alongside $q$\cite{kingma14}.

\section{Predictive Coding as cognition}\label{3.2}

Rao and Ballard proposed in 1999 that visual stimuli is processed through hierarchical generative models\cite{rao99}.
The basic building blocks were an image $\bm{O}$ consisting of $n$ pixels, being represented in the cortex as hypothetical causes $\bm{X}^{(0)} \in \Omega_{X}^{(0)}$ filtered through a non-linear map $f$.

\begin{equation}
	\bm{O} = f(U^{(0)} \bm{X}^{(0)}) + \bm{N}^{(0)}.
\end{equation}

Here, $U^{(0)}$ is a matrix of weights and $\bm{N}^{(0)}$ is Gaussian noise (with variance $\Sigma_0$).
A higher level of causes $\bm{X}^{(1)}$ then predict $\bm{X}^{(0)}$, and so forth:

\begin{equation}
	\bm{X}^l = f(U^{(l+1)} \bm{X}^{(l+1)}) + \bm{N}^{(l+1)}.
\end{equation}

This relation is purported to continue to the level of depth the given cortex permits.
At layer $0$, it is suggested that the sum of noise terms $\bm{N}^{(0)}$, $\bm{N}^{(1)}$, (normalized by their respective variances) from ``bottom up'' and ``top down'' predictions which are minimized.
The learning method for this optimization process is gradient descent on weights $U^{(0)}$, $U^{(1)}$)\cite{rao99}.\\
\\
Four years later, Friston explained that this can be derived by performing an Expectation Maximization algorithm\cite{friston03}.
The crucial step is to understand $\bm{X}^{(0)} \sim \mathcal{N}( f( U^{(1)} \bm{x}^{(1)} ), \Sigma_1 )$ as a prior distribution over possible causes of sensory input, while $\bm{O} \sim \mathcal{N}( f( U^{(0)} \bm{X}^{(0)} ), \Sigma_0 )$ acts as a predictive distribution over the sensory data itself.
From here an approximated prior $q( x \vert o ; U^{(1)}, \bm{x}^{(1)} )$ (dependent on the same parameters as $p(x)$) can be constructed mathematically, taking the form of a $\mathcal{N}( m_q, \Sigma_q )$ distribution, with parameters:

\begin{gather}
	m_q
	=
	\arg \min_{\hat{\bm{x}}^{(0)}}
	(
	\bm{o}
	-
	f( U^{(0)} \hat{\bm{x}}^{(0)} )
	)^{\top}
	\Sigma_0^{-1}
	(
	\bm{o}
	-
	f( U^{(0)} \hat{\bm{x}}^{(0)} )
	)
	\\
	\Sigma_q
	=
	(
	J^{\top}
	\Sigma_0^{-1}
	J
	+
	\Sigma_1^{-1}
	)^{-1},
\end{gather}

where $J$ is the Jacobian of the joint model with respect to $x = \bm{X}^{(0)}$.
Gradient descent on $U^{(0)}$, $U^{(1)}$, as outlined by Rao and Ballard\cite{rao99}, implements an \textit{Evidence Lower Bound (ELBO)}\cite{jordan99} maximization with approximate posterior $q$.
The negated ELBO constitutes the free energy functional\cite{friston03}.\\
\\
The concept of a \textit{recognition distribution} was introduced, which drives the selection of $m_q$ given parameters $U^{(0)}$, $U^{(1)}$, $\bm{x}^{(1)}$ and models \textit{inference.}
Slower change over time on parameters $U^{(0)}$, $U^{(1)}$ is understood to constitute \textit{learning}\cite{friston03}.\\
\\
This framing was later generalized\cite{friston05,Friston10} and has since become a dominant way to frame predictive coding\cite{buckley17,millidge21}.
Given any generative model $p(o, x ; \theta) = p( o \vert x ; \theta)p(x ; \theta)$ and approximated posterior $q( o \vert x ; \lambda )$, \textit{predictive coding} now commonly refers to the hypothesis that the brain performs gradient descent on $\theta$, $\lambda$ to maximize the ELBO of $q$\cite{Friston10}.
However, variational Bayes is not strictly necessary for the original framework\cite{rao99}.
Emprical evidence appears to support the idea that predictive coding may occur in the brain\cite{caucheteux23}.

\subsection{Hierarchical Predictive Coding}\label{3.2.1}

Key to Predictive Coding is the idea that the generative model is \textit{layered}\cite{rao99,millidge21}.
Specifically, there are $L+1$ random layers $x_{0:L} =  x_0, x_1, \ldots, x_L$ forming a model $p( x_0, \ldots, x_L )$.
The $0^{\textup{th}}$ layer $x_0 \sim \bm{O}$ are pure sensory inputs predicted by $x_1$, while each subsequent $x_{l}$ will both make predictions about the preceding layer $x_{l-1}$ and be predicted the subsequent layer $x_{l+1}$.
The final layer $x_L$ is considered purely random, making predictions about $x_{l-1}$ which are then propagated down the network.\\
\\
When we assume Bayesian Network structure for dependencies, the generative model may be expressed as:

\begin{equation}
	p( x_{0:L} )
	=
	p( x_L )
	\prod_{l=0}^{L-1}
	p( x_l \vert x_{l+1} ).
\end{equation}

When we assume a Gaussian distribution $\mathcal{N}( \hat{m}_L, \Sigma_L )$ for layer $L$, and Gaussian priors $\mathcal{N}( f_l( U^{({l+1})}, x_{l+1}) \Sigma_l ) )$ for each preceding layer $x_0, \ldots, x_{l-1}$, then by assuming a dirac-delta distribution we may initialize a parameterized approximate posterior:

\begin{gather}
	q( x_L \vert x_{L-1} ) = \delta( x_L - m_L ) \\
	q( x_{L-1} \vert x_{L-2} ) = \delta( x_{L-1} - m_{L-1} ) \\
	\ \vdots \ \ \vdots \ \ \vdots \ \\
	q( x_2 \vert x_1 ) = \delta( x_2 - m_2 ) \\
	q( x_1 \vert x_0 ) = \delta( x_1 - m_1 ).	
\end{gather}

We may define an error vector of $\epsilon_L = (m_L - \hat{m}_L)$ for layer $L$ and $\epsilon_l = ( m_l - f_l(U^{(l+1)}, m_{l+1}) )$ for lower layers.
Placed into our ELBO formula, we retrieve expected model evidence:

\begin{equation}
	\textup{ELBO}(q)
	=
	-
	L \log 2\pi
	-
	\frac{1}{2}
	\sum_{l=1}^L \left(
	\epsilon_l^{\top}
	\Sigma_l^{-1}
	\epsilon_l
	+
	\log
	\lvert
	\Sigma_l
	\rvert
	\right),
\end{equation}

where $\lvert \Sigma_l \rvert$ is the determinant on $\Sigma_l$.
When lateral conditional independence across any components of each $x_l$ is assumed, the log term will simply be $2n_l \log \sigma_l$.
Gradient descent is then performed over $m_l$ to \textit{infer} the vector $( x_1, \ldots, x_L )$ which best predicts $x_0 \sim \bm{O}$.
Meanwhile, it is assumed that gradient descent happens at slower scales over $U^{(1)}, \ldots, U^{(L)}$ (and perhaps even over each $f_l$ if we imagine it as a linear combination of polynomials) to \textit{learn} the optimal generative model\cite{millidge21}.

\subsection{Generalization and assumptions}\label{3.2.2}

The precise nature of the approximate posterior can vary depending on the choice of model.
While some frameworks use Gaussian approximate posteriors\cite{friston03}, others use Dirac-delta distributions\cite{buckley17,millidge21}.
The priors in each case are Gaussian.
When considering other types of priors, one might need to consider other algorithms\cite{yao18} for modeling variational Bayes in the brain\cite{rezende15,miller16,yao18,castillo25}.\\
\\
Alternatively, one could consider classifying biological neural networks by considering how applicable contemporary predictive coding methods are to their dynamics.
Studies on artificial networks suggest that depth and width can impact the efficacy of variational approximations\cite{foong20,sheinkman25}.
Further, artificial networks can be classified by depth-width ratio, which impact how fluctuations propagate\cite{roberts22}.
If it is assumed predictive coding occurs on a neuronal level, further research leverage existing studies of ANNs to inform perspectives on BNNs.

\section{The Free Energy Principle and Active Inference}\label{3.3}

The reconciliation of predictive coding with variational Bayes was part of a broader effort to unify theories in computational neuroscience under a shared Free Energy Principle (FEP)\cite{friston03,friston05}.
While conceived as a meta-theory of neuroscience\cite{Friston10} and biological self-organization\cite{friston06}, FEP has evolved alongside a process theory active inference\cite{friston16,pezzulo24}.
The core argument is that when a permeable boundary exists between an interior and exterior of some system, then survival of the blanket depends upon accurate performance of inference\cite{bettinger23}.\\
\\
In continuous-time, this is currently formalized through Langevin equations over the whole system (internal, external, and blanket states\cite{parr19,friston23} as well as rules governing dependency relationships, which model the separation.
Discrete-time models employ Partially Observable Markov Decision Processes (POMDPs)\cite{daCosta20}.\\
\\
Though originally developed to model learning and homeostasis\cite{friston13,friston16}, active inference has since been applied to questions of ecology\cite{mirza16}, physics\cite{sakthivadivel22,sakthivadivel23,maxwell23,friston23}, consciousness\cite{friston20}, problem-solving behavior\cite{lin17}, and psychosis\cite{adams16}.
In artificial intelligence, Active Inference has been used to improve reinforcement learning methods\cite{millidge19,tschantz20} and to propose alternative algorithms\cite{pazem24}.
Mathematically, there is some comparison between Active Inference and Soft Actor-Critic methods\cite{haarnoja18}.\footnote{
	If hidden states, internal states, approximate posteriors, and the negated energy term from Active Inference are identified with actions, system states, policies, and rewards from Soft Actor-Critic, then minimizing the \textit{generalized free energy} of strange particles is equivalent to maximizing the objective function in Soft Actor-Critic methods.
	See the appendix for proof.
}
These two have been combined in the field of robotics to simulate curiosity in robots\cite{kawahara22}.
It has been proposed as a useful framework more broadly within the field\cite{daCosta22} and shown to predict the behavior of robots post-training\cite{pio-lopez16}.
More recently, it has been used to improve the adaptability of controllers\cite{pezzato20} and comprises an ongoing area of research\cite{baioumy22}.

\subsection{FEP as an organizing principle in Theoretical Neuroscience}\label{3.3.1}

Given a generative model $p(o, x) = p(o \vert x)p(x)$ and approximate posterior distribution $q$ with parameters $\lambda$, the \textit{free energy} associated with the observation $O=o$ is defined expected (with respect to $q(\cdot \vert o ; \lambda)$) log-ratio between the approximated posterior and joint model:

\begin{equation}
	\mathcal{F}[q]
	=
	\mathbb{E}_{q( \cdot \vert o ; \lambda )} \left[
	\log
	q( X \vert o ; \lambda )
	\right]
	-
	\mathbb{E}_{q( \cdot \vert o ; \lambda )} \left[
	\log
	p( o,  X )
	\right].
\end{equation}

It acts as an upper-bound on the KL divergence between the approximate posterior $q( x \vert o ; \lambda )$ and true posterior $p(x \vert o)$, and carries the same form as Helmholtz free energy used in thermodynamics\cite{friston03}.
Accordingly, we understand it as a decomposition of energy and entropy terms\cite{millidge21}:

\begin{equation}\label{freeEnHelmholtz}
	\mathcal{F}[q]
	=
	\underbrace{
		\mathbb{E}_{q( \cdot \vert o ; \lambda )} \left[
		-
		\log
		p( o,  X )
		\right]
	}_{\textup{Energy}}
	-
	\underbrace{
		\mathbb{E}_{q( \cdot \vert o ; \lambda )} \left[
		-
		\log
		q( X \vert o ; \lambda )
		\right].
	}_{\textup{Entropy}}
\end{equation}

Thus, \textit{free energy} admits a standard learning-based interpretation while also pointing toward conceptual influence from statistical mechanics methods\cite{friston03,friston05}.
The connection to statistical mechanics need not necessarily be a physical one\cite{gottwald20} --- Friston's early work emphasized it as information theoretic in nature\cite{Friston10}, and Jaynes himself framed entropy as a statistical concept applied to the study of physical systems\cite{jaynes57,jaynes80}.
However, since brains are (among other things) thermodynamic systems, hypotheses which interpret free energy physically are carefully explored in the literature\cite{friston23}.\\
\\
The key insight when this functional was introduced was the unification of several leading methods in theoretical neuroscience and computer science at the time\cite{friston03}.
Unsupervised and supervised learning were, rather straightforwardly, framed as minimizing the free energy functional.
Barlow's redundancy principle\cite{barlow61} was re-framed in terms of free energy optimization and found to yield the \textit{InfoMax} principle\cite{friston03}.
In later work, attention, perceptual learning, optimal control, and computational motor control, were also argued to arise from minimization of the free energy functional.
Thus, differences in context or assumptions appeared to drive a deeper, underlying law\cite{Friston10}.

\subsubsection{Understanding surprisal}\label{3.3.1.1}

The free energy functional can be decomposed into accuracy and complexity terms\cite{millidge21}:

\begin{equation}\label{freeEnComp}
	\mathcal{F}[q]
	=
	\underbrace{
		\mathcal{D}\left(
		q( \cdot \vert o ; \lambda )
		\lvert
		\rvert
		p( \cdot )
		\right).
	}_{\textup{Complexity}}
	-
	\underbrace{
		\mathbb{E}_{q( \cdot \vert o ; \lambda )} \left[
		\log
		p( o \vert  x )
		\right]
	}_{\textup{Accuracy}}
	.
\end{equation}

The complexity term is an approximation of \textit{Bayesian Surprise,} which constitutes divergence between posterior and prior beliefs\cite{itti09}.
Within the FEP framework, it relates to the cost of updating one's prior beliefs\cite{Friston10,friston11}.
Meanwhile, $\mathbb{E}_{q( \cdot \vert o, \lambda )}[ \log p( o \vert X ) ]$ quantifies model accuracy by assessing the extent to which states predicted by $q$ produce the results observed in the sensory data.
The free energy of $q( X \vert o, \lambda )$ is thus minimized as its complexity (i.e. Bayesian surprisal) decreases and as its accuracy increases.
FEP thus implies that self-organization and learning entails an exploration-exploitation trade-off\cite{Friston10,friston11}.\\
\\
Another decomposition observed in the literature positions free energy as an upper bound on the posterior approximation error\cite{millidge21}:

\begin{equation}
	\mathcal{F}[q]
	=
	\mathcal{D}(
	q( \cdot \vert o ; \lambda )
	\lvert
	\rvert
	p( \cdot \vert o )
	)
	-
	\log p(o).
\end{equation}

We note that each $- \log p(o)$ does not vary with $X$ or $q$ and is always non-negative due to $p(o)$ being greater than or equal to $1$ (for continuous models, we may scale to ensure this is the case).
Thus, for any value $o \in \Omega_O$ we see that $\mathcal{F}[q](o)$ reduces with divergence between the true and approximate posterior.\\
\\
When we consider the expectation over $p(o)$, we note that $- \log p(o)$ describes the \textit{sensory surprisal} of observation(s) $o \in \Omega_O$\cite{friston11} and thus $\mathbb{E}_{p(\cdot)}[ - \log p(O) ] = \mathcal{H}^p ( O )$ is a measure of expected sensory surprisal.
Meanwhile, $\mathbb{E}_{p(\cdot)} \left[ \mathcal{D}( q( \cdot \vert O ; \lambda ) \lvert \rvert p( \cdot \vert O ) ) \right]$ describes the expected observed approximation error of our generative model\cite{millidge21}.
Thus, while these quantities cannot be directly calculated, we understand that an effective $( p(X)p(O \vert X), q( X \vert O ; \lambda ) )$-pairing will trade off reducing error against reducing sensory surprisal.
In the discretized case, larger effective state space\footnote{
	See chapter \ref{5_IIT_FEP} for discussions on perplexity.
} of $o \in \Omega_O$ will generate higher expected surprisal, indicating a relationship between FEP and considerations of redundancy\cite{barlow61}, as well as concepts in complexity theory such as the Law of Requisite Variety\cite{ashby91}.

\subsection{FEP and Active Inference}\label{3.3.2}

The innovation of active inference arises from three central primitives:
1) brains/organisms/cells are agents, as well as observers;
2) action and observation are inextricably linked because they contribute to the same outcome (survival);
3) survival and homeostasis relate\cite{friston23}.
In fact, FEP might be best understood as a study of survival, with inference and learning arising incidentally due to environmental complexity.
Active inference details proposed survival mechanisms and requirements for the interior of a Markov blanket to be an NESS\cite{parr19}.

\subsubsection{Separation}\label{3.3.2.1}

The first condition identified is \textit{separation} --- the motivation for this being two-fold.
Firstly, for ``something'' to exist in an environment, a level of functional separation must exist.
Conditional independence is the mathematical tool used to model this\cite{daCosta20,friston23}.
Secondly, separation allows for the development of internal dynamics which are distinct from external ones\cite{parr19}.\\
\\
Markov blankets are used to model separation.
To this end, we represent the internal, blanket, and external states of our system $\mathbb{S} = \{ \bm{S}^t \vert t \in \mathbb{T} \}$:

\begin{equation}
	\bm{S}^t
	=
	(
	\eta^t
	,
	\omega^t
	,
	\alpha^t
	,
	\bm{x}^t
	),
\end{equation}

with $\eta^t$ representing environmental states, $\bm{x}^t$ internal states, and $(\omega^t, \alpha^t)$ blanket states comprised of sensory data $\omega^t$ and actions $\alpha^t$\cite{parr19}.
The precise nature of dependency relationships depends on the type of system being modeled, but in general the internal and external states are both conditionally independent with respect to blanket paths $(\omega^{[0, T]}, \alpha^{[0,T]})$\cite{friston23}:

\begin{equation}
	p(
	\eta^{[0, T]}
	,
	\bm{x}^{[0, T]}
	\vert
	\omega^{[0, T]}
	,
	\alpha^{[0, T]}
	)
	=
	p(
	\eta^{[0, T]}
	\vert
	\omega^{[0, T]}
	,
	\alpha^{[0, T]}
	)
	p(	
	\bm{x}^{[0, T]}
	\vert
	\omega^{[0, T]}
	,
	\alpha^{[0, T]}
	),	
\end{equation}

For discrete-time dynamics $t = \{ 0, 1, \ldots, T \}$, the internal states $\bm{x}$ contain \textit{beliefs} about $\eta$, but cannot directly observe it.
A chain of sensory data $\omega^{0:T} = ( \omega^1, \ldots, \omega^T )$ and latent environmental states $\eta^{0:T}$ are related by a \textit{likelihood matrix} $A$.
\textit{Policies} $\alpha^{0:T}$ describe a sequence of actions.
The following equation models their relationships\cite{daCosta20}:

\begin{equation}
	\mathbb{P}(
	\omega^{0:T}
	,
	\eta^{0:T}
	,
	A
	,
	\alpha^{0:T}
	)
	=
	\mathbb{P}( \alpha^{0:T} )
	\mathbb{P}( A )
	\mathbb{P}( \eta^0 )
	\prod_{t=1}^T
	\mathbb{P}(
	\eta^t
	\vert
	\eta^{t-1}
	,
	\alpha^{1:T}
	)
	\prod_{t=0}^T
	\mathbb{P}(
	\omega^t
	\vert
	\eta^t
	,
	A
	).
\end{equation}

Meanwhile, the probability distribution of $\bm{x}^{0:T}$ depends on $( \omega^{0:T}, \alpha^{0:T} )$\cite{daCosta20}.

\subsubsection{Free Energy}\label{3.3.2.2}

Within the active inference \textit{process,} there are three types of free energy employed, defined as follows\cite{friston23}:

\begin{gather}\label{freeEns}
	\mathcal{F}[ \omega^{\gamma}, \alpha^{\gamma}, \bm{x}^{\gamma} ]
	=
	\mathbb{E}_{q( \eta^{\Gamma} \vert \omega^{\gamma}, \alpha^{\gamma} ; \bm{x}^{\gamma} )} \left[
	-
	\log
	p(
	\eta^{\Gamma}
	,
	\omega^{\gamma}
	,
	\alpha^{\gamma}
	)
	\right]
	-
	\mathbb{E}_{q( \eta^{\Gamma} \vert \omega^{\gamma}, \alpha^{\gamma} ; \bm{x}^{\gamma} )} \left[
	-
	\log
	q(
	\eta^{\Gamma}
	\vert
	\bm{x}^{\gamma}
	)
	\right]
	;
	\\
	\mathcal{E}[ \alpha^{\gamma}, \bm{x}^{\gamma} ]
	=
	\mathbb{E}_{ p( \eta^{\Gamma}, \omega^{\Gamma} \vert \alpha^{\gamma}, \bm{x}^{\gamma} ) } \left[
	-
	\log
	p(
	\eta^{\Gamma}
	,
	\omega^{\Gamma}
	)
	+
	\log
	p(
	\eta^{\Gamma}
	\vert
	\alpha^{\gamma}
	,
	\omega^{\gamma}
	)
	\right]
	=
	- 
	\log
	p(
	\alpha^{\gamma}
	,
	\omega^{\gamma}
	)
	;
	\\
	\mathcal{G}[ \omega^{\gamma}, \bm{x}^{\gamma} ]
	=
	\mathbb{E}_{ q( \eta^{\Gamma}, \alpha^{\Gamma} \vert \bm{x}^{\gamma} ) } \left[
	-
	\log
	p(
	\eta^{\Gamma}
	,
	\omega^{\gamma}
	,
	\alpha^{\Gamma}
	,
	\bm{x}^{\gamma}
	)
	\right]
	-
	\mathbb{E}_{ q( \eta^{\Gamma}, \alpha^{\Gamma} \vert \bm{x}^{\gamma} ) } \left[
	-
	\log
	q(
	\eta^{\Gamma}
	,
	\alpha^{\Gamma}
	\vert
	\mu^{\gamma}
	)
	\right]
	.
\end{gather}

Here, we use the superscripts $\Gamma$ to refer to random path ensembles and $\gamma$ to refer to specific paths.\\
\\
The $\mathcal{F}[ \omega^{\gamma}, \alpha^{\gamma}, \bm{x}^{\gamma} ]$ is \textit{variational free energy (VFE),} describing dynamical constraints over the system.
Given observed boundary states $( \omega^{\gamma}, \alpha^{\gamma} )$, there is an induced internal path $\bm{x}^{\gamma} \vert ( \omega^{\gamma}, \alpha^{\gamma} )$ which parameterizes a posterior density $q( \eta^{\Gamma} \vert \omega^{\gamma}, \alpha^{\gamma} ; \bm{x}^{\gamma} )$ over environmental paths $\eta^{\Gamma}$.
The process of \textit{inference} would arises from the fact that the internal path $\bm{x}^{\gamma}$ which parameterizes $q$ minimizes (for specific $( \omega^{\gamma}, \alpha^{\gamma} )$ pairing) being chosen\cite{friston23}.\\
\\
$\mathcal{E}[ \alpha^{\gamma}, \bm{x}^{\gamma} ]$ represents the \textit{surprisal} of a specific action-state pairing $(\alpha^{\gamma}, \mu^{\gamma})$.
It trades off expected cost over sensory paths against expected information gain over external paths, and is therefore thought to optimize decision making.
It is referred to as \textit{expected free energy}\cite{friston23}.\\
\\
\textit{Generalized free energy (GFE)} $\mathcal{G}[ \omega^{\gamma}, \bm{x}^{\gamma} ]$ exists only in a specific subset of systems where active states $\alpha^{\Gamma}$ are hidden from internal states $\mu^{\gamma}$.
It is mathematically similar to VFE, but conceptualizes something different.
VFE describes something \textit{true} about the system, in the sense that (above a certain level of complexity) a $(\omega^{\gamma}, \alpha^{\gamma})$ pairing will induce a path $\bm{x}^{\gamma}$ which minimizes $- \log p( \omega^{\gamma}, \alpha^{\gamma}, \bm{x}^{\Gamma} )$ \textit{alongside} inducing a conditional ensemble $p( \eta^{\Gamma} \vert \omega^{\gamma}, \alpha^{\gamma} )$ over paths.
Thus, there is a well-defined relationship between each internal path $\bm{x}^{\gamma}$ and external path ensemble $p( \eta^{\Gamma} \vert \omega^{\gamma}, \alpha^{\gamma} )$, motivating a free energy functional which describes externally observable dynamics.
Conversely, GFE describes something ``subjective'' about the system, in the sense that an active path $\alpha^{\gamma}$ exists but is not observed by $\bm{x}^{\gamma}$.
Thus, for any $( \omega^{\gamma}, \alpha^{\gamma} )$ pairing there will exist a VFE functional $\mathcal{F}[ \omega^{\gamma}, \alpha^{\gamma}, \bm{x}^{\gamma} ]$ which describes the true dynamics of the system, as well as a GFE functional $\mathcal{G}[ \omega^{\gamma}, \bm{x}^{\gamma} ]$ which describes an internal model.
While they are not equal, they are both minimized by the same $\bm{x}^{\gamma}$\cite{friston23}.\\
\\
In the discrete case, the functionals used are conceptually similar however are expressed though different formulas.
GFE is not defined here, while VFE and expected free energy are defined for each $t \in \mathbb{T} = \{ 0, 1, \ldots, T \}$ by the following formulae\cite{daCosta20}:

\begin{gather}\label{freeEnsDisc}
	\mathcal{F}[q]
	=
	\mathbb{E}_{q( \eta^{0:T}, A, \alpha^{0:T} \vert \omega^{0:t} )} \left[
	-
	\log
	p(
	\omega^{0:t}
	,
	\eta^{0:T}
	,
	A
	,
	\alpha^{0:T}
	)
	\right]
	-
	\mathbb{E}_{q( \eta^{0:T}, A, \alpha^{0:T} \vert \omega^{0:t} )} \left[
	-
	\log
	q(
	\eta^{0:T}
	,
	A
	,
	\alpha^{0:T}
	\vert
	\omega^{0:t}
	)
	\right]
	;
	\\
	\mathcal{E}[ \alpha^{0:T} ]
	=
	\mathbb{E}_{ q( \eta^{0:T}, A \vert \alpha^{0:T} ) \mathbb{P}( \bm{x}^{0:T} \vert \omega^{0:T}, A ) } \left[
	-
	\log
	\mathbb{P}(
	\bm{x}^{0:T}
	,
	\omega^{0:T}
	,
	A
	)
	\right]
	-
	\mathbb{E}_{ q( \eta^{0:T}, A \vert \alpha^{0:T} ) \mathbb{P}( \bm{x}^{0:T} \vert \omega^{0:T}, A ) } \left[
	-
	\log
	q(
	\eta^{0:T}
	,
	A
	\vert
	\alpha^{0:T}
	)
	\right]
	.
\end{gather}

We note that the requisite approximate posterior distribution over policies $Q(\alpha^{0:T})$ is a softmax function over negative expected free energy: $Q( \alpha^{0:T} ) = \sigma( - \mathcal{E}( \alpha^{0:T} ) )$\cite{daCosta20}.

\subsubsection{Action and inference}\label{3.3.2.3}

Internal states $\bm{x}$ and external states $\eta$ are thought to be coupled.
In particular, due to the dependency structure outlined in section \ref{3.3.2.1}, Active Inference asserts that the internal path of least action $\bm{x}^{\gamma}$ (which in most steady states is realized\cite{friston23}) parameterizes a posterior distribution $q(\eta \vert \alpha, \omega ; \bm{x}^{\gamma})$ over external states\cite{sakthivadivel22}.
Initially it was argued that selection bias ensures surviving states will appear to perform inference\cite{friston13}.
It has since been discovered that when a Markov blanket structure is assumed, then inference naturally follows under appropriate (though fairly general) physical assumptions\cite{friston23}.\\
\\
Perception is thought to arise from minimizing the variational free energy functional, while action relies on minimizing expected free energy\cite{friston11}.
The precise way that this is done is modeled in a variety of ways.
For example, predictive coding can be derived from active inference when gaussian generative models are assumed and dirac-delta priors are\cite{buckley17}.
Alternatively, in the discrete case, categorical and dirichlet distributions might be used\cite{daCosta20}.
In the continuous-state context, when a Langevin equation governing the joint state space is assumed, minimizing \textit{generalized free energy} will drive both action and perception simultaneously\cite{friston23}.

%% file: 4-info-dev/4_info_dev.tex
\chapter{Maximum-Caliber techniques as a unifying framework}\label{4_info_dev}

\textsc{Profoundly different questions} in theoretical neurobiology are addressed by IIT and FEP.
The former is a process theory, rooted in empirical Bayesian techniques\cite{friston03,sakthivadivel22}.
It models perception, action, and self-organization from an onlooker's perspective\cite{friston06,friston23}.
The latter is an ontology rooted in information theory\cite{tononi04,tononi08}, modeling sentience from an internal point of view\cite{albantakis2023,oizumi2014,tononi04}.\\
\\
Yet, beneath this apparent divergence lies a surprising commonality:
their quantities of interest each relate to deviations from MaxCal ensembles of trajectories\cite{maxwell23}.\\
\\
In particular, IIT's trajectories (in its current forms act over a single discrete step forward in time\cite{oizumi2014,albantakis2023}.
These can be represented by probabilistic graphical models (PGMs) termed \textit{transition networks.}
Unconstrained cause and effect repertoires are the inputs and outputs when these networks are imbued with a MaxCal ensemble.
\textit{Information} is a difference imposed on some observable of interest when this ensemble is replaced with a constrained MaxCal distribution.\\
\\
FEP is described over discrete or continuous timescales of longer intervals\cite{daCosta20,friston23}.
For continuous variables, it has been shown to be dual to constrained MaxEnt problems.
While not proven for general MaxCal problems, it has been shown to hold under modest assumptions and is an active area of research.
The \textit{excess free energy} (beyond its minimum) of an internal state given blanket conditions, appears to be the system's deviation from a constrained MaxEnt (and in many cases MaxCal) state\cite{sakthivadivel23,maxwell23}.\\
\\
This opens scope for interpreting ``information'' as an excess of free energy, though precise assumptions are required and so this must be approached with care.
Here, we \textit{prove} IIT's association with MaxCal methods and \textit{suggest} further research into its relationship with the \textit{process theory} active inference, via application of statistical mechanics techniques\cite{presse13}.
In chapter \ref{5_IIT_FEP} (sections \ref{5.4}--\ref{5.5}) we \textit{formally propose} reconciliation of IIT and predictive processing theories, via \textit{inspired}\cite{leung23} MaxCal analysis and a novel form of Friston's \textit{unifying framework,} FEP\cite{friston03}.
While we might hope for full reconciliation, it is currently important to separate these threads.

\section{Mathematical prerequisites and techniques}\label{4.1}

\subsection{Dynamic Bayesian Networks}\label{4.1.1}

Graphs are mathematical objects which represent relationships.
In this text, we assume familiarity with basic graph theory, however where required an overview can be found in \textit{Bondy and Murty (2008)}\cite{bondy08}.
We employ the use of \textit{Probabilistic Graphical Models (PGMs),} which are graphs in which nodes are random variables and edges represent conditional dependency relationships.
When the graph is ordered, edges encode factorization of the generative model at hand\cite{koller2009}.
In our analysis, it will be necessary to distinguish nodes which act as sources from nodes which act as relays or recipients, which motivates the following terminology:

\begin{dfn}[Primary; Ancillary]
	Let $\mathcal{G} = (\mathcal{V}, \mathcal{E})$ be a Probabilistic Graphical Model, with $\mathcal{E}$ containing ordered pairs.
	Then, we refer to $V \in \mathcal{V}$ as a \textup{primary node} if and only if there exist no $W \in \mathcal{V}$ such that $(W, V) \in \mathcal{E}$.
	If $V \in \mathcal{V}$ is not primary, then we refer to it as an \textup{ancillary node.}
\end{dfn}

A form of PGMs which will be of particular interest to us are Bayesian Networks\cite{rao14}, because they encode the directed dependency relationships which are implicit in Integrated Information Theory\cite{oizumi2014}.
Of particular interest will be Dynamic Bayesian Networks, which model hierarchical cause-effect relationships across time\cite{neapolitan07}.
For completeness, we provide their definitions below.

\begin{dfn}[Bayesian Network]\label{baysNet}
	Let $\mathcal{G}$ be a directed, acyclic probabilistic graphical model if and only if its nodes, $V_1, \ldots, V_n$, encode the following dependency relationships:
	\begin{equation*}
		\text{For each variable}
		\
		V_i:
		\
		\left(
		\
		V_i
		\
		\bot
		\
		\text{NonDescendents}(V_i)
		\
		\vert
		\
		\mathcal{I}(V_i)
		\right).
	\end{equation*}
	In other words, conditional on the value of its inputs, every variable $V_i$ in the PGM $\mathcal{G}$ is independent of all of its non-descendents\cite{rao14}.
\end{dfn}

\begin{dfn}[Dynamic Bayesian Network]\label{DBN}
	A \textup{Dynamic Bayesian Network} is a Bayesian Network consisting of the variables that combine the $T$ random vectors $\mathbf{X}^t$ and is determined by the following specifications:
	\begin{enumerate}
		\item An initial Bayesian Network $(\mathcal{G}^0, \rho)$ consisting of: (a) an initial directed acyclic graph $\mathcal{G}^0$ containing the variables $\mathbf{X}^0 = (X_1^0, \ldots, X_n^0)$; (b) an initial probability distribution $\rho$ of these variables.
		\item A transition Bayesian Network $(\mathcal{G}_\rightarrow, \bm{P})$ which acts as a template consisting of: (a) a transition directed acyclic graph $\mathcal{G}_\rightarrow$ containing the variables $\mathbf{X}^t, \mathbf{X}^{t+1}$ for each $t \in \mathbb{T}$;
		(b) a transition probability distribution, $\bm{P}$, which assigns a conditional probability to every value of $\mathbf{X}^{t+1}$ given every value of $\mathbf{X}^t$.
		That is, for every $\mathbf{x}^t \in \Omega_X$ there is a a probability distribution $\mathbf{X}^{t+1} \vert (\mathbf{X}^t = \mathbf{x}^t) \sim \bm{P} (\mathbf{x}^t, \cdot)$.
		\item The Dynamic Bayesian Network $(\mathcal{G}, \mathcal{E})$ consists of:
		(a) the directed acyclic graph $\mathcal{G} = \mathcal{G}^0 \sqcup \mathcal{G}^1 \sqcup \ldots \sqcup \mathcal{G}^T$ composed of the union of the initial and (each iteration of the) transition Bayesian Networks;
		(b) the joint probability distribution $\rho \times \bm{P}^T = \rho(\bm{X}^0) \prod_{t=1}^T \bm{P}(\bm{X}^t , \bm{X}^{t+1})$ resulting from combining the initial distribution $\rho$ with each iteration of the conditional distribution $\bm{P}$\cite{neapolitan07}.
	\end{enumerate}
\end{dfn}

We reiterate here that $\bm{P}( \bm{x}^t, \bm{x}^{t+1} ) = \mathbb{P}( \bm{X}^{t+1} = \bm{x}^{t+1} \vert \bm{X}^t = \bm{x}^t )$ has been written in this form to emphasize that they constitute entries of a transition probability matrix.
It does not refer a joint probability distribution.
We note also that conditional dependencies within $\bm{X}^{t+1}$ are not precluded provided the network overall is acyclic.\\
\\
The entropy $\mathcal{H}( \mathcal{G} )$ of a Bayesian Network can be expressed in terms of its factorization structure, representing a major computational advantage\cite{rao14}.
Here, we state the standard result for computing entropy and leave readers to see Koller and Friedman (2009) \cite{koller2009} for the proof.

\begin{thm}[Entropy of a Bayesian Network]\label{BNEnt}
	Let $\mathcal{G} = (\mathcal{V}, \mathcal{E})$ be a Bayesian Network associated with the random variables $\bm{V}$.
	Let $\mathcal{W}$ be its set of primary nodes, $\mathcal{A} = \mathcal{V} \setminus \mathcal{W}$ be the set of ancillary nodes, and $\mathcal{I}( A )$ represent the inputs of each ancillary node $A \in \mathcal{A}$.
	\textup{Then,} the entropy of $\mathcal{G}$ can be expressed as:
	\begin{equation}
		\mathcal{H}(
		\mathcal{V}
		)
		=
		\sum_{W \in \mathcal{W}}
		\mathcal{H}(
		W
		)
		\
		+
		\
		\sum_{A \in \mathcal{A}}
		\mathcal{H}(
		A
		\vert
		\mathcal{I}(A)
		).
	\end{equation} 
\end{thm}

\subsection{An overview of maximum-caliber techniques}\label{4.1.2}

The Principle of Maximal Entropy \textit{(MaxEnt)} was introduced by Jaynes in 1957 to solve inference problems in statistical mechanics.
The premise is that when making inferences about a system, one should use information available in the least biased way possible\cite{jaynes57}.
In physics, the constraints are physical (e.g. expected total energy) and the technique recovers the underlying distributions of equilibrium systems\cite{giffin09}.
\textit{Maximum Caliber} (MaxCal) assesses the path entropy of systems\cite{jaynes80} and is applied in broader contexts, including non-equilibrium systems\cite{presse13,dixit18}.\\
\\
Both principles have been applied across broader disciplines\cite{giffin09,caticha10}, including machine learning\cite{berger96,bishop06}, ecology\cite{giffin09,kumar2021}, and optimal control theory\cite{gottwald20}.
Jaynes's perspective when introducing the theory is that it was an application of information theory to physics, rather than being fundamentally physical itself\cite{jaynes57}.
However, couplings between information and thermodynamic processes\cite{szilard29,landauer61} complicate this reading\cite{floridi10}.\\
\\
The method involves maximizing the expected surprisal $- \log p( \bm{x} )$ (which we shall measure in \textbf{nats}) of some variable $\bm{X}$ by altering its probability distribution $p$ over some state space $\Omega_{\bm{X}}$.
When completely unconstrained over finite state spaces, we retrieve a uniform distribution $\textup{Unif}( \Omega_{\bm{X}} )$.
However, typically some fundamental constraints will be in place, and these are represented as expected values $\bar{C}_1, \ldots, \bar{C}_r$ over some statistics $C_1( \bm{X} ), \ldots, C_r ( \bm{X} )$.
Letting $\bm{\lambda} = ( \lambda_1, \ldots, \lambda_r )$ represent a vector of parameters, this equates to maximizing the Lagrange function\cite{gottwald20}:

\begin{equation}\label{MaxEntLagrange}
	\mathcal{L}( p ; \bm{\lambda} )
	=
	- \sum_{ \bm{x} }
	p( \bm{x} )
	\log p( \bm{x} )
	+
	\sum_{i=1}^r
	\left[
	\lambda_i
	\left(
	\bar{C}_i
	-
	\sum_{\bm{x}}
	p( \bm{x} )
	C_i ( \bm{x} )
	\right)
	\right]
\end{equation}

In the case of an equilibrium system with temperature $1 / \beta$, we retrieve a Gibbs distribution:

\begin{equation}
	p^*( \bm{x} )
	=
	\dfrac{
		e^{- \beta \bm{E}( \bm{x} )}
	}{
		Z( \beta )
	},
\end{equation}

where $p^* = \arg \min_p \mathcal{L}( p ; \bm{\lambda} )$, $\bm{E}( \bm{x} )$ is an energy of each \textit{microstate,} and $Z( \beta )$ is a \textit{partition function}\cite{jaynes57}.
In continuous spaces, we optimize over integrals and can retrieve any exponential family by choosing its sufficient statistics as the appropriate controls\cite{giffin09}.
Notable examples include retrieving exponential distributions $\alpha e^{-\alpha x}$ by maximizing entropy over $( 0, \infty )$ given a mean $1 / \alpha$, as well as normal distributions $\mathcal{N}( \mu ; \sigma^2 )$ by performing MaxEnt over $( - \infty, \infty )$ given a mean and variance.\\
\\
In the case of MaxCal, we path entropy over a stochastic process $( \bm{X}^t )_{t \in \mathbb{T}}$\cite{jaynes80}.
It is common for a reference distribution $\pi$ to be used when entropy over the entire path ensemble is intractable.
In these cases, we maximize over a caliber function which is the negative relative entropy of the two\cite{presse13,dixit18}:

\begin{equation}
	\mathcal{C}[p]
	=
	-
	\mathcal{D}(
	p
	\lvert
	\rvert
	\pi
	).
\end{equation}

We note that in finite state spaces $\Omega_{\bm{X}}$, maximization of ``pure'' entropy $\mathcal{H}( \bm{X} )$ is equivalent to using the uniform distribution as a reference distribution.

\section{The Maximum-Caliber foundations of IIT}\label{4.2}

We will soon show how IIT implicitly uses Dynamic Bayesian Networks to model causal relations\cite{oizumi2014}.
From here we can analyze how random noise perturbations, in IIT 3.0's deterministic context, represent various constrained MaxCal problems.
The key insight here is not the mathematical formalism, but rather how we might view IIT's methods if we view information in mobile, rather than static, terms.

\subsection{Transition Networks as Dynamic Bayesian Networks}\label{4.2.1}

To represent causal relations across time, IIT employs time-unfolded digraphs.
More than mere metaphors, these graphs encode dependence relationships born from the operational properties of each node.
To describe this mathematically we employ PGMs, which are graphs for which every vertex is a random variable.
We will formally refer to these as \textit{transition networks,} and they shall become our primary object of focus.

\begin{dfn}[Transition network]\label{trans_Ntwk}
	Let $\mathbb{X} = \{ \bm{X}^t \vert t \in \mathbb{T} \}$ be a homogeneous Markovian stochastic process, indexed by a discrete set $\mathbb{T}$ and taking values in discrete $\Omega_{\bm{X}}$.
	For any given $t$, we refer to $\mathcal{G} = ( \bm{X}^t \sqcup \bm{X}^{t+1}, \bm{P} )$ as a \textup{transition network.}
	It consists of \textup{inputs} $\bm{X}^t = ( X_1^t, \ldots, X_n^t )$, \textup{outputs} $\bm{X}^{t+1} = ( X_1^{t+1}, \ldots, X_n^{t+1} )$, as well as a transition probability matrix $\bm{P}$ which determines directed edges $\mathcal{E}( \bm{P} )$.
	In particular, $( X_i^t, X_j^{t+1} ) \in \mathcal{E}( \bm{P} )$ if and only if $X_j^{t+1}$ depends conditionally on $X_i^t$.
	In non-deterministic settings, we might also include non-directed edges between the input nodes $\bm{X}^t$.
\end{dfn}

We shall note that from the perspective of any single transition $\bm{X}^t \rightarrow \bm{X}^{t+1}$, it does not matter whether or not $\mathbb{X}$ is homogeneous.

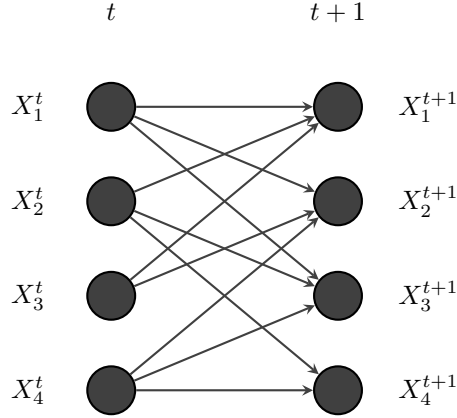
\begin{figure}[h]
	\centering
	\begin{tikzpicture}
		\node at (-3, 2.5) {$t$};
		\node at (0, 2.5) {$t+1$};
		
		\definecolor{input1}{RGB}{234, 67, 53}
		\definecolor{input2}{RGB}{251, 188, 184}
		\definecolor{output1}{RGB}{52, 168, 83}
		\definecolor{output2}{RGB}{129, 201, 149}
		
		\node[circle, fill=darkgray, draw=black, thick, minimum size = 18pt] (X1t) at (-3, 1.25) {};
		\node[circle, fill=darkgray, draw=black, thick, minimum size = 18pt] (X2t) at (-3, 0) {};
		\node[circle, fill=darkgray, draw=black, thick, minimum size = 18pt] (X3t) at (-3, -1.25) {};
		\node[circle, fill=darkgray, draw=black, thick, minimum size = 18pt] (X4t) at (-3, -2.5) {};
		
		\node[circle, fill=darkgray, draw=black, thick, minimum size = 18pt] (X1t1) at (0, 1.25) {};
		\node[circle, fill=darkgray, draw=black, thick, minimum size = 18pt] (X2t1) at (0, 0) {};
		\node[circle, fill=darkgray, draw=black, thick, minimum size = 18pt] (X3t1) at (0, -1.25) {};
		\node[circle, fill=darkgray, draw=black, thick, minimum size = 18pt] (X4t1) at (0, -2.5) {};
		
		\draw [-stealth, thick, darkgray] (X1t) -- (X1t1);
		\draw [-stealth, thick, darkgray] (X1t) -- (X2t1);
		\draw [-stealth, thick, darkgray] (X1t) -- (X3t1);
		\draw [-stealth, thick, darkgray] (X2t) -- (X1t1);
		\draw [-stealth, thick, darkgray] (X2t) -- (X3t1);
		\draw [-stealth, thick, darkgray] (X2t) -- (X4t1);
		\draw [-stealth, thick, darkgray] (X3t) -- (X1t1);
		\draw [-stealth, thick, darkgray] (X3t) -- (X2t1);
		\draw [-stealth, thick, darkgray] (X4t) -- (X2t1);
		\draw [-stealth, thick, darkgray] (X4t) -- (X3t1);
		\draw [-stealth, thick, darkgray] (X4t) -- (X4t1);
		
		\node at (-4.1, 1.25) {$X_1^t$};
		\node at (-4.1, 0) {$X_2^t$};
		\node at (-4.1, -1.25) {$X_3^t$};
		\node at (-4.1, -2.5) {$X_4^t$};
		
		\node at (1.2, 1.25) {$X_1^{t+1}$};
		\node at (1.2, 0) {$X_2^{t+1}$};
		\node at (1.2, -1.25) {$X_3^{t+1}$};
		\node at (1.2, -2.5) {$X_4^{t+1}$};
		
	\end{tikzpicture}
	\caption[Transition network]{Here we have a transition network which shows non-independent relationships between the inputs $\bm{X}^t = ( X_1^t, X_2^t, X_3^t, X_4^t )$ and conditionally independent outputs $\bm{X}^{t+1} = ( X_1^{t+1}, X_2^{t+1}, X_3^{t+1}, X_4^{t+1} )$. The system is deterministic and thus our PGM is a dynamic bayesian network.}
	\label{fig:trans_ntwk}
\end{figure}

In IIT 3.0, the transition networks employed are Dynamic Bayesian Networks, because the determinism of the states and uniformity of perturbations ensures that inputs are always independent\cite{oizumi2014}, though in non-deterministic settings we connect these edges to allow for the possibility of non-deterministic inputs.
Employing Theorem \ref{BNEnt}, we see that $\mathcal{H}( \mathcal{G} )$ is equal to $\sum_{i=1}^n \mathcal{H}( X_i^t ) + \sum_{j=1}^{n} \mathcal{H}( X_j^{t+1} \vert \mathcal{I}_j )$, where $\mathcal{I}_j$ are the inputs of $X_j^{t+1}$.
The determinism of $\bm{P}$ ensures that each $\mathcal{H}( X_j^{t+1} \vert \mathcal{I}_j )$ is null.
Thus, we retrieve our identity:

\begin{equation}\label{transitionEnt}
	\mathcal{H}(
	\mathcal{G}
	)
	=
	\mathcal{H}(
	\bm{X}^t
	).
\end{equation}

\subsection{Retrieving (non-integrated) cause and effect information}\label{4.2.2}

From equation \ref{transitionEnt}, it straightforwardly follows that placing each input $X_i^t$ into a uniform distribution will maximize the path entropy (caliber) of $\mathcal{G}$.
We also see that if we fix some subset to have fixed values, $\bm{Y}_{\bot}^t = \bm{y}_{\bot}^t$, then placing each of the remaining $\bm{Y}^t$ into a uniform distribution maximizes the entropy of the residual transition network, since $\mathcal{H}( \mathcal{G} \vert \bm{Y}_{\bot}^t = \bm{y}_{\bot}^t ) = \mathcal{H}( \bm{Y}^t )$.
Thus, we can retrieve IIT's \textit{effect information} methods\cite{oizumi2014} through two key steps.\\
\\
Firstly, we choose a subsidiary transition network $\mathcal{G}_{\bm{Y}} = ( \bm{Y}^t \sqcup \bm{Y}^{t+1}, \bm{P}_{\bm{Y}}( \cdot, \cdot \vert \bm{y}_{\bot}^{t} ) )$ by fixing background conditions $\bm{Y}_{\bot}^t = \bm{y}_{\bot}^t$ and marginalizing over corresponding outputs $\bm{Y}_{\bot}^{t+1}$ (which is equivalent to creating a conditional probability matrix over the subset $\bm{Y}^t$ of interest).
Then, we examine the remaining network in a MaxCal state and compare its outputs $\bm{Y}^{t+1}$ against those of the realization, $\bm{Y}^t = \bm{y}^t$.\\
\\
When identifying core effects, we follow a similar procedure.
We set background conditions the same way, however will marginalize over some other set of outputs $\bm{V}_{\bot}^{t+1}$.
From here, we calculate ``effect information'' in precisely the same way.\\

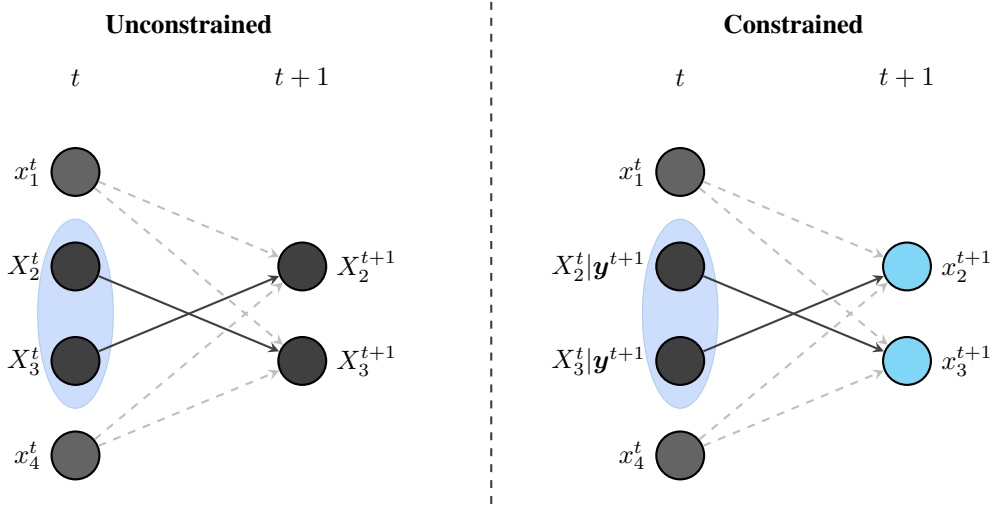
\begin{figure}[h]
	\centering
	\begin{tikzpicture}
		
		\definecolor{input1}{RGB}{234, 67, 53}
		\definecolor{input2}{RGB}{251, 188, 184}
		\definecolor{output1}{RGB}{52, 168, 83}
		\definecolor{output2}{RGB}{129, 201, 149}
		\definecolor{fixed}{RGB}{100, 100, 100}
		\definecolor{past2}{RGB}{154, 190, 247}
		
		\tikzset{tn/.style={circle, fill=darkgray, inner sep=0pt, minimum size=10pt}}
		
		\node[font=\bfseries] at (-4.5, 3.2) {Unconstrained};
		
		\node at (-6, 2.5) {$t$};
		\node at (-3, 2.5) {$t+1$};
		
		\draw [fill=past2, fill opacity=0.5, draw=past2, draw opacity=0.8] (-6, -0.625) ellipse (0.5cm and 1.25cm);
		
		\node[circle, fill=fixed, draw=black, thick, minimum size = 18pt] (LA1) at (-6, 1.25) {};
		\node[circle, fill=darkgray, draw=black, thick, minimum size = 18pt] (LA2) at (-6, 0) {};
		\node[circle, fill=darkgray, draw=black, thick, minimum size = 18pt] (LA3) at (-6, -1.25) {};
		\node[circle, fill=fixed, draw=black, thick, minimum size = 18pt] (LA4) at (-6, -2.5) {};
		
		\node[circle, fill=darkgray, draw=black, thick, minimum size = 18pt] (LB2) at (-3, 0) {};
		\node[circle, fill=darkgray, draw=black, thick, minimum size = 18pt] (LB3) at (-3, -1.25) {};
		
		\draw [-stealth, thick, dashed, lightgray] (LA1) -- (LB2);
		\draw [-stealth, thick, dashed, lightgray] (LA1) -- (LB3);
		\draw [-stealth, thick, darkgray] (LA2) -- (LB3);
		\draw [-stealth, thick, darkgray] (LA3) -- (LB2);
		\draw [-stealth, thick, dashed, lightgray] (LA4) -- (LB2);
		\draw [-stealth, thick, dashed, lightgray] (LA4) -- (LB3);
		
		\node[anchor=east] at (LA1.west) {$x_1^t$};
		\node[anchor=east] at (LA2.west) {$X_2^t$};
		\node[anchor=east] at (LA3.west) {$X_3^t$};
		\node[anchor=east] at (LA4.west) {$x_4^t$};
		
		\node[anchor=west] at (LB2.east) {$X_2^{t+1}$};
		\node[anchor=west] at (LB3.east) {$X_3^{t+1}$};
		
		\draw[thick, dashed, darkgray] (-0.5, 3.5) -- (-0.5, -3.2);
		
		\node[font=\bfseries] at (3.5, 3.2) {Constrained};
		
		\node at (2, 2.5) {$t$};
		\node at (5, 2.5) {$t+1$};
		
		\draw [fill=past2, fill opacity=0.5, draw=past2, draw opacity=0.8] (2, -0.625) ellipse (0.5cm and 1.25cm);
		
		\node[circle, fill=fixed, draw=black, thick, minimum size = 18pt] (RA1) at (2, 1.25) {};
		\node[circle, fill=darkgray, draw=black, thick, minimum size = 18pt] (RA2) at (2, 0) {};
		\node[circle, fill=darkgray, draw=black, thick, minimum size = 18pt] (RA3) at (2, -1.25) {};
		\node[circle, fill=fixed, draw=black, thick, minimum size = 18pt] (RA4) at (2, -2.5) {};
		
		\node[circle, fill=cyan, fill opacity=0.5, draw=black, thick, minimum size = 18pt] (RB2) at (5, 0) {};
		\node[circle, fill=cyan, fill opacity=0.5, draw=black, thick, minimum size = 18pt] (RB3) at (5, -1.25) {};
		
		\draw [-stealth, thick, dashed, lightgray] (RA1) -- (RB2);
		\draw [-stealth, thick, dashed, lightgray] (RA1) -- (RB3);
		\draw [-stealth, thick, darkgray] (RA2) -- (RB3);
		\draw [-stealth, thick, darkgray] (RA3) -- (RB2);
		\draw [-stealth, thick, dashed, lightgray] (RA4) -- (RB2);
		\draw [-stealth, thick, dashed, lightgray] (RA4) -- (RB3);
		
		\node[anchor=east] at (RA1.west) {$x_1^t$};
		\node[anchor=east] at (RA2.west) {$X_2^t \vert \bm{y}^{t+1}$};
		\node[anchor=east] at (RA3.west) {$X_3^t \vert \bm{y}^{t+1}$};
		\node[anchor=east] at (RA4.west) {$x_4^t$};
		
		\node[anchor=west] at (RB2.east) {$x_2^{t+1}$};
		\node[anchor=west] at (RB3.east) {$x_3^{t+1}$};
		
	\end{tikzpicture}
	\caption[Constraining a transition network]{Input nodes $\bm{Y}_{\bot}^t = ( X_1^t, X_4^t )$ have been fixed to background conditions $\bm{y}_{\bot}^t$, and output nodes $\bm{Y}_{\bot}^{t+1} = ( X_1^{t+1}, X_4^{t+1} )$ have been marginalized over, to select the subsidiary network $\mathcal{G}_{\bm{Y}}$ over $\bm{Y} = ( X_2, X_3 )$. On the left we have the unconstrained case in which entropy across $\bm{Y}^t \sqcup \bm{Y}^{t+1}$ has been maximized. On the right hand side we have maximized entropy subject to the conditions $\bm{Y}^{t+1} = \bm{y}^{t+1}$, retrieving the constrained cause repertoire.}
	\label{fig:cause_effect_ntwk}
\end{figure}

For cause information the picture is slightly more complex.
We choose our subsidiary network in precisely the same way (conditioning on $\bm{y}_{\bot}^t$ and marginalizing over $\bm{Y}_{\bot}^{t+1}$), and apply a uniform distribution over $\bm{Y}^t$ to attain our unconstrained MaxCal, but this time looking at input distributions.
With regard to the constrained cause repertoire, IIT uses a uniform distribution over the support of our realization, $\textup{Supp}( \bm{y}^{t+1} ) = \{ \bm{y}^t : \bm{P}( ( \bm{y}^t, \bm{y}_{\bot}^t ), \bm{y}^{t+1} ) = 1 \}$.
In doing this, we maximize our constrained entropy $\mathcal{H}( \mathcal{G}_{\bm{Y}} \vert \bm{y}^{t+1} ) = \mathcal{H}( \bm{Y}^t \vert \bm{Y}^{t+1} = \bm{y}^{t+1} ; \bm{Y}_{\bot}^t = \bm{y}^t_{\bot} )$, however it is worth considering the nuances of our posterior and prior distributions for later non-deterministic cases.\\
\\
At the time in which $\bm{Y}^t$ takes its occupied value, it does not yet have information about the value of $\bm{Y}^{t+1}$, so we have two possible perspectives.
In the first, we use our realization $\bm{Y}^{t+1} = \bm{y}^{t+1}$ to retrieve information about the inputs $\bm{Y}^t$, and so we choose a MaxEnt posterior distribution (the uniform distribution of the support).
In this view, information is predictive.\\
\\
Under a more ontic view, we might say that the observation $\bm{Y}^{t+1} = \bm{y}^{t+1}$ necessarily rules out any states $\hat{\bm{y}}$ which do not lead to $\bm{y}^{t+1}$ from our state space $\Omega_{\bm{Y}}$ (to form a new $\Omega_{\bm{Y}}'$).
From here, we can use Bayes' Theorem to note that $\mathbb{P}( \bm{Y}^t = \bm{y}^t \vert \bm{y}^{t+1} ; \bm{y}_{\bot}^t ) \mathbb{P}( \bm{Y}^{t+1} = \bm{y}^{t+1} )$ is equal to $\mu( \bm{y}^t ) \bm{P}( \bm{y}^t, \bm{y}^{t+1} ; \bm{y}_{\bot}^t )$, where $\mu( \bm{y}^t )$ is some prior distribution.
Since we are working with deterministic $\bm{P}$, it happens to be the case that $\mathbb{P}( \bm{Y}^{t+1} = \bm{y}^{t+1} )$ and $\bm{P}( \bm{y}^t, \bm{y}^{t+1} ; \bm{y}_{\bot}^t )$ are both always equal to $1$ over our \textit{applicable} state space $\Omega_{\bm{Y}}' = \textup{Supp}( \bm{y}^{t+1} )$, and so our posterior $\mathbb{P}( \bm{Y}^t = \bm{y}^t \vert \bm{y}^{t+1} ; \bm{y}_{\bot}^t )$ and prior $\mu( \bm{y}^t )$ are both equal.
Thus, we might say we maximize \textit{epistemic} entropy $\mathcal{H}( \mathcal{G}_{\bm{Y}} \vert \bm{y}^{t+1} ; \bm{y}_{\bot}^t )$ to derive a ``best guess'' of the prior $\mu( \bm{y}^t )$ which induced the observation $\bm{Y}^{t+1} = \bm{y}^{t+1}$.
Though immaterial in deterministic cases, this difference impacts our approach in non-deterministic settings.

\subsection{Integrated information}\label{4.2.3}

To assess integration, we compare the same background conditions against counterfactual partitions.
The analogy used in IIT is that connections are injected with random noise\cite{oizumi2014}, however we shall visualize it differently here.\\
\\
Consider a partition $\mathcal{P} = \{ \bm{Z}^t \sqcup \bm{V}^{t+1}, \bm{Z}_{\bot \vert \bm{Y}}^t \sqcup \bm{V}_{\bot \vert \bm{Y}}^{t+1} \}$.
In IIT we would ``cut'' information across the partition, however for notational ease here we will disrupt it \textit{within} the partition (which makes no difference to our end result, as IIT examines every possible partition).
Thus, for effect information we are interested in the causal information that $\bm{z}_{\bot \vert \bm{Y}}^t$ has on $\bm{V}^{t+1}$ when the signals from $\bm{Z}^t$ are disrupted.
We can model this as applying a background condition $\bm{Z}_{\bot \vert \bm{Y}}^t = \bm{z}_{\bot \vert \bm{Y}}^t$ and marginalizing over $\bm{V}_{\bot \vert \bm{Y}}$, to produce a new network from $\bm{Z}^t$ toward $\bm{V}^{t+1}$.
In MaxCal terms, disrupting signals from $\bm{Z}^t$ within this network is equivalent to equipping the marginalized network with a MaxCal path ensemble.
The same is true with regard to conditioning on $\bm{Z}^t = \bm{z}^t$, marginalizing over $\bm{V}^{t+1}$, and then disrupting signals from $\bm{Z}_{\bot \vert \bm{Y}}^t$ toward $\bm{V}_{\bot \vert \bm{Y}}^{t+1}$.\\

\begin{figure}[h]
	\centering
	\begin{tikzpicture}
		
		\definecolor{input1}{RGB}{234, 67, 53}
		\definecolor{input2}{RGB}{251, 188, 184}
		\definecolor{output1}{RGB}{52, 168, 83}
		\definecolor{output2}{RGB}{129, 201, 149}
		\definecolor{fixed}{RGB}{100, 100, 100}
		\definecolor{past2}{RGB}{154, 190, 247}
		
		\tikzset{tn/.style={circle, fill=darkgray, inner sep=0pt, minimum size=10pt}}
		
		\node at (-4.5, 3.2) {$\mathcal{G}_{\bm{Z} \vert \bm{Y}}^{\bm{y}^t}$};
		
		\node at (-6, 2.5) {$t$};
		\node at (-3, 2.5) {$t+1$};
		
		\draw [fill=past2, fill opacity=0.5, draw=past2, draw opacity=0.8] (-6, -0.625) ellipse (0.5cm and 1.25cm);
		
		\node[circle, fill=fixed, draw=black, thick, minimum size = 18pt] (LA1) at (-6, 1.25) {};
		\node[circle, fill=darkgray, draw=black, thick, minimum size = 18pt] (LA2) at (-6, 0) {};
		\node[circle, fill=fixed, draw=black, thick, minimum size = 18pt] (LA3) at (-6, -1.25) {};
		\node[circle, fill=fixed, draw=black, thick, minimum size = 18pt] (LA4) at (-6, -2.5) {};
		
		\node[circle, fill=darkgray, draw=black, thick, minimum size = 18pt] (LB2) at (-3, 0) {};
		
		\draw [-stealth, thick, dashed, lightgray] (LA1) -- (LB2);
		\draw [-stealth, thick, dashed, lightgray] (LA3) -- (LB2);
		\draw [-stealth, thick, dashed, lightgray] (LA4) -- (LB2);
		
		\node[anchor=east] at (LA1.west) {$x_1^t$};
		\node[anchor=east] at (LA2.west) {$X_2^t$};
		\node[anchor=east] at (LA3.west) {$x_3^t$};
		\node[anchor=east] at (LA4.west) {$x_4^t$};
		
		\node[anchor=west] at (LB2.east) {$X_2^{t+1}$};
		
		\draw[thick, dashed, darkgray] (-0.5, 3.5) -- (-0.5, -3.2);
		
		\node at (3.5, 3.2) {$\mathcal{G}_{\bm{Z}_{\bot} \vert \bm{Y}}^{\bm{y}^t}$};
		
		\node at (2, 2.5) {$t$};
		\node at (5, 2.5) {$t+1$};
		
		\draw [fill=past2, fill opacity=0.5, draw=past2, draw opacity=0.8] (2, -0.625) ellipse (0.5cm and 1.25cm);
		
		\node[circle, fill=fixed, draw=black, thick, minimum size = 18pt] (RA1) at (2, 1.25) {};
		\node[circle, fill=fixed, draw=black, thick, minimum size = 18pt] (RA2) at (2, 0) {};
		\node[circle, fill=darkgray, draw=black, thick, minimum size = 18pt] (RA3) at (2, -1.25) {};
		\node[circle, fill=fixed, draw=black, thick, minimum size = 18pt] (RA4) at (2, -2.5) {};
		
		\node[circle, fill=darkgray, draw=black, thick, minimum size = 18pt] (RB3) at (5, -1.25) {};
		
		\draw [-stealth, thick, dashed, lightgray] (RA1) -- (RB3);
		\draw [-stealth, thick, dashed, lightgray] (RA2) -- (RB3);
		\draw [-stealth, thick, dashed, lightgray] (RA4) -- (RB3);
		
		\node[anchor=east] at (RA1.west) {$x_1^t$};
		\node[anchor=east] at (RA2.west) {$x_2^t$};
		\node[anchor=east] at (RA3.west) {$X_3^t$};
		\node[anchor=east] at (RA4.west) {$x_4^t$};
		
		\node[anchor=west] at (RB3.east) {$X_3^{t+1}$};
		
	\end{tikzpicture}
	\caption[Partitioned transition network]{In this case, $\bm{Y} = ( X_2, X_3 )$ and our partition is $\mathcal{P} = \{ X_2^t \sqcup X_2^{t+1}, X_3^t \sqcup X_3^{t+1} \}$. The subsidiary network $\mathcal{G}_{\bm{Z} \vert \bm{Y}}^{\bm{y}^t}$ conditions on $X_3^t = x_3^t$ and applies a MaxCal path ensemble over the network. In this case, $X_2^t$ should have a uniform distribution over its state space $\Omega_2$ while the value $x_2^{t+1}$ of $X_2^{t+1}$ should be fixed by the background conditions. For our network $\mathcal{G}_{\bm{Z}_{\bot} \vert \bm{Y}}^{\bm{y}^t}$ we condition on $X_2^t = x_2^t$ and then apply a MaxCal ensemble. Likewise, we should see that $X_3^{t+1}$ is fixed by background conditions while $X_3^t$ has a uniform distribution. Integrated effect information in this case is zero, because the partitioned networks in a MaxCal ensemble produced the same outputs as a constrained transition network with fixed inputs $\bm{Y}^t = \bm{y}^t$.}
	\label{fig:int_info_ntwk}
\end{figure}
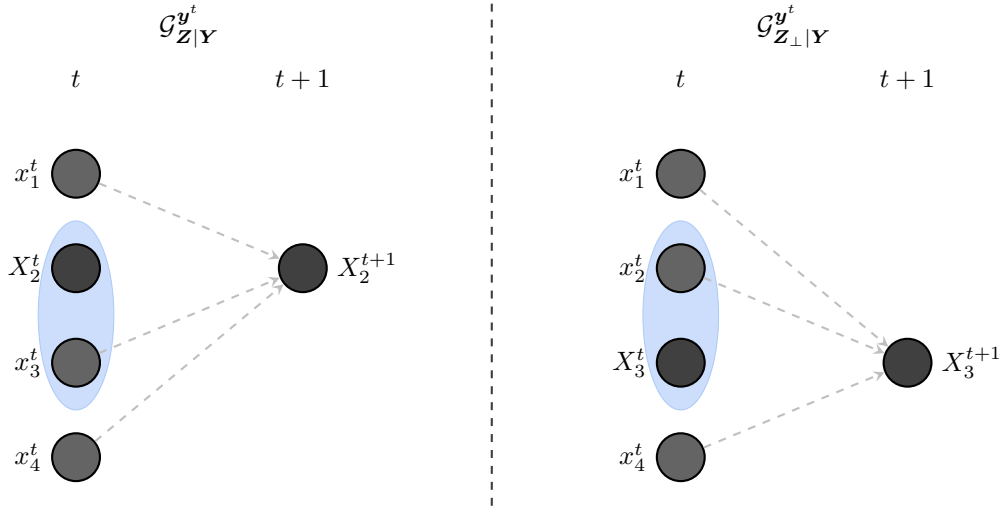

Mathematically, we can understand this as creating a new partitioned matrix $\bm{P}_{\bm{Y}, \mathcal{P}}^{\bm{y}^t}$ with transition probabilities defined by the following equation:

\begin{equation}
	\begin{split}
		\bm{P}_{\bm{Y}, \mathcal{P}}^{\bm{y}^t}( \hat{\bm{y}}^t, \hat{\bm{y}}^{t+1} )
		&=
		\bm{P}_{\bm{Y}}( ( \hat{\bm{z}}^t, \bm{z}_{\bot \vert \bm{Y}}^t ), \hat{\bm{v}}^{t+1} )
		\bm{P}_{\bm{Y}}( ( \bm{z}^t, \hat{\bm{z}}_{\bot \vert \bm{Y}}^t ), \hat{\bm{v}}_{\bot \vert \bm{Y}}^{t+1} )
		\\
		&=
		\bm{P}( ( \hat{\bm{z}}^t, \bm{z}_{\bot \vert \bm{Y}}^t, \bm{y}_{\bot}^t ), \hat{\bm{v}}^{t+1} )
		\bm{P}( ( \bm{z}^t, \hat{\bm{z}}_{\bot \vert \bm{Y}}^t, \bm{y}_{\bot}^t ), \hat{\bm{v}}_{\bot \vert \bm{Y}}^{t+1} ).
	\end{split}
\end{equation}

We then apply a MaxCal ensemble to it and see if we lose any effect information.\\
\\
With regard to integrated causal information over $\mathcal{P}$ for $\bm{Y}^{t+1} = \bm{y}^{t+1}$, IIT marginalizes over the output nodes $\bm{V}_{\bot \vert \bm{Y}}^{t+1}$ and then replaces each $\bm{Z}_{\bot \vert \bm{Y}}^t$ with individuated ``virtual elements''.
These virtual elements are then marginalized over to retrieve cause information about $\bm{Z}^t$.
Mathematically, the MaxCal analogue is less clear because introduction of virtual elements may introduce non-determinism into transitions from $\bm{Z}^t$ to $\bm{V}^{t+1}$.
However, since our MaxCal view is trajectory-based, rather than state-based, only the effect integration procedure is relevant for our generalization.

\subsection{System-level integration}\label{4.2.4}

Now that we understand IIT's algorithm through a MaxCal lens, we can infer some necessary (though insufficient) conditions for a deterministic system $\bm{X}^t$ in a particular state $\bm{x}^t$ to produce a positive $\Phi$ value.
Firstly, there must be a collection of mechanisms $\mathbb{M} = \{ \bm{M}_1, \ldots, \bm{M}_R \}$ wherein if the background condition $(\bm{M}_r)_{\bot} = (\bm{m}_r)_{\bot}$ is applied, placing the residual network $\mathcal{G}_{r} = \mathcal{G}_{\bm{M}_r}$ into a MaxCal ensemble causes the outputs $\bm{M}_r^{t+1}$ to lose information.
In this case, this looks like them being non-deterministic.
In addition, when any conditioned partition of the network $\mathcal{G}_r$ (as described in the previous section) is placed into a MaxCal ensemble, it cannot produce deterministic outputs.
In effect, partitioning these mechanisms must make them less precise and cause the space of available futures to expand.\\
\\
Further, ``cutting'' the whole system $\bm{X}$ should meaningfully alter the number of mechanisms or their effects on the system.
If it reduces the number of mechanisms, then this should imply that areas prone to losing precision when partitioned are interspersed throughout the system.\\
\\
It therefore seems that $\bm{X}^t = \bm{x}^t$ reduces available futures of $\bm{X}$ on many levels.
It is not simply that the available space over $\bm{X}^{t+1}$ overall reduces, but also that within a comprehensive non-separable set $\mathbb{M}$ of ``parts'', the state $\bm{M}_r^t = \bm{m}_r^t$ maximally reduces available future (given background conditions).
Intuitively, we might see this as a very intricate push spanning the whole system toward precision.

\subsection{Interpretation}\label{4.2.5}

The straightforward interpretation of our results is that key quantities from IIT 3.0\cite{oizumi2014} can be derived from variational MaxCal principles --- in particular, with intrinsically generated observables (dynamical laws; states) governing constraints on the MaxCal functional.
While many of the methods used in IIT are not commonly used MaxCal inference\cite{dixit18}, the procedure itself is associated.\\
\\
Where interpretation becomes less clear is whether this is a feature or coincidence.
IIT centers its ontology around cause-effect power\cite{oizumi2014,albantakis2023}, and so one could interpret this as purely mathematical.
However, when reviewing $\Phi$'s relationship with complexity measures\cite{rosas22} and free energy\cite{mayama25}, it is worth considering whether mathematical ``coincidences'' may play a role.\\
\\
In addition, one might question why a theory rooted in phenomenological axioms produces methods which are commonly used in physics.
If a bridge can be made between IIT's ontology and MaxCal procedures, the theory might stand to become more widely applicable.

\section{Free Energy Principle and Maximum-Entropy techniques}\label{4.3}

The Free Energy Principle can be linked to MaxEnt and MaxCal techniques on one of two grounds:
in the first, there is a direct computational equivalence between constrained MaxEnt (CMEP) problems and free energy minimization ones.
Thus, Friston's Free Energy Principle\cite{friston03} can be re-discussed as a CMEP principle on epistemic grounds\cite{gottwald20}.\\
\\
Secondly, Bayesian mechanics developed for active inference\cite{parr19} can be derived through thermodynamic MaxEnt methods\cite{sakthivadivel22,maxwell23,sakthivadivel23}.
In doing this, the authors ground functional interpretations of ``belief'' and ``inference'' in physical dynamics\cite{friston23}.

\subsection{Variational inference as the principle of maximum entropy}\label{4.3.1}

As shown in equation \ref{MaxEntLagrange}, constrained entropy maximization involves maximizing a Lagrangian.
In addition to retrieving the form of $p( \bm{x} )$, we retrieve Lagrange multipliers $\bm{\lambda}$ which reflect the constraints.
If we let $\beta_i$ represent $\arg \max_{\lambda_i} \mathcal{L}$, then our constrained MaxEnt $p^*$ given that $\mathbb{E}[C_i] = \bar{C}_i$ for each $i$ minimizes a free energy functional\cite{gottwald20}:

\begin{equation}
	\mathcal{F}[p]
	=
	\sum_{i=1}^r
	\beta_i
	\langle
	C_i
	\rangle
	-
	\mathcal{H}(p).
\end{equation}

We can express restrictions of the domain $\Omega_{\bm{X}}$ through these means too by defining $C'( \bm{x} ) = 1$ for all $\bm{x}$ in some $\Omega' \subsetneq \Omega_{\bm{X}}$ and then letting $C'( \bm{x} ) = 0$ otherwise.
When we set the expected value of $C'$ to be $1$ over $p$, we restrict the domain of our probability distribution.\\
\\
When considering variational inference, we can consider $\mathcal{E}( \bm{x} ) = - \log p( o, \bm{x} )$ to be an energy functional.
The variational free energy in this case takes the form\cite{gottwald20}:

\begin{equation}
	\mathcal{F}_V[q]
	=
	\langle
	\mathcal{E}
	\rangle
	-
	\mathcal{H}(q).
\end{equation}

From a MaxCal perspective, a parameterized distribution $q( \bm{x} \vert o ; \bm{\lambda} )$ which minimizes this would maximize entropy over $\bm{X} \vert o$ subject to derived statistics $\mathbb{E}_q[ - \log p( o, \bm{x} ) ]$.
We can also understand the functional $\mathcal{F}_V$ itself to be a function of relative entropy\cite{gottwald20}.

\subsection{Active inference as a dual of maximum-caliber physics}\label{4.3.2}

In the case of active inference, we begin with a continuous stochastic system $\mathbb{X} = \{ \bm{X}^t \vert t \in \left[ 0, T \right] \}$ characterized by a Langevin equation:

\begin{equation}\label{langevin}
	d \bm{X}^t
	=
	f( \bm{X}^t, t )
	dt
	+
	\sqrt{2D}
	d \bm{W}^t.
\end{equation}

The term $f(\bm{X}^t, t)$ refers to \textit{drift} and $\bm{W}^t$ is a \textit{Wiener process}.
From here, we apply a decomposition $\bm{X}^t = ( \eta^t, \omega^t, \alpha^t, \lambda^t )$ in which $\eta^t, \lambda^t$ are conditionally independent with respect to $(\omega^{[0, T]}, \alpha^{[0, T]})$\cite{sakthivadivel23}.\\
\\
From this point, a \textit{Gauge Theory} is created to describe \textit{constrained MaxEnt} problems over $\lambda$.
A key feature of this construction is treating \textit{connections} (a generalization of derivatives) which preserve action as parallel transport.
Meanwhile, gradient ascent toward MaxEnt constitutes a gauge force\cite{maxwell23}.
This is shown to be the dual adjoint of the free energy in active inference when described geometrically\cite{maxwell23}.
In certain cases, this also applies to MaxCal inference\cite{sakthivadivel23}.
See \textit{Ramstead et al. (2023)} \cite{maxwell23} for further mathematical details.\\
\\
The notable result of this procedure is that minimizing free energy given a set of blanket states $(\omega, \alpha)$ ``looks like'' gradient ascent toward constrained MaxEnt value.
The same appears to be true for constrained path entropy functionals in idealized equilibrium systems\cite{sakthivadivel23}, as well as any for which asymptotic MaxCal equates to MaxEnt\cite{maxwell23}.
In particular, this drives a process which underlies sentient behavior\cite{parr19,pezzulo24}.

\section{Discussion}\label{4.4}

The different modeling choices impose some challenges in interpreting these results.
However, for those interested in unifying FEP and IIT (or IIT-inspired frameworks\cite{leung23}), in particular through entropy-based means such as IMWT\cite{safron20,safron22}, the above results might point toward a general direction of inquiry.\\
\\
Active inference associates sentient behavior with the constrained drive toward maximum entropy\cite{pezzulo24}, while over longer timescales seeking to associate it with a drive toward maximizing path entropy\cite{maxwell23}.
Meanwhile IIT, over shorter timescales, appears to associate sentience in the perceptual sense with changes imposed on a system's dynamics when it is brought away from its MaxCal ensemble.
Several lines of questioning open up from here.\\
\\
If we understand active inference to be taking place at the same speed that perception does, then these two objectives would appear to be opposed.
Active inference, under this view, would serve the goal of reducing the volume of conscious content a system ``needs'' to experience.
This is not necessarily at odds with Safron's view that variational autoencoders drive the IMWT process\cite{safron22} or general principles of redundancy in information theory\cite{barlow61}.
Conceptually, it proposes a model where sentient behavior might exist to regulate, rather than induce, sentient perceptual content.\\
\\
On the other hand, when we believe active inference to be taking place over a longer timescale than perception (e.g. over the course of a lifespan), a question of trade-offs between long and short term entropy maximization is opened up.
Complex dynamics such as metastability have been shown to arise in random walks when they are optimized for long term, rather than short term, path entropy maximization\cite{burda10}.
Thus, one could explore whether perception (driven by extremely high volumes of internal orchestration and organization) arises in sufficiently complex systems when long and short term entropy maximization objectives oppose.\\
\\
Finally, since active inference may occur at many different scales, one could investigate whether spatial trade-offs underwrite this.
If it were the case that CMEP was adhered to globally better when there is a local, deviant region driving updates, then one could arrive at a functional perspective on cognition from IIT-based first principles.\\
\\
To formalize these ideas more precisely, at least three key challenges exist.
Firstly, compatibility of a more comprehensive MaxCal-based methodology would need to be assessed against IIT's ontological principles.
Secondly, whether motivated by IIT or by thermodynamic perspectives on cognition\cite{plenz21,buxton25}, one would need to reconcile the discrete mathematics of our PGMs with the continuous mathematics of Langevin equations (or, alternatively, develop some discrete approximations of the MaxEnt dualism).
Finally, a fuller reconciliation of active inference and MaxCal techniques would be needed to fully bridge this gap.\\
\\
Nonetheless, when we choose a MaxCal-based methodology, extracted from the structure of IIT, we see some results which appear promising.
When applying Friston's unifying framework FEP (which is hypothesized to be driven by active inference) to ANN-inspired models that employ a long-trajectory version of the MaxCal methodology, we retrieve an association between prediction error and information (understood as the ``substance'' of conscious experience).
Further, the mathematics used to reach this result is that of large deviations (LDP), which in physics is used to model fluctuations in a system\cite{touchette12}.
Thus, while not conclusive, there appears to be some mathematical grounds for investigating the crossover of these perspectives.

%% file: 5-IIT-FEP/5_IIT_FEP.tex
\chapter{Information as prediction error}\label{5_IIT_FEP}

\textsc{Empirical investigations} have indicated a relationship between IIT and FEP\cite{mayama25,olesen23,albantakis14}, while conceptual work has proposed mechanisms\cite{safron20,safron22,friston20}.
However, a precise mathematical relationship is yet to be formalized.
In this chapter, we use MaxCal as a unifying principle to explore one.\\
\\
``Information'' here is defined as deviation from a MaxCal ensemble of paths, and ``integration'' is defined accordingly.
We emphasize the difference to IIT's structural definitions rooted in cause/effect structure\cite{oizumi2014,albantakis2023,prentner23}, while recognizing the structural basis explored in chapter \ref{4_info_dev}.
By applying the insights of Kleiner and Tull (2021), we may understand this as \textit{an} ``integrated information theory''\cite{tull21}.
``FEP'' refers to the framework proposed by Friston\cite{friston03,friston05}.
We remain agnostic on whether active inference is the process by which it is implemented.\\
\\
In taking this approach, we find an interesting convergence:
``integration'' relates to how ``irreducibly large'' a fluctuation is across the system, echoing the thermodynamic-oriented FTD theories\cite{buisan25}.\\
\\
Further, our results indicate a nuanced relationship between these ``integrated fluctuations'' and FEP, aligning with the intuition it models many non-conscious systems\cite{friston13,friston23} in addition to the brain.
In particular, we predict ``integration'' occurs only when ``justified'' by proportionate gains to model accuracy.
Further, when IIT is applied otherwise-faithfully under our lens, it predicts that sentience relates to a maximal and irreducible combination of ``integrated fluctuations'', working to produce deviations which exceed that of individuated parts.\\
\\
We begin this chapter by introducing our IIT-inspired\cite{leung23} MaxCal deviations framework, which generalizes the method we observed in chapter \ref{4_info_dev} and then applies it to non-deterministic systems.
However, instead of modeling information as an interaction between a static state and the system's structure\cite{prentner23}, we model it in trajectory-based terms.
In section \ref{5.2} we interpret this under the lens of path space compression, while in section \ref{5.3} we demonstrate how it can be modeled as a free energy functional.\\
\\
In sections \ref{5.4}--\ref{5.5}, we progress from modeling information over a single time-step, to over arbitrarily long paths.
First we examine this from a central limit (CLT) view, and then from a large deviations (LDP) view.
In the latter, we find a convergence between MaxCal inference and adjustment of weights and bias terms, which underlies an interpretation of VFE as an L2-regularized loss function under LeCun's energy-based methods\cite{lecun06}.

\section{An IIT-inspired MaxCal deviations framework}\label{5.1}

By applying our MaxCal perspective over stochastic systems, we retrieve that unconstrained cause repertoires $\mu$ are Gibbs distributions\cite{jaynes57} with conditional entropy $h(\bm{x}^t) = \mathcal{H}( \bm{X}^{t+1} \vert \bm{X}^t = \bm{x}^t )$ being negated to form ``internal energy'' $\bm{E}( \bm{x}^t ) = -h( \bm{x}^t )$.
However, rather than emphasizing the input cause repertoire $\mu$ or output effect repertoire $\mu \bm{P}$, we focus attention on the \textit{transition network MaxCal} $(\mu \times \bm{P})( \bm{X}^t, \bm{X}^{t+1} ) = \mu( \bm{X}^t ) \bm{P}( \bm{X}^t, \bm{X}^{t+1} )$ as our primary reference point for ``no information''.\\
\\
``Caliber'' is defined as path entropy.
When we subtract that from an ``informed'' transition network $\mathcal{G} \sim \rho \times \bm{P}$ from the ``uninformed'' ensemble $\mu \times \bm{P}$, we retrieve a KL divergence term $\psi( \rho \times \bm{P} ) = \mathcal{D}( \rho \lvert \rvert \bm{P} )$.
We may also consider the action of some observed path $\bm{x}^t \rightarrow \bm{x}^{t+1}$ and see it is equal to a difference in expected and realized log-probabilities.\\
\\
In applying this procedure, there are many considerations, such as whether to interpret the probabilities ontically or epistemically\cite{floridi10}, as well as which properties we preserve when conditioning.
We do not attempt to answer these all conclusively here, but do provide a general exploratory framework.

\subsection{Defining transition networks}\label{5.1.1}

The stochastic process $\mathbb{X}$ representing our system is a single-step Markov chain $\{ \bm{X}^t, \bm{X}^{t+1} \}$, parameterized by its transition probability matrix (TPM) $\bm{P}$.
We do this so that information ``lives in the present'' at some maximally informative grain of time, as suggested in IIT 3.0\cite{oizumi2014}.
Our difference is that we understand ``the moment'' as occurring in the movement between these frames, rather than inside a fixed point.

\begin{dfn}\label{transNtwk}
	Let $\bm{X}^t \rightarrow \bm{X}^{t+1}$ be some Markov process parameterized by transition probability matrix $\bm{P}$.
	Then the \textup{transition network} $\mathcal{G} = (\bm{X}^t \sqcup \bm{X}^{t+1}, \bm{P})$, in this instance, is simply the coupling of the time-indexed variables and the transition matrix $\bm{P}$.
\end{dfn}

We will no longer assume these are Dynamic Bayesian Networks\cite{neapolitan07} to allow for non-independent input distributions over $\bm{X}^t$.
However, while our methods generalize, we shall sometimes investigate conditional independence of $\bm{X}^{t+1}$ with respect to $\bm{X}^t$ when this lends us toward parsimonious interpretation.

\subsection{Deriving unconstrained MaxCal ensembles}\label{5.1.2}

Given a variable path ensemble $\rho \times \bm{P}$ over $\mathcal{G}$, our task is to find the one $\mu \times \bm{P}$ which will maximize caliber (path entropy\cite{jaynes80}).
To this end, we apply the chain rule for entropy to express this as the sum of $\mathcal{H}(\rho)$ and a conditional entropy term $h( \rho ) = \mathcal{H}( \bm{X}^{t+1} \vert \bm{X}^t \sim \rho )$.

\begin{equation}\label{optimization}
	\mathcal{H}^{\rho}( \mathcal{G} )
	=
	\mathcal{H}(\rho)
	+
	h( \rho )
\end{equation}

Our transition probabilities $\bm{P}$ form a set of constraints on the values $h(\rho)$ may take.
Thus, this is equivalent to the constrained MaxEnt problems discussed in section \ref{4.2}.
By applying Lagrangian calculus we retrieve that $\mathcal{H}^{\rho}( \mathcal{G} )$ is maximized by a softmax distribution over conditional entropies, or equivalently a Gibbs distribution with inverse temperature $\beta = 1$ and energy functions $\bm{E}( \bm{x}^t ) = -h( \bm{x}^t )$:

\begin{equation}\label{maxEnt}
	\mu(
	\bm{x}^t
	)
	=
	\dfrac{
		e^{
			h( \bm{x}^t )
		}
	}{
		\kappa
	},
	\qquad
	\kappa = \sum_{\bm{x}} e^{ h( \bm{x} ) }
\end{equation}

Each conditional distribution $\bm{P}( \bm{x}^t, \cdot )$ may also be thought of as a Gibbs distribution over its support $\{ \bm{x}^{t+1} : \bm{P}( \bm{x}^t, \bm{x}^{t+1} ) > 0 \}$, with inverse-temperature $\beta = -1$ and energy terms $- \log \bm{P}( \bm{x}^t, \bm{x}^{t+1} )$.
Thus, when we consider the \textit{action functionals} characterizing our joint distribution $\mu \times \bm{P} \equiv \textup{diag}(\mu) \bm{P}$, while recalling that $h( \bm{x}^t )$ is an expectation over negative log-probabilities, we may interpret the \textit{action} $\mathcal{A}( \bm{x}^t, \bm{x}^{t+1} )$ as a gap between expected and realized energies\footnote{
	We have subtracted the constant $\log \kappa$ from our action terms.
}:

\begin{equation}\label{action}
	\mathcal{A}(
	\bm{x}^t, \bm{x}^{t+1}
	)
	=
	-
	\log
	\bm{P}(
	\bm{x}^t
	,
	\bm{x}^{t+1}
	)
	-
	\mathbb{E}_{\bm{X}^{t+1} \sim \bm{P}( \bm{x}^t, \cdot )} \left[
	-
	\log
	\bm{P}(
	\bm{x}^t
	,
	\bm{X}^{t+1}
	)
	\right].	
\end{equation}

Thus, $\mu \times \bm{P}$ favors paths which are ``less costly'' than the starting point $\bm{x}^t$ would imply.
With regard to IIT, $\bm{X}^t \sim \mu$ and $\bm{X}^{t+1} \sim \mu \bm{P}$ are analogous to the unconstrained cause and effect repertoires\cite{oizumi2014}.

\subsection{Constrained MaxCal ensembles}\label{5.1.3}

Applying constraints to our MaxCal ensembles is equivalent to further constraining the MaxEnt problem discussed in the previous section.
Constraints may be applied to inputs $c \equiv \bm{Y}^t = \bm{y}^t$, outputs $c' \equiv \bm{V}^{t+1} = \bm{v}^{t+1}$, or combined $d \equiv c \land c'$.
We consider these separately to reflect IIT's time-asymmetric approach toward modeling information.\\
\\
As alluded to in section \ref{5.1.1}, assuming conditional independence of $\bm{X}^{t+1}$ with respect to $\bm{X}^t$ provides us with a more physically interpretable model.
Thus, while our methods do generalize we shall focus on the case $\mathcal{H}( \bm{X}^{t+1} \vert \bm{X}^t ) = \sum_{i=1}^n \mathcal{H}( X_i^{t+1} \vert \bm{X}^t )$ when handling effective constraints $c'$.

\subsubsection*{Constrained inputs}

Applying some $c \equiv \bm{Y}^t = \bm{y}^t$ restricts the domain of our input space to $\{ \bm{y}^t \} \times \Omega_{\bm{Z}} \subset \Omega_{\bm{X}}$.
Equivalently, we may apply an indicator function $\bm{I}( \bm{x} ) = \bm{1} \{ \bm{x} \vert_{\bm{Y}} = \bm{y}^t  \}$ and set its expected value in our constrained MaxEnt (CMEP)\cite{gottwald20} problem to be equal to $1$.
In any case, over the variable nodes $\bm{Z}^t$ we retrieve a similar Gibbs distribution, though with a smaller partitioning constant.

\begin{equation}\label{maxEntInput}
	\hat{\mu}_c(
	\bm{z}^t
	)
	=
	\dfrac{
		e^{
			h_c( \bm{z}^t )
		}
	}{
		\kappa_c
	},
	\qquad
	\kappa_c = \sum_{\bm{z}} e^{ h( \bm{z} ; \bm{y}^t ) }.
\end{equation}

The constrained MaxCal input $\mu_c$ over $\bm{X}^t$ is therefore the tensor product of $\hat{\mu}_c$ and a Kronecker-Delta distribution $\bm{Y}^t \sim \delta( \bm{y}^t )$: $\mu_c \sim \delta( \bm{y}^t ) \otimes \hat{\mu}_c$.
The overall ensemble is $\mu_c \times \bm{P}$, while its output marginal is $\mu_c \bm{P}$.

\subsubsection*{Constrained outputs}

In considering constrained outputs $c' \equiv \bm{V}^{t+1} = \bm{v}^{t+1}$, we discern between \textit{observed} and \textit{inevitable} outcomes, as well as the \textit{generation} and \textit{receipt} of information.
We propose that epistemic posteriors $\mu_{c'}( \cdot \vert \bm{v}^{t+1} )$ estimate information received, while derived priors $\mu_{c'}( \cdot )$ estimate information generated.
Discussion of their differences can be found in section \ref{4.2.2}.\\
\\
The CMEP optimization from equation \ref{optimization} takes the form of optimization over a posterior $\mu_{c'}( \cdot \vert \bm{v}^{t+1} )$.
In this setting, we treat the outcome of $c'$ as certain (from an ontic point of view, this would be inevitable rather than observed).

\begin{equation}
	\mathcal{H}^{\rho}( \mathcal{G} \vert \bm{v}^{t+1} )
	=
	\mathcal{H}( \bm{X}^t \vert \bm{v}^{t+1} )
	+
	\mathcal{H}( \bm{W}^{t+1} \vert \bm{X}^t ).
\end{equation}

This retrieves an optimized posterior $\mu_{c'} \propto e^{h_{\bm{W}}( \bm{x}^t )}$, with $h_{\bm{W}}$ defined as one would expect.
It is defined over the support $\textup{Supp}( \bm{v}^{t+1} ) = \{ \bm{x}^t : \bm{P}( \bm{x}^t, \bm{v}^{t+1} ) > 0 \} \subseteqq \Omega_{\bm{X}}$ and so generally not equivalent to a na\"{i}ve maximization over $\mathcal{H}( \bm{X}^t ) + h_{\bm{W}}( \bm{x}^t )$.
Nonetheless, we can apply Bayes' Theorem to produce a \textit{derived prior} which we propose relates to an ``energy'' term generated at $\bm{X}^t$:

\begin{equation}
	\mu_{c'}( \bm{x}^t )
	=
	\dfrac{
		e^{
			\left[h_{\bm{W}}( \bm{x}^t ; \bm{v}^{t+1} )
			-
			\log \bm{P}( \bm{x}^t, \bm{v}^{t+1} )\right]
		}
	}{
		\kappa_{c'}		
	},
	\qquad
	\kappa_{c'} = \sum_{\bm{x}} \dfrac{ e^{ h_{\bm{W}}( \bm{x} ) } }{ \bm{P}( \bm{x}, \bm{v}^{t+1} ) }.
\end{equation}

When we combine our posterior $\mu_{c'}( \cdot \vert \bm{v}^{t+1} )$ with a conditional $\bm{P}_{\bm{W}}( \cdot, \cdot \vert \bm{v}^{t+1} )$ (which, in our conditionally independent case, is equivalent to the marginal $\bm{P}_{\bm{W}}$), we retrieve action functionals of the form used in equation \ref{action} with log probabilities over $\bm{W}^{t+1}$ instead of $\bm{X}^{t+1}$.
For our derived prior $\mu_{c'}$ being combined with $\bm{P}$, each transition $\bm{x}^t \rightarrow \hat{\bm{x}}^{t+1}$ will have an additional term $- \log \bm{P}( \bm{x}^t, \hat{\bm{v}}^{t+1} ) + \log \bm{P}( \bm{x}^t, \bm{v}^{t+1} )$.

\subsubsection*{Combined constraints}

By combining the methods of constrained inputs and outputs, we can obtain generalizations of the derived priors over a support $\{ \bm{y}^t \} \times \textup{Supp}_{\bm{W}}( \bm{v}^{t+1} )$ in the case of combined constraints $d \equiv c \land c'$.
We first state this in Gibbs form, before describing $\mu_d$ as a perplexity function.
To aide this, we define $\bm{P}_{c'}$ as the conditional transition matrix derived by applying Bayes' Theorem to each entry of $\bm{P}$.

\begin{thm}[MaxCal prior for non-deterministic networks]\label{MaxCalThm}
	Let $\mathcal{G} = ( \bm{X}^t \sqcup \bm{X}^{t+1}, \bm{P} )$ represent a transition network and let $d = c \land c'$ represent a set of combined constraints.
	\textup{Then,} our derived prior $\mu_d$ may be expressed as a Gibbs distribution over the space $\textup{Supp}(d) = \{ \bm{y}^t \} \times \textup{Supp}_{\bm{Z}}( \bm{v}^{t+1} )$ with temperature parameter $\beta = -1$ and energy function $\bm{E}_d( \bm{x}^t ) = \log \bm{P}( \bm{x}^t, \bm{v}^{t+1} ) - h_{\bm{W}}( \bm{x}^t )$:
	\begin{equation}\label{MaxEnt}
		\mu_d (
		\bm{x}^t
		)
		=
		\dfrac{
			e^{\left[h_{\bm{W}}(
				\bm{z}^t
				;
				\bm{y}^t
				)
				-
				\log \bm{P}( \bm{x}^t, \bm{v}^{t+1} )\right]
			}
		}{
			\kappa_d
		},
		\qquad
		\kappa_d
		=
		\sum_{\bm{x}} \dfrac{ e^{h_{\bm{W}}( \bm{x} )} }{ \bm{P}( \bm{x}, \bm{v}^{t+1} ) }.
	\end{equation}
\end{thm}

\begin{crl}[Derived priors in perplexity terms]\label{MaxCalLaw}
	Assume the conditions outlined in Theorem \ref{MaxCalThm}.
	\textup{Then}, the derived prior $\mu_d$ is proportional over $\textup{Supp}(d)$ to the perplexity of a distribution $\bm{P}_d( \bm{x}^t, \cdot )$ under a model $\bm{P}( \bm{x}^t, \cdot )$\footnote{
		See \cite{jurafsky26} for a discussion on perplexity functions.
	}.
	\begin{equation}
		\mu_d(
		\bm{x}^t
		)
		\propto
		\prod_{\bm{x}^{t+1}} \left(
		\dfrac{
			1
		}{
			\bm{P}(
			\bm{x}^t, \bm{x}^{t+1}
			)^{
				\bm{P}_d(
				\bm{x}^t, \bm{x}^{t+1}
				)
			}
		}
		\right).
	\end{equation}
\end{crl}

\subsubsection*{Exploring an energy-based interpretation}

Our use of Gibbs distributions and action functions allows one to use energy-based analogies in interpreting these results.
When we, as is common in Ising models\cite{glauber63}, interpret $\mathcal{E}( \bm{x}^t, \bm{x}^{t+1} ) = -\log \bm{P}( \bm{x}^t, \bm{x}^{t+1} )$ as some cost associated with moving from $\bm{x}^t$ toward $\bm{x}^{t+1}$, we may interpret $\texttt{constant} + \log \bm{P}( \bm{x}^t, \bm{x}^{t+1} )$ as a residual.
In particular, when each $- \log \bm{P}( \bm{x}^t, x_i^{t+1} )$ is modeled as a cost of $\bm{x}^t$ inducing the state $x_i^{t+1}$ in node $X_i^{t+1}$, we can think of $\tfrac{1}{n} \texttt{constant} + \log \bm{P}( \bm{x}^t, x_i^{t+1} )$ as its own residual.\\
\\
Turning our attention toward our unconstrained ensembles $\mu \times \bm{P}$, we see that action $\mathcal{A}( \bm{x}^t, \bm{x}^{t+1} )$ is determined by a deficit or excess of cost $- \log \bm{P}( \bm{x}^t, \bm{x}^{t+1} )$, from an expected residual value $- \log \mu( \bm{x}^t ) = \log \kappa + \mathbb{E}_{\bm{P}( \bm{x}^t, \cdot )} \left[ \log \bm{P}( \bm{x}^t, \bm{X}^{t+1} ) \right]$.
Thus, it would certainly be possible to propose a model in which $h( \bm{x}^t ) = \mathbb{E}_{\bm{P}( \bm{x}^t, \cdot )} \left[ - \log \bm{P}( \bm{x}^t, \bm{X}^{t+1} ) \right]$ represents some ``burst'' of ``energy'' released, while $\log \kappa - h( \bm{x}^t )$ models its deduction a reserve.
The value $- \log \bm{P}( \bm{x}^t, \bm{x}^{t+1} )$ would represent what in practice is demanded by movement from the system toward $\bm{x}^{t+1}$.\\
\\
Our unconstrained input marginals produce $\mathcal{H}(\mu) + h(\mu) = \log \kappa$, and thus would represent a scenario (viewing $\bm{X}^t$ ontically) in which all the reserves are used, or a volume of reserves which are necessary in the absence of information.\\
\\
Meanwhile, when we examine our constrained prior $\mu_d$, the expected burst $- \log \bm{P}_{\bm{W}}( \bm{x}^t, \bm{v}^{t+1} ) - \mathbb{E}_{\bm{P}( \bm{x}^t, \cdot )}\left[ \log \bm{P}( \bm{x}^t, \bm{W}^{t+1} ) \right]$ is deducted from a reserve of $\log \kappa_d \leqslant \log \kappa$.
Thus, perhaps $- \log \bm{P}( \bm{x}^t, \bm{v}^{t+1} )$ would represent some refined knowledge of energy which must have been expended in the transition from $\bm{X}^t$ toward $\bm{X}^{t+1}$.

\subsection{Defining information and integration}\label{5.1.4}

``Information'' in IIT is a distance between observables from a constrained and unconstrained MaxCal transition network.
While further procedures are applied to derive cause, effect, integrated, and conceptual information\cite{oizumi2014}, the unifying procedure can be summarized by this principle.\\
\\
We retain this core structure, while treating transitions instead of states as our locus.
Further, we study the MaxCal ensembles directly instead of using other observables.
Therefore, we refer to our approach as IIT-inspired\cite{leung23}.
Information, under this frame, is a reduction in caliber from the maximal value.

\begin{dfn}[Information (stochastic)]\label{stocInfo}
	Let $\mathcal{G} = ( \bm{X}^t \sqcup \bm{X}^{t+1}, \bm{P} )$ be a transition network equipped with a MaxCal path ensemble $\mu \times \bm{P}$ and placed into a distribution $\rho \times \bm{P}$.
	Then, the \textup{information} $\psi$ of $\rho \times \bm{P}$ is defined as its entropy subtracted from that of $\mu \times \bm{P}$:
	\begin{equation}\label{info}
		\psi(
		\rho \times \bm{P}
		)
		=
		\mathcal{H}^{\mu}(
		\mathcal{G}
		)
		-
		\mathcal{H}^{\rho}(
		\mathcal{G}
		).
	\end{equation}
\end{dfn}

Since $\mathcal{H}^{\mu}( \mathcal{G} )$ is equal to $\log \kappa$, this may be expressed as $\log \kappa - h(\rho) - \mathcal{H}( \rho )$.
Noting that $- h( \rho ) = - \sum_{\bm{x}} \rho( \bm{x} )h( \bm{x} )$, and that each $h( \bm{x}^t )$ may be expressed as $\log e^{h( \bm{x} )} = \log \frac{e^{h(\bm{x})}}{\kappa} + \log \kappa$, we see that $\log \kappa - h( \rho )$ is the cross-entropy term $\mathcal{H}( \rho, \mu )$.
Accordingly, we may express our information $\mathcal{H}( \rho, \mu ) - \mathcal{H}( \rho )$ as a KL divergence term.

\begin{equation}\label{info_kl_div}
	\psi( \rho \times \bm{P} )
	=
	\mathcal{D}( \rho \lvert \rvert \mu ).
\end{equation}

Since this depends only on $\rho$ and $\mu$, we shall express information $\psi(\rho \times \bm{P})$ as $\psi( \rho )$.\\
\\
To assess integration, we generalize the partitioning procedure discussed in section \ref{4.2.3}.
Hence, we formalize our definition of a \textit{partitioned network.}

\begin{dfn}[Partitioned network (stochastic)]\label{partitionedNetworks}
	Let $\mathcal{G} = ( \bm{X}^t \sqcup \bm{X}^{t+1}, \bm{P} )$ be a transition network equipped with a distribution $\rho \times \bm{P}$, and let $\mathcal{P} = \{ \bm{Y}^t, \bm{V}^{t+1} \} \sqcup \{ \bm{Z}^t, \bm{W}^{t+1} \}$ be some partition.
	\textup{Then,} we construct our \textup{partitioned network} $\mathcal{G}_{\mathcal{P}}^{\rho}$ by adhering to the following procedure:
	\begin{enumerate}
		\item Construct a conditional transition matrix $\bm{P}^{\rho}_{\bm{Y} \rightarrow \bm{V}}$ with entries $\bm{P}^{\rho}_{\bm{Y} \rightarrow \bm{V}}( \bm{y}^t, \bm{v}^{t+1} ) = \mathbb{P}( \bm{V}^{t+1} = \bm{v}^{t+1} \vert \bm{Y}^t = \bm{y}^t , \bm{Z}^t \sim \rho_{\bm{Z}} )$.
		Then, construct an analogous conditional transition matrix $\bm{P}^{\rho}_{\bm{Z} \rightarrow \bm{W}}$.
		\item Construct a \textup{partitioned transition matrix} $\bm{P}^{\rho}_{\mathcal{P}}$ by taking the tensor product of the two conditioned ones:
		\begin{equation*}
			\bm{P}^{\rho}_{\mathcal{P}} = \bm{P}^{\rho}_{\bm{Y} \rightarrow \bm{V}} \otimes \bm{P}^{\rho}_{\bm{Z} \rightarrow \bm{W}}.
		\end{equation*}
		\item Define our \textup{partitioned network} by equipping our nodes with the partitioned transition matrices: $\mathcal{G}_{\mathcal{P}}^{\rho} = ( \bm{X}^t \sqcup \bm{X}^{t+1}, \bm{P}^{\rho}_{\mathcal{P}} )$.
	\end{enumerate}
\end{dfn}

The partitioned matrix $\bm{P}^{\rho}_{\mathcal{P}}$ sends input $\bm{x}$ to the conditional distributions of $\bm{V}^t$ given each $\bm{y}^t$, while $\bm{Z}^t \sim \rho_{\bm{Z}}$ is treated as a background condition, tensor-multiplied by the analogous distributions of $\bm{W}^t$ with respect to $\rho_{\bm{Y}}$ and each $\bm{z}^t$.
Thus, $\bm{X}^{t+1}_{\mathcal{P}} \sim \rho \bm{P}^{\rho}_{\mathcal{P}}$ will produce an output that factors out interactions between $\bm{Y}^t$, $\bm{Z}^t$ (and thus is equal to $\rho \bm{P}$ for $\bm{Y}^t \perp_{\rho} \bm{Z}^t$).\\
\\
A measure more faithful to IIT's canonical formulations would be comparison of the outputs $\mu^{\rho}_{\mathcal{P}} \bm{P}^{\rho}_{\mathcal{P}}$ and $\rho \bm{P}$, similar distributions representing a lack of integration.
However, this is hard to interpret within our MaxCal framework and so we will instead adapt IIT's algorithm\cite{tull21} to how we have defined integration.
Since ``information'' for us means deviation from a MaxCal ensemble, we will define ``integration'' as whether there exists more, an equal amount, or less of it across some partition.
In essence, we understand there to be one transition distribution $\rho \times \bm{P}$ and are interested in identifying where it ``stands out'' against na\"{i}ve assumptions the most.\\
\\
To this end, we define conditional entropies $h^{\rho}_{\mathcal{P}}( \bm{x}^t ) = h^{\rho}_{\bm{V}}(\bm{y}^t) + h^{\rho}_{\bm{W}}(\bm{z}^t)$, partition constants $\kappa^{\rho}_{\mathcal{P}} = \kappa^{\rho}_{\bm{Y} \rightarrow \bm{V}} \cdot \kappa^{\rho}_{\bm{Z} \rightarrow \bm{W}}$, and MaxCal distributions $\mu^{\rho}_{\mathcal{P}} = \mu^{\rho}_{\bm{Y} \rightarrow \bm{V}} \otimes \mu^{\rho}_{\bm{Z} \rightarrow \bm{W}}$ for our transition network.
The \textit{partitioned information} $\psi^{\rho}_{\mathcal{P}}( \rho )$ is just the information of $\rho \times \bm{P}^{\rho}_{\mathcal{P}}$ in our partitioned network $\mathcal{G}_{\mathcal{P}}^{\rho}$.
We can now define integration.

\begin{dfn}[Integration $\mathcal{P}$]\label{intInfoPartition}
	Let $\mathcal{G} = ( \bm{X}^t \sqcup \bm{X}^{t+1}, \bm{P} )$ be a transition network equipped with a path ensemble $\rho \times \bm{P}$.
	Let $\mathcal{G}^{\rho}_{\mathcal{P}}$ be its partitioned network.
	Then, we define \textup{integration} as the difference in partitioned and non-partitioned information:
	\begin{equation}\label{intInfoPartitionEqn}
		\phi^{\mathcal{P}}(
		\rho
		)
		=
		\psi(
		\rho
		)
		-
		\psi^{\mathcal{P}}(
		\rho
		).
	\end{equation}
\end{dfn}

We note that integration $\phi^{\mathcal{P}}(\rho)$ across any partition depends only on the conditional entropy $h(\rho)$ and MaxCal values $\kappa$, $\kappa^{\rho}_{\mathcal{P}}$.
It does not depend on the entropy $\mathcal{H}( \rho )$ of inputs $\bm{X}^t$.

\begin{equation}\label{int_info_conditional}
	\phi^{\rho}_{\mathcal{P}}( \rho )
	=
	\log
	\frac{ \kappa }{ \kappa^{\rho}_{\mathcal{P}} }
	-
	h( \rho )
	+
	h^{\rho}_{\mathcal{P}}( \rho )
\end{equation}

Taking inspiration from the principle of maximal existence\cite{albantakis2023}, we define the \textit{integration} of $\rho \times \bm{P}$ overall to be the minimum of $\phi^{\mathcal{P}}( \rho )$ across all partitions:

\begin{equation}\label{integration}
	\phi( \rho )
	=
	\min_{\mathcal{P}}
	\phi^{\mathcal{P}}( \rho ).
\end{equation}

\subsection{Comparison to canonical IIT methods}\label{5.1.5}

While we have largely adopted the structure of IIT\cite{tull21}, we have deviated from its ontological definition of information\cite{oizumi2014,albantakis2023}.
Nonetheless, we are investigating a structure implicit in IIT's formalism\cite{oizumi2014}, including our MaxCal structures and the partitioning procedure.
The differences lie in how these structures are applied, with IIT capturing intrinsic causal structure and our framework focusing on the restriction of movement.\\
\\
In adopting a ``MaxCal-first'' formalism, we have been able to derive physically intuitive interpretations of cause and effect repertoires, as well as theoretical links toward free energy based methods (discussed in later sections).
Thus, while the precise relationship between $\Phi$ and MaxCal deviations remains unproven, we propose that it may underlay some of the empirical results relating integrated information to FEP measures\cite{albantakis14,olesen23,mayama25}.\\
\\
In the upcoming sections, we build an understanding of information as restriction of path space over transitions, before proposing theoretical relationships between free energy, integration, and information.

\section{Information as path-space compression}\label{5.2}

In the previous section, we developed a formalism for what information \textit{is.}
Here, we examine what it \textit{does.}\\
\\
\textit{Perplexity} is a metric used to assess the effective state space size of a distribution\cite{jurafsky26}.
We may employ it here to develop an intuition of information as an effective ``compression factor'' on the path space.
In particular, $\psi(\rho)$ measures how much the effective number of paths over $\bm{X}^t \rightarrow \bm{X}^{t+1}$ shrinks when $\bm{X}^t$ is distributed as $\rho$ rather than $\mu$.

\subsection{Introducing perplexity}\label{5.2.1}

Perplexity was first introduce in language modeling to assess the complexity of speech recognition tasks\cite{jelinek77}.
It has since been widely adopted to assess the performance of language models, quantifying the effective branching factor of a model\cite{jozefowicz16,cui25}.
Intuitively, it reflects the fact that a Kronecker-Delta distribution will place its mass on just one element, while a uniform distribution will traverse the entire state space.\\
\\
It is defined as the inverted geometric mean of a probability mass function\cite{jurafsky26}:

\begin{equation}\label{perp}
	\textup{Perp}( \rho )
	=
	\prod_{\bm{x} \in \Omega}
	\rho( \bm{x} )^{
		-
		\rho( \bm{x} )
	}.
\end{equation}

Equivalently, perplexity is an exponentiation of the entropy functional: $\textup{Perp}(\rho) = e^{\mathcal{H}( \rho )}$.
Sometimes, the cross-entropy $\mathcal{H}( \rho, \mu )$ can be used to define a ``cross perplexity'' term $\textup{Perp}( \rho, \mu ) = e^{\mathcal{H}( \rho, \mu )}$, which equivalently is the inverted geometric mean of $\mu$ weighted by $\rho$.
In these instances, we interpret it as the volume of state space a model $\mu$ must traverse to produce the distribution $\rho$.\\
\\
Our analysis is largely built around conditional entropy functions $h(\bm{x}^t)$, and so we will use the term $\textup{PP}( \bm{x}^t )$ to refer to the perplexity of our conditional distribution $\bm{P}( \bm{x}^t, \cdot )$.
In fact, shall generalize this further:

\begin{equation}\label{cond_perp}
	\textup{PP}( \rho )
	=
	\textup{Perp}( \bm{X}^{t+1} \vert \bm{X}^t \sim \rho ).
\end{equation}

\subsection{MaxCal as a uniform measure over paths}\label{5.2.2}

Our MaxCal ensemble $\mu \times \bm{P}$ has an input marginal $\mu$, and we now see each $\mu( \bm{x}^t )$ is proportional to $\textup{PP}( \bm{x}^t )$.
Straightforwardly, we observe each $\bm{x}^t$ is weighted by its branching capacity.\\
\\
The partition constant $\kappa = \sum_{\bm{x}} \textup{PP}( \bm{x} )$ is the sum over effective path spaces, but it is also the maximal value of $\textup{Perp}( \bm{X}^t, \bm{X}^{t+1} ) = e^{\mathcal{H}( \bm{X}^t, \bm{X}^{t+1} )}$ and therefore itself quantifies the maximal --- and therefore, total available --- volume of paths.\\
\\
Thus, by weighting each $\bm{x}^t$ by its effective number of available paths, we ensure that probability mass is spread as evenly as possible across $\bm{X}^t \rightarrow \bm{X}^{t+1}$.
This serves the same function as applying a uniform distribution to some unconstrained state space $\Omega$.
We simply have factored in the fundamental constraints $\bm{P}$.

\begin{equation}\label{mu_as_uniform}
	\mu( \bm{x}^t )
	\approx
	\dfrac{
		\#\{
		\textup{paths}
		\
		\bm{x}^t
		\rightarrow
		\bm{X}^{t+1}
		\}
	}{
		\#\{
		\textup{paths}
		\
		\bm{X}^t
		\rightarrow
		\bm{X}^{t+1}
		\}
	}.
\end{equation}

With regard to our derived priors, corollary \ref{MaxCalLaw} tells us that each $\mu_d( \bm{x}^t )$ is weighted by the effective volume of state space a model governed by $\bm{P}( \bm{x}^t, \cdot )$ must inhabit to encode the conditional transition probability $\bm{P}_d( \bm{x}^t, \cdot )$.

\subsection{Information as shrinkage of path space}\label{5.2.3}

The information of an ensemble $\delta( \bm{x}^t ) \times \bm{P}$ is precisely the inverse of the interpretation expressed in equation \ref{mu_as_uniform}.
Since $\mathcal{D}( \delta( \bm{x}^t ) \lvert \rvert \mu ) = -\log \mu( \bm{x}^t )$, we retrieve that $\psi( \bm{x}^t )$ is equal to $\kappa / \textup{PP}( \bm{x}^t )$ --- the total effective number of paths divided by the effective path space from $\bm{x}^t$.\\
\\
We can generalize this further when examining $\psi( \rho )$.
Since information is defined as $\mathcal{H}^{\mu}( \mathcal{G} ) - \mathcal{H}^{\rho}( \mathcal{G} )$, we retrieve that exponentiating information provides us with an ``inverted scale factor'' over the effective number of paths.

\begin{equation}\label{info_log_ratio}
	e^{
		\psi( \rho )
	}
	\
	=
	\
	\dfrac{
		\kappa
	}{
		\textup{Perp}(
		\rho
		\times
		\bm{P}
		)
	}
	\
	\approx
	\
	\dfrac{
		\#\{
		\textup{paths}
		\
		\bm{X}^t
		\rightarrow
		\bm{X}^{t+1}
		\}
	}{
		\#\{
		\textup{paths}
		\
		\rho
		\rightarrow
		\rho
		\bm{P}
		\}
	}.
\end{equation}

We can also understand $\psi( \rho )$ as a difference in entropy terms: $\mathcal{D}( \rho \lvert \rvert \mu ) = \mathcal{H}( \rho , \mu ) - \mathcal{H}( \rho )$, and therefore of $e^{\psi(\rho)}$ as a ratio of perplexity terms:

\begin{equation}
	e^{
		\psi(\rho)
	}
	\
	=
	\
	\dfrac{
		\textup{Perp}(
		\rho
		,
		\mu
		)
	}{
		\textup{Perp}(
		\rho
		)
	}.
\end{equation}

Thus, within this single-step context, $e^{\psi( \rho )}$ is related to the additional volume of state space which must be searched to produce the probability distribution $\rho$ when this is done by a model governed by $\mu$.

\subsection{Integration as a maximum compression-factor}\label{5.2.4}

Since exponentiated information $e^{\psi(\rho)}$ is the shrinkage factor over $\mathcal{G}$ imposed by $\bm{X}^t \sim \rho$, we see that $e^{\psi^{\mathcal{P}}(\rho)}$ must govern the effective shrinkage factor across $\mathcal{P}$ imposed by $\bm{X}^t \sim \rho$.
Hence, when $\phi( \rho ) > 0$ we observe in $\mathcal{G}$ a local maximum of ``paths destroyed'' by the input distribution $\bm{X}^t \sim \rho$.\\
\\
An equivalent way to interpret some $e^{\phi^{\mathcal{P}}( \rho )}$ is as a measure of whether partitioning destroys total or realized path space more.
$\kappa^{\rho}_{\mathcal{P}}$ quantifies the effective number of available paths in our partitioned network $\mathcal{G}^{\rho}_{\mathcal{P}}$, while $\textup{PP}^{\rho}_{\mathcal{P}}( \rho )$ is that given the input distribution $\rho$.
When $\textup{PP}^{\rho}_{\mathcal{P}}( \rho ) / \textup{PP}( \rho )$ is larger than $\kappa^{\rho}_{\mathcal{P}} / \kappa$, integration across some partition $\mathcal{P}$ is positive.

\begin{equation}
	\begin{split}
		e^{\phi^{\mathcal{P}}}( \rho )
		\
		&=
		\
		\dfrac{ \kappa }{ \kappa^{\rho}_{\mathcal{P}} }
		\cdot
		\dfrac{
			\textup{Perp}(
			\rho \times \bm{P}^{\rho}_{\mathcal{P}}
			)
		}{
			\textup{Perp}(
			\rho \times \bm{P}
			)
		}
		\\
		&\approx
		\
		\dfrac{
			\#\{
			\textup{paths}
			\
			\bm{X}^t
			\overset{\bm{P}}\rightarrow
			\bm{X}^{t+1}
			\}
		}{
			\#\{
			\textup{paths}
			\
			\bm{X}^t
			\overset{\bm{P}^{\rho}_{\mathcal{P}}}\rightarrow
			\bm{X}^{t+1}
			\}
		}
		\cdot
		\dfrac{
			\#\{
			\rho
			\overset{\bm{P}^{\rho}_{\mathcal{P}}}\rightarrow
			\rho
			\bm{P}^{\rho}_{\mathcal{P}}
			\}
		}{
			\#\{
			\rho
			\overset{\bm{P}}\rightarrow
			\rho
			\bm{P}
			\}
		}.
	\end{split}
\end{equation}

\section{MaxCal deviations as free energy}\label{5.3}

When we extend the energy-based interpretation explored in section \ref{5.1.3}, we may conceive of information as a free energy functional.
In particular, when we let our ``residual'' determine an ``internal energy'' function $\bm{E}( \bm{x}^t ) = \log \kappa - h( \bm{x}^t )$, and then subtract entropy from its expected value, we obtain a Helmholtz-style free energy functional:

\begin{equation}\label{maxcal_freeEn}
	\mathcal{F}[\rho]
	=
	\langle
	\bm{E}
	\rangle_{\rho}
	-
	\mathcal{H}( \rho ),
	\qquad
	\bm{E}( \bm{x}^t ) = \log \kappa - h( \bm{x}^t )
\end{equation}

This is a result of the CMEP and free energy equivalence described by Gottwald and Braun\cite{gottwald20}.
It allows us to interpret ``integration'' as a global maximum of such a functional:

\begin{equation}
	\phi^{\mathcal{P}}( \rho )
	=
	\mathcal{F}[\rho]
	-
	\mathcal{F}^{\rho}_{\mathcal{P}}[\rho].
\end{equation}

This does not automatically imply an equivalence to variational free energy functionals\cite{gottwald20}, however there are conditions in which one might be proposed, which we specify below.
To achieve this, we carefully study Friston's FEP framework\cite{friston03}.\\
\\
The generative model $p( \bm{\eta}, \gamma ) = p( \bm{\eta} )p( \gamma \vert \bm{\eta} )$ exists over the external environment and in this case is modeled by an internal ``inverse model'' $\hat{p}( \bm{x}, \gamma )$.
There is no requirement that the inverted model follows the same factorization structure as $p$.\\
\\
Moreover, while the definition of intractability is clear in computational terms, it is less obvious how this presents itself in physical/biological terms.
If we take a working definition that ``intractable'' represents a physically inaccessible state, we may apply this to our free energy functional.\\
\\
From an internal perspective, if (as is the case within our partitioned networks) we understand $\bm{P}$ as dependent on some blanket path $\gamma \equiv \bm{b}^t \rightarrow \bm{b}^{t+1}$, then we might consider $\mu_{\gamma} \times \bm{P}_{\gamma}$ to constitute part of an internal model $\hat{p}( \bm{x}, \gamma ) = \hat{p}( \gamma ) (\mu_{\gamma} \times \bm{P}_{\gamma})( \bm{X}^t, \bm{X}^{t+1} \vert \gamma )$ over the internal paths $\bm{X}^t \rightarrow \bm{X}^{t+1}$.
Conversely, $\hat{q} \equiv \rho \times \bm{P}_{\gamma}$ could represent a physically realized trajectory which occurred due to the inputs $\bm{X}^t \sim \rho$ combining with the observations $\gamma$ via the conditional TPM $\bm{P}_{\gamma}$ --- in effect playing the role of an approximate posterior.
From here, we can relate $\mathcal{F}$ to a variational functional which relates to our stated internal model:

\begin{equation}
	\begin{split}
		\mathcal{F}^{\textup{int}}[\hat{q}]
		&= \mathbb{E}_{\rho \times \bm{P}}[
		- \log \hat{p}( \gamma ) ( \mu_{\gamma} \times \bm{P}_{\gamma} )( \bm{X}^t, \bm{X}^{t+1} )
		]
		-
		\mathbb{E}_{\rho \times \bm{P}}[
		- \log ( \rho \times \bm{P} )( \bm{X}^t, \bm{X}^{t+1} )
		]
		\\
		&= \mathcal{F}[\rho]
		-
		\log \hat{p}( \gamma ).		
	\end{split}
\end{equation}

The final assumption required to relate this to the free energy principle would be to assume the existence of external trajectories $\bm{\eta}^t \rightarrow \bm{\eta}^{t+1}$ which are conditionally independent of $\bm{X}^t \rightarrow \bm{X}^{t+1}$ with respect to $\gamma$, and then to assume that $\bm{P}_{\gamma}$ is well-calibrated enough for $\hat{p}( \bm{X}^t \rightarrow \bm{X}^{t+1}, \gamma ) = \hat{p}( \gamma ) (\mu_{\gamma} \times \bm{P}_{\gamma})( \bm{X}^t, \bm{X}^{t+1} )$ to sufficiently invert/approximate $p( \bm{\eta}, \gamma ) = p( \bm{\eta} )p( \gamma \vert \bm{\eta} )$\cite{friston03,friston05}.
Simultaneously, the realized path ensemble $\hat{q}( \bm{X}^t \rightarrow \bm{X}^{t+1} \vert \gamma ) \equiv ( \rho \times \bm{P}_{\gamma} )( \bm{X}^t, \bm{X}^{t+1} )$ would need to characterize an approximate posterior $q( \bm{\eta}^t \rightarrow \bm{\eta}^{t+1} \vert \gamma )$ over external paths --- which could be a product of the same relationship that relates $\hat{p}$ to $p$.\\
\\
These assumptions combined would provide us with a \textit{variational free energy} functional $\mathcal{F}_V$ which is related to our MaxCal free energy $\mathcal{F}$:

\begin{equation}
	\mathcal{F}_V[q]
	\approx
	\mathcal{F}[ \rho ]
	-
	\log p( \gamma ).
\end{equation}

``Inference'' would correspond to being able to appropriately generate the correct distributions $\rho$ (which, in the presence of reasonably consistent sensory input, could be stationary distributions), while ``learning'' would entail performance is highest for the commonly encountered sensory paths $\gamma$.

\section{Information at the central limit}\label{5.4}

The cortex contains approximately 16 billion neurons\cite{herculano_houzel09}, each of which fire (in a wakeful state) at the order of $15.7$ Hz\cite{steriade01}.
When we look inside neurons toward ion gates and action potentials, there volume of activity per second significantly increases\cite{hodgkin52}.
Conversely, experiments indicate that conscious intention takes approximately $0.2$s--$1.4$s to form\cite{matsuhashi08}, while a flicker of visual perception will form at approximately $200$ ms\cite{rutiku16}.\\
\\
It therefore seems reasonable to explore models in which ``information'' accumulates over many interactions, which we shall express with Markov chains $\{ \bm{X}^t : t \in \mathbb{T} \}$ over arbitrarily long timescales $\mathbb{T} = \{ 0, 1, \ldots, T \}$.
Rather than analyzing these chains directly, we shall examine these under two coarse-grained views.
The first shall be under central limit theorem for Markov chains\cite{billingsley61}, which models the empirical distribution with a normal approximation.\\
\\
We find that, under this view, ``information'' can be interpreted as Bayesian surprise within a predictive coding VFE functional.

\subsection{Defining our generative model}\label{5.4.1}

Central Limit Theorem for Markov chains states that homogenous, irreducible, aperiodic $\mathbb{X}$ over finite state space can be approximated by a normal distribution about its stationary distribution $\pi$\cite{billingsley61,ricci21}.
In particular, our empirical distribution $\rho^T = \sum_{t=0}^T \delta( \bm{X}^t )$ is normally distributed about a mean $\pi$ and a covariance matrix $\Sigma$ which shrinks with $T$:

\begin{equation}\label{clt_markov_statement}
	\rho^T
	\approx
	\mathcal{N}(
	\pi
	,
	\tfrac{1}{T+1} \Sigma_x
	),
	\qquad
	\pi = \pi( \bm{P} ),
	\quad
	\Sigma_x = \Sigma_x( \bm{P} ).
\end{equation}

In considering how this relates to observations, as explored in section \ref{5.3}, we might consider $\bm{X} \sim \pi$ to be predicted by a set of internal parameters characterized by transition matrix $\bm{P}$.
These parameters could generate a stationary distribution $\pi$ and covariance factor $\Sigma_x$, or be optimized to align smoothly with these.
In an isolated system, the dynamics governing $\rho^T$ should be solely determined by those which produce $\pi$.
However, here $\bm{X}$ receives sensory data from some blanket variable $\bm{B} \sim \delta(b)$.\\
\\
We shall consider a conditional transition matrix $\bm{P} \vert b$ to still be characterized as internal parameters, which allows $\hat{p}( \rho^T ) \sim \mathcal{N}( \rho^T ; \pi, \tfrac{1}{T+1}\Sigma_x )$ to functionally be framed as predictive of $\bm{B}$, assuming they are calibrated so that $\hat{p}( b \vert \rho^T ) \sim \mathcal{N}( b ; f( \rho^T ), \Sigma_b )$.
We then might consider the empirical distribution $\rho$ which is actually realized to be an approximate posterior $\hat{q}( \rho^T \vert b ) \sim \delta(\rho)$.
In doing this, we would retrieve the standard predictive coding toolkit\cite{buckley17,millidge21}, and can define a VFE functional:

\begin{equation}\label{freeEn_clt}
	\mathcal{F}^{\textup{clt}}_V[ q ]
	=
	\frac{1}{2}
	\left(
	\lvert \Omega_{\bm{X}} \rvert
	+
	\lvert \Omega_{\bm{B}} \rvert
	\right)
	\log( 2\pi )
	+
	\frac{1}{2}
	\log
	\lvert
	\Sigma_b
	\Sigma_x^{(T+1)}
	\rvert
	+
	\frac{1}{2}
	\epsilon_b
	\Sigma_b^{-1}
	\epsilon_b^{\top}
	+
	\frac{T+1}{2}
	\epsilon_x
	\Sigma_x^{-1}
	\epsilon_x^{\top}.
\end{equation}

Here, we have that $\epsilon_b = f( \rho^T ) - b$ is a prediction error over sensory data and $\epsilon_x = \rho - \pi$ is a prediction error over the generative model.
We note that $\epsilon_x$ in practice should quantify the ability of the paths of $\bm{x}$ to be altered by sensory data.\\
\\
To reconcile this with Friston's FEP, we would need to assume that $\hat{p}( \bm{x} )$ appropriately models some latent variable $p( \bm{\eta} )$ and that $p( \bm{\eta} )p( \bm{b} \vert \bm{\eta} ) \approx \hat{p}( \bm{x} ) \hat{p}( \bm{b} \vert \bm{x} )$\cite{friston03}.

\subsection{Information as MaxEnt deviation}\label{5.4.2}

Within this framework, we need to understand ``information'' as an internal property and also as a property of states, due to using the empirical distribution.
We note that $\rho^T \sim \mathcal{N}( \pi, \tfrac{1}{T+1}\Sigma_x )$ represents a MaxEnt distribution given mean $\pi$ and variance $\tfrac{1}{T+1} \Sigma_x$.
More pertinently, it is approximates the most unbiased empirical distribution given only knowledge of $\bm{P}$.\\
\\
From an internal perspective, the realized path $( \bm{x}^0, \bm{x}^1, \ldots, \bm{x}^T )$ is a full set of constraints over our timescale $\mathbb{T}$.
Since we are coarse-graining, the resulting empirical distribution $\rho^T$ characterizes our fully constrained network.\\
\\
Since probabilities $\bm{P}$ are epistemic in this framing, we should use action to characterize our deviation (otherwise information would be the same for all $\rho^T$).
Under this frame, we retrieve an expression for information:

\begin{equation}\label{info_clt}
	\psi_{\textup{clt}}( \delta( \rho ) )
	=
	\frac{ \lvert \Omega_{\bm{X}} \rvert }{ 2 }
	\log( 2\pi )
	+
	\frac{1}{2(T+1)}
	\log
	\lvert
	\Sigma_x
	\rvert
	+
	\frac{T+1}{2}
	\epsilon_x
	\Sigma_x^{-1}
	\epsilon_x^{\top}.
\end{equation}

The consistent way to approach ``integration'' in this setting would be to, for a partition $\mathcal{P} = \{ \mathbb{Y}, \mathbb{Z} \}$, would be to use conditional stationary distributions $\pi_y^{\rho}$, $\pi_z^{\rho}$ and covariance functions $\Sigma_y^{\rho}$, $\Sigma_z^{\rho}$ and combine them into a partitioned distribution $\mathcal{N}( \rho^T ; \pi^{\rho}_{\mathcal{P}}, \Sigma^{\rho}_{\mathcal{P}} ) = \mathcal{N}( \pi^{\rho}_y, \Sigma^{\rho}_y ) \otimes \mathcal{N}( \pi^{\rho}_z, \Sigma^{\rho}_z )$.
From here, we can construct a partitioned information function $\psi_{\textup{clt}}^{\mathcal{P}}( \delta(\rho) )$ and then use it to construct integration $\phi_{\textup{clt}}^{\rho}$:

\begin{equation}\label{integration_clt}
	\phi^{\mathcal{P}}_{\textup{clt}}( \delta( \rho ) )
	=
	\frac{1}{2( T+1 )}
	\log
	\lvert
	\Sigma_x
	(\Sigma^{\rho}_{\mathcal{P}})^{-1}
	\vert
	+
	\frac{T+1}{2}
	\left(
	\epsilon_x
	\Sigma_x^{-1}
	\epsilon_x^{\top}
	-
	\epsilon_x^{\mathcal{P}}
	(\Sigma^{\rho}_{\mathcal{P}})^{-1}
	\epsilon_x^{\mathcal{P}, \top}
	\right).
\end{equation}

\subsection{Information as model complexity}\label{5.4.3}

We may combine equations \ref{freeEn_clt} and \ref{info_clt} to develop an interpretation of the role information plays under this regime:

\begin{equation}\label{freeEn_clt_psi}
	\mathcal{F}^{\textup{clt}}_V[q]
	=
	\psi_{\textup{clt}}( \delta(\rho) )
	+
	\frac{1}{2}
	\epsilon_b
	\Sigma_b^{-1}
	\epsilon_b^{\top}
	+
	\frac{1}{2}
	\log
	\lvert
	\Sigma_b
	\rvert
	+
	\frac{ \lvert \Omega_{\bm{B}} \rvert }{ 2 }
	\log( 2\pi ).
\end{equation}

In effect, information is the Bayesian surprise of our free energy functional, which quantifies model complexity.
We can use this to propose how information (as a MaxEnt deviation) may evolve under a predictive coding scheme.
If information is high while $\mathcal{F}^{\textup{clt}}_V$ is low, this would imply it is justified by sufficient reductions in model accuracy, $\epsilon_b \Sigma_b^{-1} \epsilon_b^{\top}$.
This could arise when sensory data is complex and therefore the model must accurately predict it.
However, we also see that given any particular level of accuracy, VFE is minimized when $\psi_{\textup{clt}}$ is low.
Thus, our model would not ``want'' information to be higher than it needs to be.\\
\\
On the topic of integration, which requires us to partition systems, it is worth creating some corresponding definitions for free energy.
If given a partition $\mathcal{F}$ we can assume existence of analogous free energy functionals $\mathcal{F}^{y, \textup{clt}}_V$, $\mathcal{F}^{z, \textup{clt}}_V$ which relate to marginal probabilities and observed empirical distributions.
When we subtract these from $\mathcal{F}^{\textup{clt}}_V$, we retrieve a series of difference terms:

\begin{equation}\label{freeEn_diff_clt_psi}
	\Delta_{\mathcal{P}} \mathcal{F}^{\textup{clt}}_V
	=
	\phi^{\mathcal{P}}_{\textup{clt}}( \delta(\rho) )
	+	
	\frac{1}{2} \Delta_{\mathcal{P}}( \epsilon_b )
	+
	\frac{1}{2} \Delta_{\mathcal{P}} \log \lvert \Sigma_b \rvert
	+
	\frac{1}{2}	\Delta_{\mathcal{P}} \lvert \Omega \rvert \log 2 \pi.
\end{equation}

Here, our integrated information functional $\phi^{\mathcal{P}}_{\textup{clt}}( \delta(\rho) )$ represents the difference between partitioned and non-partitioned internal prediction errors.
Our $\Delta_{\mathcal{P}}( \epsilon_b ) = \epsilon_b \Sigma_b^{-1} \epsilon_b^{\top} - \epsilon_{b, z \vert y} \Sigma_{b, z \vert y}^{-1} \epsilon_{b, z \vert y}^{\top} - \epsilon_{b, y \vert z} \Sigma_{b, y \vert z}^{-1} \epsilon_{b, y \vert z}^{\top}$ is the difference between the capacity of $\bm{X} \sim \rho$ to predict blanket states $\bm{B}$, and $\bm{Y} \sim \rho_Y$, $\bm{Z} \sim \rho_{\bm{Z}}$ to predict their respective blanket states $\bm{B} \cup \bm{Z} \vert_{\bm{Y}}$, $\bm{B} \cup \bm{Y} \vert_{\bm{Z}}$.\\
\\
The $\Delta_{\mathcal{P}} \log \lvert \Sigma_b \rvert = \log \lvert \Sigma_b \Sigma_{b, z \vert y}^{-1} \Sigma_{b, y \vert z}^{-1} \rvert$ term signals a difference in complexity between the predictions made by $\bm{X}$ and those made by $\bm{Y}$, $\bm{Z}$, while $\Delta_{\mathcal{P}} \lvert \Omega \rvert$ is a difference in state space size between $\lvert \Omega_{\bm{X}} \rvert + \lvert \Omega_{\bm{B}} \rvert$ over $\bm{X}$, and that $\lvert \Omega_{\bm{Y}} \rvert + \lvert \Omega_{\bm{B} \vert y} \Omega_{\bm{Z} \vert y} \rvert$ over $\bm{Y}$ and an analogous one over $\bm{Z}$.\\
\\
While a complicated formula, we can extract some predictions around the relationship between VFE and integration under this model.
To achieve this, we assume that VFE is minimized across whichever partition of $\mathcal{F}^{\mathcal{P}, \textup{clt}}_V$ is largest, and that this over time should push the system to oscillate around $\Delta_{\mathcal{P}} \mathcal{F}^{\textup{clt}}_{\mathcal{P}} = 0$.
So, $\Delta_{\mathcal{P}} \mathcal{F}^{\textup{clt}}_V > 0$ should indicate global updating, while $\Delta_{\mathcal{P}} \mathcal{F}^{\textup{clt}}_V < 0$ should indicate local updating.
With this in mind, we note that partitioning increases the volume of blanket predictions and blanket prediction errors, while reducing the volume of internal states, so when we discount integration (i.e. set $\phi^{\mathcal{P}}_{\textup{clt}}( \delta(\rho) ) = 0$) we should see a natural trade-off between complexity of internal and blanket inference if the system oscillates around $\Delta_{\mathcal{P}} \mathcal{F}^{\textup{clt}}_{\mathcal{P}} = 0$.\\
\\
When a system achieves this balance, sensory data $b$ which produces a path $\phi^{\mathcal{P}}_{\textup{clt}}( \delta(\rho) ) > 0$ could cause $\Delta_{\mathcal{P}} \mathcal{F}^{\textup{clt}}_{\mathcal{P}} > 0$ and push the system toward global updating, increasing global coordination.
Conversely, if $\Delta_{\mathcal{P}} \mathcal{F}^{\textup{clt}}_{\mathcal{P}} < 0$ while $\phi^{\mathcal{P}}_{\textup{clt}}( \delta(\rho) ) > 0$, this should indicate that the ability of $\bm{X} \sim \delta(\rho)$ to explain $\bm{B} \sim \delta(b)$ exceeds the ability of internal parameters $\bm{P}$ to explain $\bm{X} \sim \rho$, and in particular the ability of $\bm{Y}$ and $\bm{Z}$ to explain each other.
Thus, this should specifically drive updating across the partition.\\
\\
Finally, if the system is integrated and stable, with $\Delta_{\mathcal{P}} \mathcal{F}^{\textup{clt}}_{\mathcal{P}} = 0$ while $\phi^{\mathcal{P}}_{\textup{clt}}(\delta( \rho )) > 0$, we would infer that this is required due to the structure of blanket predictions.
If high mutual information exists between $\bm{B}_{y}$ and $\bm{B}_z$, then a determinant of $\Sigma_b$ which is lower than that of $\Sigma_{b, z \vert y} \otimes \Sigma_{b, y \vert z}$ could occur when the distributions $\rho_{\bm{Y}}$, $\rho_{\bm{Z}}$ are insufficient for predicting each other and knowledge of non-independent $\rho$ is required to predict $\bm{B}$.
This would mean that components $\bm{Y}$, $\bm{Z}$ of the system predict each other (from an internal) poorly, but that their combined state $\rho$ predicts blanket states well.\\
\\
From a larger perspective, we can echo the hill-shaped trajectory observed by Mayama et al.\cite{mayama25}.
The system starts in some equilibrium well-aligned with its sparse structure.
Upon introduction of sensory shocks introduces global internal prediction error, causing the system to reorganize.
Once it has found a partition (if one exists) which is sufficient for predicting sensory data in a distributed fashion, it reduces $\phi^{\mathcal{P}}( \delta( \rho ) )$ to the minimum volume necessary required to balance the accuracy of global blanket predictions against that of local ones.
High $\phi^{\mathcal{P}}( \delta(\rho) )$ should generally indicate high(er) $\psi^{\mathcal{P}}( \delta(\rho) )$, which in our model is equivalent to Bayesian surprise.\\
\\
Another high-level outcome of our model is that integration should never be higher than is required --- a finding broadly observed in studies of integrated information and Bayesian inference\cite{albantakis14,olesen23}.\\
\\
Nonetheless, we note that our predictions remain speculative and require empirical validation.
Thus, we present this as a proposed bridge between the domains which could explain existing results\cite{albantakis14,olesen23,mayama25}, rather than a ground truth.

\section{Large deviations in neuronal networks}\label{5.5}

While our CLT view provides a model of how integration and free energy might relate at the state level, it does not assist us with modeling trajectories.
Further, its biological plausibility is unclear.\\
\\
In this section, we build a trajectory-based model through the methods of large deviations\cite{touchette12}.
Inspired by artificial neural networks\cite{roberts22}, we parameterize the transition probabilities in $\bm{P}$ using bias and weight variables which produce lognormal\cite{buzsaki14,petersen16} firing strengths and logistic normal\cite{marreiros08} activations.\\
\\
Through this approach, we can identify the system performing MaxCal with it performing linear updates to the bias and weight variables.
Minimization of VFE trades off this objective against the costs of performing this computation.
Further, it aligns with optimization of an L2-regularized loss function under LeCun's energy-based methods\cite{lecun06}.

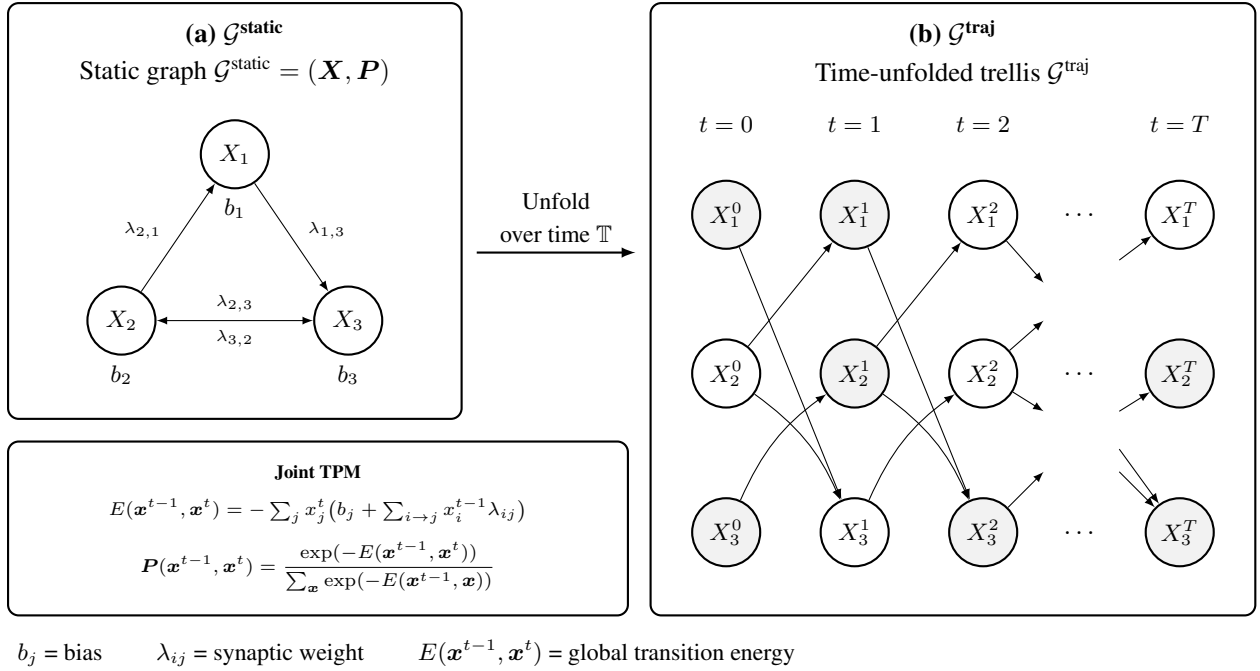
\begin{figure}[htbp]
	\centering
	\begin{tikzpicture}[
		>=latex,
		node distance=2cm and 1.8cm,
		state/.style={circle, draw, thick, minimum size=0.9cm, inner sep=1pt, font=\small},
		every edge/.append style={thick, -latex},
		every label/.append style={font=\small}
		]
		
		\node[font=\bfseries\large, align=center] at (7.75, 7.5) 
		{Static graph $\mathcal{G}^{\text{static}}$ and time-unfolded trellis $\mathcal{G}^{\text{traj}}$ (LDP view)};
		
		\draw[thick, rounded corners] (0, 1.3) rectangle (6, 6.8);
		\node[font=\bfseries] at (3, 6.4) {(a) $\mathcal{G}^{\text{static}}$};
		\node at (3, 5.9) {Static graph $\mathcal{G}^{\text{static}} = (\bm{X}, \bm{P})$};
		
		\node[state, label=below:{$b_1$}] (x1) at (3, 4.8) {$X_1$};
		\node[state, label=below:{$b_2$}] (x2) at (1.5, 2.6) {$X_2$};
		\node[state, label=below:{$b_3$}] (x3) at (4.5, 2.6) {$X_3$};
		
		\draw[->] (x2) -- node[above left, font=\tiny, pos=0.4] {$\lambda_{2,1}$} (x1);
		\draw (x1)[->] -- node[above right, font=\tiny, pos=0.6] {$\lambda_{1,3}$} (x3);
		\draw (x2)[<->] -- node[above, font=\tiny] {$\lambda_{2,3}$} node[below, font=\tiny] {$\lambda_{3,2}$} (x3);
		
		\draw[thick, -latex] (6.2, 3.5) -- node[above, align=center, font=\small] {Unfold\\over time $\mathbb{T}$} (8.3, 3.5);
		
		\draw[thick, rounded corners] (8.5, -1.3) rectangle (16.5, 6.8);
		\node[font=\bfseries] at (12.5, 6.4) {(b) $\mathcal{G}^{\text{traj}}$};
		\node at (12.5, 5.9) {Time-unfolded trellis $\mathcal{G}^{\text{traj}}$};
		
		\foreach \t/\pos in {0/9.5, 1/11.2, 2/12.9} {
			\node[font=\small] at (\pos, 5.2) {$t=\t$};
		}
		\node[font=\small] at (15.5, 5.2) {$t=T$};
		
		\node[state, fill=gray!10] (top0) at (9.5, 4.0) {$X_1^0$};
		\node[state]               (mid0) at (9.5, 1.9) {$X_2^0$};
		\node[state, fill=gray!10] (bot0) at (9.5, -0.2) {$X_3^0$};
		
		\node[state, fill=gray!10] (top1) at (11.2, 4.0) {$X_1^1$};
		\node[state, fill=gray!10]               (mid1) at (11.2, 1.9) {$X_2^1$};
		\node[state] (bot1) at (11.2, -0.2) {$X_3^1$};
		
		\node[state] (top2) at (12.9, 4.0) {$X_1^2$};
		\node[state]               (mid2) at (12.9, 1.9) {$X_2^2$};
		\node[state, fill=gray!10] (bot2) at (12.9, -0.2) {$X_3^2$};
		
		\node at (14.2, 4.0) {$\dots$};
		\node at (14.2, 1.9) {$\dots$};
		\node at (14.2, -0.2) {$\dots$};
		
		\node[state] (topT) at (15.5, 4.0) {$X_1^T$};
		\node[state, fill=gray!10] (midT) at (15.5, 1.9) {$X_2^T$};
		\node[state,fill=gray!10] (botT) at (15.5, -0.2) {$X_3^T$};
		
		\foreach \t/\next in {0/1, 1/2} {
			\draw[->] (mid\t) -- (top\next);
			
			\draw[->] (top\t) -- (bot\next);
			
			\draw[->] (mid\t) to[bend left=15] (bot\next);
			\draw[->] (bot\t) to[bend left=15] (mid\next);
		}
		
		\draw[->] (top2) -- (13.7, 3.1);
		\draw[->] (mid2) -- (13.7, 2.6);
		\draw[->] (mid2) -- (13.7, 1.4);
		\draw[->] (bot2) -- (13.7, 0.6);
		
		\draw[<-] (topT) -- (14.7, 3.4);
		\draw[<-] (midT) -- (14.7, 1.4);
		\draw[<-] (botT) -- (14.7, 0.6);
		\draw[<-] (botT) -- (14.7, 0.9);

		\node[draw, thick, minimum width=8.2cm, minimum height=2.3cm, align=center, font=\scriptsize, fill=white, rounded corners]
		at (4.1, -0.15) 
		{\textbf{Joint TPM}\\[0.6em]
			$E(\bm{x}^{t-1}, \bm{x}^t) = -\sum_j x_j^t \big(b_j + \sum_{i \to j} x_i^{t-1} \lambda_{ij}\big)$\\[0.8em]
			$\bm{P}(\bm{x}^{t-1}, \bm{x}^t) = \dfrac{\exp(-E(\bm{x}^{t-1}, \bm{x}^t))}{\sum_{\bm{x}} \exp(-E(\bm{x}^{t-1}, \bm{x}))}$};
		
		\node[align=left, font=\small, anchor=north west] at (0, -1.5) 
		{$b_j$ = bias \qquad $\lambda_{ij}$ = synaptic weight \qquad $E( \bm{x}^{t-1}, \bm{x}^t )$ = global transition energy};
		
	\end{tikzpicture}
	
	\caption[MaxCal on an Ising model (LDP view)]{We understand our generative model as a graph with a series of biases and weights which combine to produce a transition probability distribution. Each node $x_j^t$ at time $t \geq 1$ has an intrinsic bias term $b_j$ and receives a set of sensory signals $\bm{\lambda}_j( \bm{X}^{t-1} ) = \sum_{i \rightarrow j} X_i^{t-1} \lambda_{ij}$ from the preceding set of nodes $\bm{X}^{t-1}$. When a logistic-sigmoid activation is applied to these to define the parameter $p( \bm{x}^{t-1}, x_j^t ) = \sigma( b_j + \bm{\lambda}_j( \bm{x}^{t-1} ) )$ of a Bernoulli distribution, conditioned on $\bm{X}^{t-1} = \bm{x}^{t-1}$, we retrieve our joint TPM $(\bm{P}( \bm{x}^{t-1}, \bm{x}^t ))_{\Omega_{\bm{X}} \times \Omega_{\bm{X}}}$. Each row $\bm{P}( \bm{x}^{t-1}, \cdot )$ is a softmax distribution over the tuple of $\bm{\lambda}( \bm{x}^{t-1}, \bm{x}^t ) + \bm{b}( \bm{x}^t )$ functions, defined in the text below. Equivalently, it is a Gibbs-style distribution over transition energies $\bm{E}( \bm{x}^{t-1}, \bm{x}^t )$, with $\bm{E} = - \bm{\Lambda} - \bm{1}\bm{b}^{\top}$. Maximising the calibre of a path is equivalent to maximising the accuracy of predictions, under this framework. When each synaptic weight $\lambda_{ij}$ and bias $b_j$ is equipped with a normal distribution we, in addition to retrieving biologically plausible lognormal distributions over synaptic strengths and uptake probabilities, find that minimizing VFE is mathematically equivalent to minimizing an L2-regularized log loss function.}
	\label{fig:ldp}
\end{figure}

\subsection{Introducing an ANN-inspired network}\label{5.5.1}

In this view, we model neurons as nodes $X_i$ with an internal variable $b_i$, and synapses as directed edges $(X_i, X_j) \in \mathcal{E}$ with a variable $\lambda_{i,j}$.
Following this, we shall ``unfold'' our graph $\mathcal{G}^{\textup{static}} = ( \bm{X}, \mathcal{E} )$ over some time interval $\mathbb{T} = \{ 0, 1, \ldots, T \}$ to produce a graph $\mathcal{G}^{\textup{traj}} = ( \bm{X}^0 \sqcup \bm{X}^1 \sqcup \ldots \sqcup \bm{X}^T, \bigsqcup_{t=1}^T \{ ( X_i^{t-1}, X_j^t ) : ( X_i, X_j ) \in \mathcal{E} \} )$.
We shall model $\lambda_{ij}$ and biases $b_i$ as static over $t \in \mathbb{T}$ and thus assume $\mathbb{T}$ constitutes some period of data collection prior to the model updating itself.\\
\\
We shall also assume that each $X_j^t$ is a Bernoulli variable with a parameter that is conditionally dependent on $\bm{X}^{t-1}$.
In particular, with $\sigma$ denoting the logistic-sigmoid function we let $X_j^t \sim \textup{Ber}( \sigma( b_j + \sum_{i : (X_i, X_j) \in \mathcal{E}} x_i \lambda_{ij} ) )$ --- i.e. we assume that at each timestep $X_i^{t-1}$ sends a signal to $X_j^t$ depending on whether it is ``on'' or ``off'' ($X_i^{t-1}$ results in a signal $\lambda_{ij}$ being sent, and no signal is sent otherwise); these then combine with the internal ``bias'' parameter $b_j$ to impact the probability that $X_j^t$ fires.\\
\\
Finally, we shall assume that each synapse $\lambda_{ij}$ and bias $b_j$ is normally distributed, with respective means $\bar{\lambda}_{ij}$, $\bar{b}_j$ and variances $\tfrac{1}{n}\sigma_{ij}^2$, $\omega_j^2$.
In doing this, we adopt the same baseline architecture as random (RNNs) and artificial (ANNs) neural networks\cite{roberts22}, while inducing biologically plausible lognormal distributions over synaptic weights $e^{\lambda_{ij}}$, baseline and baseline firing rates $e^{b_j}$, as well as logistic-normal distributions of spike-transfer probabilities $e^{b_j + \sum_{i \ \textup{input}} x_i^{t-1} \lambda_{ij}} / ( 1 + e^{b_j + \sum_{i \ \textup{input}} x_i^{t-1} \lambda_{ij}} )$\cite{marreiros08,buzsaki14,petersen16}.

\subsection{Defining a global internal model}\label{5.5.2}

We combine the parameter priors with the graph structure to characterize our generative model, which defines a transition probability matrix $\bm{P}$ that predicts pairwise empirical distributions.
The logistic-sigmoid distributions combine to produce softmax ones.
In particular, if we let $\bm{\lambda}( \bm{x}^{t-1}, \bm{x}^t ) = \sum_j x_j^t \sum_{i \ \textup{input of} \ j} x_i^{t-1} \lambda_{ij}$ and $b( \bm{x}^t ) = \sum_j x_j^t b_j$, then the Bernoulli variables $X_j^t \vert \bm{X}^{t-1}$ combine to produce a TPM with rows determined by softmax distributions:

\begin{equation}
	\bm{P}( \bm{x}^{t-1}, \cdot )
	\sim
	\bm{\sigma}( \bm{\lambda}( \bm{x}^{t-1}, \bm{x}^t ) + b( \bm{x}^t ) : \bm{x}^t \in \Omega_{\bm{X}} ),
\end{equation}

with $\bm{\sigma}( \cdot )$ representing our softmax function.\\
\\
On a practical level, all we are doing here is identifying the bias $b_j$ of each node $X_j^t$, identifying all of the signals $\sum_{i \ \textup{input of} \ j} x_i^{t-1} \lambda_{ij}$ each receive from the state $\bm{x}^{t-1}$, and then assessing how compatible each outcome $\bm{x}^t$ is with receipt of these signals.\\
\\
We may understand each $\bm{\lambda}( \bm{x}^{t-1}, \bm{x}^t )$ as constituents of an adjacency matrix $\bm{\Lambda} = ( \bm{\lambda}( \bm{x}^{t-1}, \bm{x}^t ) )_{(\bm{x}^{t-1}, \bm{x}^t) \in \Omega_{\bm{X}} \times \Omega_{\bm{X}}}$ and each $b( \bm{x}^t )$ as components of a vector $\bm{b} = ( b( \bm{x}^t ) )_{ \bm{x}^t \in \Omega_{\bm{X}} }$.
Since these directly parametrise our transition probabilities, while directly incorporating the structure of $\mathcal{G}^{\textup{static}}$ and $\mathcal{G}^{\textup{traj}}$ (which determines transition probabilities as much as the weights themselves do), we shall consider the induced distribution $\mathcal{N}( \bm{\Lambda} ; \bar{\bm{\Lambda}}, \Sigma_{\lambda} ) \mathcal{N}( \bm{b} ; \bar{\bm{b}}, \Sigma_b )$ to be our prior generative model.

\subsection{Bias and weight updates as a MaxCal problem}\label{5.5.3}

To understand how our model may solve MaxCal problems, we shall consider a general Markovian stochastic process $\mathbb{X}$ over a time horizon  $\mathbb{T} = \{ 0, 1, \ldots, \bm{T} \}$.
Since $T$ is arbitrarily large, we shall use our stationary distribution $\pi \times \bm{P}^T$ as a reference ensemble.
The caliber of some path distribution $\Gamma$ is thus defined as the negative relative entropy (KL divergence) against our reference ensemble\cite{jaynes80,dixit18}:

\begin{equation}\label{caliber}
	\mathcal{C}[\Gamma]
	=
	-
	\mathcal{D}(
	\Gamma
	\lvert
	\rvert
	\Pi
	).
\end{equation}

Clearly, the unconstrained MaxCal ensemble of this process is $\pi \times \bm{P}^T$.
Now suppose that we have constraints on the expected pairwise empirical distribution $\nu^T( \bm{x}, \hat{\bm{x}} ) = \sum_{t=1}^T \bm{1} \{ \bm{X}^{t-1} = \bm{x}, \bm{X}^t = \hat{\bm{x}} \}$.
In particular, suppose that for each $( \bm{x}, \hat{\bm{x}} )$ pairing we have a constraint $\mathbb{E}[\nu^T( \bm{x}, \hat{\bm{x}} )] = \nu( \bm{x}, \hat{\bm{x}} )$.
Now also let $\rho( \bm{x} ) = \sum_{\hat{\bm{x}}} \nu( \bm{x}, \hat{\bm{x}} )$ be the empirical distribution, and $\bm{Q}( \bm{x}, \hat{\bm{x}} ) = \nu( \bm{x}, \hat{\bm{x}} ) / \rho( \bm{x} )$ be its empirical transition probabilities.
Then, when we apply these conditions to our caliber function we obtain the Lagrangian:

\begin{equation}
	\mathcal{L}( \Gamma; \bm{M} )
	=
	-
	\mathcal{D}(
	\Gamma
	\lvert
	\rvert
	\Pi
	)
	+
	\sum_{\bm{x}, \hat{\bm{x}}}
	\bm{m}( \bm{x}, \hat{\bm{x}} )
	\left[
	\sum_{\gamma} \Gamma( \gamma ) \nu^T( \bm{x}, \hat{\bm{x}} )( \gamma )
	-
	\nu( \bm{x}, \hat{\bm{x}} )
	\right].
\end{equation}

When we differentiate this with respect to $\Gamma(\gamma)$ for some path $\gamma \equiv ( \bm{x}^0, \ldots, \bm{x}^T )$ and then set the derivative equal to zero, we find that $\log \Gamma( \gamma )$ must equal, up to some constant, $\log \pi( \bm{x}^0 ) + \sum_{t=1}^T \log \bm{P}( \bm{x}^{t-1}, \bm{x}^t ) + \sum_{\bm{x}, \hat{\bm{x}}} \bm{m}( \bm{x}, \hat{\bm{x}} ) \sum_{t=1}^T \bm{1}\{ \bm{x}^{t-1} = \bm{x}, \bm{x}^t = \hat{\bm{x}} \}$.
Since the sums of the rightmost term can be switched, the indicator functions will select $\sum_{t=1}^T \bm{m}( \bm{x}^{t-1}, \bm{x}^t )$.
Thus, we retrieve that our MaxCal ensemble is a product of tilted transition probabilities.

\begin{equation}
	\Gamma(\gamma)
	\propto
	\pi( \bm{x}^0 )
	\prod_{t=1}^T
	\bm{P}( \bm{x}^{t-1}, \bm{x}^t )
	e^{ \bm{m}( \bm{x}^{t-1}, \bm{x}^t ) }.
\end{equation}

We can retrieve from further analysis that $\rho \times \bm{Q}^T$ is our constrained MaxCal path ensemble.
However, first note something more conceptual:
since our Lagrangian is solved by tilting the transition probabilities by some multipliers $\bm{m}( \bm{x}, \hat{\bm{x}} )$, we can understand adjustment of $\bar{\bm{\lambda}}( \bm{x}, \hat{\bm{x}} ) + \bar{\bm{b}}( \bm{x} )$ toward some $(\bar{\bm{\lambda}} + \epsilon_{\lambda})( \bm{x}, \hat{\bm{x}} ) + (\bar{\bm{b}} + \epsilon_b)( \bm{x} )$ as a process of Lagrangian MaxCal optimization, via adjustment of model parameters.\\
\\
From the perspective of our deviation methods, it is important to note that the Caliber $\psi( \nu )$ of some distribution constrained by the empirical pair count $\nu$ produces the following identity:

\begin{equation}\label{info_LDP}
	\psi_{\textup{ldp}}( \nu )
	=
	\mathcal{D}( \rho \lvert \rvert \pi )
	+
	T
	\mathcal{D}(
	\nu
	\lvert
	\rvert
	\rho \times \bm{P}
	).
\end{equation}

The rightmost term, $T \mathcal{D}( \nu \lvert \rvert \rho \times \bm{P} )$, can be recognized from the mathematics of large deviations\cite{touchette12}.
In particular, the surprisal of the empirical pair count $\nu^T( \gamma )$ is approximated by $T$ multiplied by a rate function, $\mathcal{D}( \nu \lvert \rvert \rho \times \bm{P} )$\cite{adams23}.
Thus, we obtain the following relationship:

\begin{equation}\label{info_as_dev_ldp}
	\psi_{\textup{ldp}}( \nu )
	\approx
	\log
	\mathbb{P}(
	\nu^T( \gamma )
	\asymp
	\nu
	).
\end{equation}

From the perspective of integration across some partition $\bm{Y}$, $\bm{Z}$, we discount synaptic weights $\bm{\Lambda}_{y \rightarrow z}$, $\bm{\Lambda}_{z \rightarrow y}$ across the partition and then apply analogous methods.
Using the correspondence between deviations and rate functions, we derive the following expression:

\begin{equation}\label{integration_as_dev_ldp}
	\phi_{\textup{ldp}}( \nu )
	\approx
	\log
	\left(
	\dfrac{
		\mathbb{P}( \nu^T( \gamma ) \asymp \nu )
	}{
		\mathbb{P}( \nu^T_y( \gamma_y ) \asymp \nu_y )
		\mathbb{P}( \nu^T_z( \gamma_z ) \asymp \nu_z )
	}
	\right).
\end{equation}

An interesting outcome of this approach is that (lack of) alignment across synaptic weights appears to drive integration.
One can imagine a system in which the bias terms $b_j$ of each node $X_j$ are predicting its pairwise empirical distribution perfectly.
However, if the relationships between nodes are not effectively modeled by $\bm{\Lambda}$, integration will be high.

\subsection{Free Energy as regularized MaxCal inference}\label{5.5.4}

We can now conceive of a situation in which the purpose of our model $\mathcal{N}( \bm{\Lambda} ; \bar{\bm{\Lambda}}, \Sigma_{\lambda} ) \mathcal{N}( \bm{b} ; \bar{\bm{b}}, \Sigma_b )$, composed of both the parameters $\lambda_{ij}$, $b_j$ and underlying graph structure $\mathcal{G}^{\textup{static}}$, is to predict the pairwise empirical pair distribution $\nu^T$ via MaxCal methods.
This might be motivated by the need for incoming signals to travel with as little friction as possible.\\
\\
Under such a view, $\bm{\Lambda}$, $\bm{b}$ act as our internal variables while $\nu^T$ acts as an observation.
Thus, $\hat{p}( x ) \hat{p}( o \vert x ) = \mathcal{N}( \bm{\Lambda} ; \bar{\bm{\Lambda}}, \Sigma_{\lambda} ) \mathcal{N}( \bm{b} ; \bar{\bm{b}}, \Sigma_b ) \mathbb{P}( \nu^T \asymp \nu )$ would compose our internal model.
Of course, in practice our empirical pair count $\nu$ would rely not just on internal parameters but the behavior of adjacent neurons $\bm{B}$, which constitute our sensory data.
Thus, while $\nu^T$ is generated from internal parameters, the realization $\nu$ is an observation.\\
\\
We then assume that, given $\nu$, the parameters $\bm{\Lambda}$, $\bm{b}$ adjust to some specific value, encoding a $\delta( \bm{\Lambda}_q ) \delta( \bm{b}_q )$ distribution.
We now have a familiar predictive coding structure\cite{millidge21} and can calculate our variational free energy functional:

\begin{equation}
	\mathcal{F}^{\textup{ldp}}_V[q]
	=
	\mathbb{E}_{ \delta( \bm{\Lambda}_{\nu}, \bm{b}_{\nu} ) } \left[
	-
	\log
	\mathcal{N}( \bm{\Lambda} ; \bar{\bm{\Lambda}}, \bm{\Sigma}_{\lambda} )
	\mathcal{N}( \bm{b} ; \bar{\bm{b}}, \bm{\Sigma}_{b} )
	\mathbb{P}( \nu^T( \Gamma ) \asymp \nu )
	\right]
	-
	\mathcal{H}( \delta( \bm{\Lambda}_{\nu}, \bm{b}_{\nu} ) ).
\end{equation}

Upon calculating, we retrieve that VFE is a balance of MaxCal deviation alongside cost of altering the generative model:

\begin{equation}\label{freeEn_ldp}
	\mathcal{F}^{\textup{ldp}}_V[q]
	=
	2^{n-1} (2^n - 1)
	\log( 2\pi )
	+
	\frac{1}{2}
	\log
	\lvert
	\Sigma_{\lambda}
	\rvert
	\lvert
	\Sigma_{b}
	\rvert
	+
	\frac{1}{2}
	\epsilon_{\lambda}
	\Sigma_{\lambda}^{-1}
	\epsilon_{\lambda}^{\top}
	+
	\frac{1}{2}
	\epsilon_{b}
	\Sigma_{b}^{-1}
	\epsilon_{b}^{\top}
	+
	\psi_{\textup{ldp}}( \nu ),
\end{equation}

with $\epsilon_{\lambda}$ and $\epsilon_b$ being defined as one would expect.
Under this scheme, the cost of our prior distribution producing the parameters $\bm{\Lambda}$, $\bm{b}$ characterizes complexity, while information acts as a penalty for inaccurate accounting of the posterior over observations.
Thus, VFE under this framework would aim to compute MaxCal inference problems with bounds on tractable computation.
\textit{Information} would arise precisely when complete accounting is intractable.\\
\\
We can similarly create partitioned free energy functionals to examine the relationship with integrated information.
In this setting, we will ignore the $2^{n-1} (2^n - 1)$ terms as we focus on the deviation of VFE away from its minimal value:

\begin{equation}
	\Delta_{\mathcal{P}}
	\mathcal{F}^{\textup{ldp}}_V
	=
	\frac{1}{2}
	\Delta_{\mathcal{P}} \log \lvert \Sigma_{\lambda} \rvert \lvert \Sigma_b \rvert
	+
	\frac{1}{2}
	\Delta_{\mathcal{P}}( \epsilon_{\lambda} )
	+
	\frac{1}{2}
	\Delta_{\mathcal{P}}( \epsilon_{b} )
	+
	\phi_{\textup{ldp}}(\phi),
\end{equation}

with our difference terms defined as expected.
We see that $\phi_{\textup{ldp}}(\phi) > 0$ should co-occur with $\Delta_{\mathcal{P}} \mathcal{F}^{\textup{ldp}}_V \approx 0$ precisely when there is a justifiable increase in efficiency with regard to updating the model.

\subsection{Connection to energy-based learning methods}\label{5.5.5}

The bias vector $\bm{b}$ and adjacency matrix $\bm{\Lambda}$ may combine to produce a transition energy matrix $\bm{E} = - \bm{\Lambda} - \bm{1} \bm{b}^{\top}$ (here, $\bm{1}$ refers to the vector of $1$'s) under LeCun's framework, while our information (i.e.) deviation function $\psi( \nu )$ serves as a log-loss function.
This means our VFE term can be understood as an L2-regularized loss function between pairwise empirical observations (``test data'') and posterior weights (``network weights''), creating a theoretical link between minimization of a VFE functional and optimization of an objective function (other than the ELBO).\\
\\
From a learning perspective, regularization of the loss function avoids overfitting to a particular dataset (set of sensory inputs), retaining generalizability, which in a dynamic environment may serve to reduce future loss incurred.\\
\\
We note that the sparser $\mathcal{G}^{\textup{static}}$ is, the less regularized it is, however this also reduces the range of TPMs achievable by our network.
This trade-off directly echoes LeCun's emphasis on how the choice of architecture (e.g. sparse graph-transformer networks and factor graphs in sections 6-7) control flexibility of the energy function\cite{lecun06}.
Thus, a question opens up with regard to the role of sparsity in learning.\\
\\
We also note that the mean values $\bar{\bm{\Lambda}}$, $\bar{\bm{b}}$ (which might model baseline synaptic strengths and potentials) impact the expected value of our regularizer.
Following Friston's original distinction between maximisation (inference) and expectation (learning)\cite{friston03}, one might posit that tuning posteriors $\bm{\Lambda}_q$, $\bm{b}_q$ is a process of inference while choosing means $\bar{\bm{\Lambda}}$, $\bar{\bm{b}}$ is a process of learning.

\section{Discussion, limitations, and further research directions}\label{5.6}

Through understanding IIT's structure as one of maximum-caliber inference, we have proposed a relationship between integration and neural theories of variational inference.
In particular, chapter \ref{4_info_dev} explored speculative links with active inference, while here we explored bespoke-models involving energy-based inference.\\
\\
In particular, section \ref{5.1} generalized the MaxCal method within non-deterministic generalizations of IIT's time-unfolded digraphs\cite{oizumi2014}.
We identified, in section \ref{5.1.3}, similarities between the derived quantities we'd derived and those used in Ising models\cite{glauber63}, which motivated here and Ising models, which motivated the development of a free energy functional in section \ref{5.3} that is equivalent to ``information''.
We stated the required conditions for it to be formally connected to FEP, though note existence of such a system remains unproven.\\
\\
In sections \ref{5.4}--\ref{5.5} we extended our methods to arbitrarily long trajectories using two different, though perhaps complementary, approaches.
``Information'' under our CLT view measures difference between a system's internal conditional prediction of its state and that actually realized through direct interaction with the boundary path.
In effect, it measures how redundant sensory data becomes (or doesn't) after it is used as a boundary condition.
``Integration'' measures the extent to which this is true between internal partitions of the system.
Following an assumption that, in the context of an interconnected system in which Markov blankets may be arbitrarily drawn, VFE reduces wherever a decrease is most sharply required, we predicted that complex sensory data might initially increase information and integration, however will only remain as high as required by the system.
Further, information was identified with Bayesian surprise.\\
\\
Under our LDP framework, ``information'' is a MaxCal deviation between the system's entirely interior predictions and trajectories imposed by sensory information --- which is computationally equal (up to a small constant) to the magnitude of a deviation.
Our model possesses an ANN-inspired structure\cite{roberts22}, while producing biologically plausible lognormal strengths over biases $e^{b_j}$ and synaptic weights $e^{\lambda_{ij}}$\cite{buzsaki14}.
We deduced that MaxCal inference, in this context, can be recast as a parameter updating problem, and that VFE minimization corresponds to solving this with constraints placed on computational costs.
Further, when we understand our parameters to be $\lambda_{ij} = \bar{\lambda}_{ij} + \mathcal{N}( \epsilon_{ij} ; 0, \sigma_{ij}^2 )$, $b_j = \bar{b}_j + \mathcal{N}( \varepsilon_j ; 0, \omega_j^2 )$ variables we retrieve that VFE can be interpreted as an L2-regularized loss function\cite{lecun06}, with inference\cite{friston03} corresponding to optimization.
Additionally, we retrieve that $\psi_{\textup{ldp}}$ should only minimize as far as computational complexity is not exceeded.\\
\\
Our results, broadly speaking, appear to match empirical observations.
Albantakis and Tononi et al. (2014) observed that animats may develop conceptual information $\Phi$ in response to requirements to environmental demands, but no more than is necessary to complete complex tasks\cite{albantakis14}.
Further, Olesen and Waade et al. (2023) confirmed this, while noting that $\Phi$ and sensory surprisal fluctuate together\cite{olesen23}.
In addition, studies on dissociated neuronal cultures performed by Mayama et al. (2025) observed strong correlations between $\Phi_R^{\textup{mc}}$\cite{rosas22,tegmark2016} and Bayesian surprise, and a more complicated correlation with accuracy\cite{mayama25}.
A ``hill-shaped trajectory'' was proposed to characterize the learning process, which involved an initial ``gas-like'' state in which connections are weak, transition to a ``liquid-like'' state as $\Phi$ increases while it explores global updates, followed by a subsequent decrease as it solidifies into an optimal partition.\\
\\
Nonetheless, the explanatory power of our proposal here remains unconfirmed.
Experimental or theoretical work must be conducted to ascertain either a direct relationship MaxCal and FEP inference, or to connect the metrics observed here to those in IIT.\\
\\
Further research could focus on testing or updating this model, as well as broader theoretical links between FEP, IIT, and MaxCal principles.
A key area of consideration is how ``competing'' incentives might combine within a highly structured system.
If individual neurons\cite{harrison05} and whole brain regions\cite{mirza16} are both proposed to engage in active inference, which is further proposed to relate to physical MaxCal and MaxEnt optimization\cite{maxwell23,sakthivadivel23}, then understanding of how (and when) these either complement or contrast against each other, is key.

%% file: 6-complexity/6_complexity.tex
\chapter{An analysis of information and complexity}\label{6_complexity}

\textsc{In recent years,} there has been demand to model cognition through the language of complexity\cite{ashby47,ashby91}.
This has arisen from empirical and theoretical research in neuroscience\cite{hancock22,alderson18}, as well as advances in machine learning\cite{tishby15,poole16,roberts22}.\\
\\
On the artificial side, modeling deep network limits as features of an information bottleneck has allowed for generalization and representation learning to be understood via the information bottleneck principle\cite{tishby99}, which has allowed for optimal architecture to be understood as bifurcation points within this structure\cite{tishby15}.
Meanwhile, geometric approaches to signal propagation have allowed for them to be modeled as nonlinear dynamical systems, with transient chaos emerging at depth\cite{poole16}.
These perspectives have allowed for networks to be studied from the lens of criticality, with complex dynamics emerging at specific depth-to-width ratios\cite{roberts22}.
Recent advancements have modeled trained networks as trading off information complexity of network parameters against their perturbational vulnerability\cite{achille20}, while Quantum Field Theory has been employed to quantify how finite-width corrections can drive ``edge of chaos'' behavior\cite{grosvenor22}.\\
\\
Biologically (as we will explore further in section \ref{6.1.2}), the brain has also been modeled as a critical system as the edge of chaos\cite{cocchi17}, with complex behaviors such as metastability observed\cite{hancock22} and associated with healthy functioning\cite{hellyer15,alderson18}.
In FEP, exploitation-exploration trade-offs form a direct part of the theory\cite{friston23,pezzulo24}, while the $\Phi$ metric from IIT has been observed to coincide with both combinatorial and dynamical complexity\cite{mediano22,rosas22}.\\
\\
Here, we investigate MaxCal deviations $\psi$ and their association with complexity, to build working hypotheses around the relationship.
Section \ref{6.1} is an overview of complexity science and its role in computational neuroscience.
In section \ref{6.2}, we build a working understanding of the information $\psi( \pi )$ of a stationary distribution $\pi$ in a homogenous, memoryless, ergodic Markov chain.
The (as per our terminology) MaxCal ensemble $\mu \rightarrow \mu \bm{P}$ is its most disordered trajectory over a single step of the chain, while $\pi \rightarrow \pi$ are the long term dynamics formed due to $\bm{P}$ iterating (and is also the MaxCal set of paths as $T \rightarrow \infty$\cite{jaynes80}).
We investigate their differences first via an analysis of path space, and then via examples of permutative TPMs and random walks.\\
\\
Experiments are described and discussed in section \ref{6.3}, where we observe that $\psi$ tracks state space topology.
In section \ref{6.4}, we form preliminary conclusions regarding the relationship between MaxCal deviations and complexity.

\section{A brief introduction to complexity science}\label{6.1}

\subsection{Terms and definitions}\label{6.1.1}

\textit{Complexity science} studies the behavior of systems which exhibit emergent properties, irreducible to individual parts\cite{comfort94}.
This can relate to the information-theoretic properties of its state space\cite{adami02}, or to dynamical properties\cite{bak95}.
Markers of complexity include \textit{emergence,} which is the existence of non-centrally-coordinated global structures\cite{lizier08}, as well as \textit{distributed computation} which for our purposes might be thought of as emergent computation\cite{andrews00}.
Dynamical markers focus on critical phase transitions\cite{bak87}, as well as \textit{metastable} evolution (moving between various quasi-stable saddle points)\cite{shanahan10}.\\
\\
\textit{Self-organization} is a \textit{process} by which complexity is proposed to emerge\cite{ashby47}.
It is distinct from complexity in the sense that self-organization is ``a proposed journey'', while complexity is ``the destination''\cite{gershenson21}.
While precise definitions vary across the literature\cite{gershenson25}, a set of attributes which appear to be commonly discussed in the literature include\cite{turkheimer22}:

\begin{enumerate}\label{selfOrganization}
	\item \textit{Exploitation-exploration balance:} the system strikes a balance between being predictable, static, and precise (exploitation), and being dynamic, unpredictable, and random (exploration).
	\item \textit{Non-linearity:} the strength of a system's input is not proportional to its output.
	\item \textit{Synergy:} a compound of small, local interactions combine to produce large-scale organization.
	\item \textit{Consumption:} the system internally ``resists'' a tendency toward disorder by maintaining free energy and exporting entropy to its environment.
\end{enumerate}

Taken together, these behaviors propose the processes by which structured, adaptive behavior emerges\cite{comfort94}.
This is a very obvious point of intersection with active inference\cite{friston13,pezzulo24}, with much literature in the area actively proposing a connection\cite{friston23,mayama25}.\\
\\
A very specific type\cite{frigg03} of self-organization models the spontaneous emergence of critical states.
It is termed, \textit{self-organized criticality (SOC)}.
Specific dynamical, such as power-law distributions and scale-invariance, characterize the laws of such systems\cite{bak95,shiner00}.

\subsection{Complexity and the brain}\label{6.1.2}

A growing body of literature proposes a convergence between these concepts with empirical and theoretical neuroscience.
The methods of complexity science have been proposed as a useful lens for interpreting neuroimaging data\cite{turkheimer22}, while metastability has been suggested as a useful lens for characterizing brain function\cite{hancock24}.
In computational psychiatry, it has been used to predict symptoms of schizophrenia\cite{lee18,hancock22} and to assist in explaining the development of Alzheimer's disease\cite{alderson18}.
This sits within a broader study of neural dynamics, with researchers suggesting it arises from the mapping of connections between neurons\cite{hellyer15}.\\
\\
Other bodies of research have proposed that the brain is a critical system, characterized by phase transitions, avalanches, and operation on the edge of chaos\cite{cocchi17,jerbi22}.
Signatures of criticality have been shown to optimize conditions for efficient coding, providing a proposed benefit of the brain exhibiting this behavior\cite{safavi24}.
Meanwhile, SOC has been suggested as an underlying mechanism\cite{plenz21} and speculated as a neural correlate of consciousness\cite{walter22}.
\\
With regard to IIT, the PCI acts as a computational bridge between the ontology and complexity science\cite{casarotto16,xu24,molina25}, while empirical tests have demonstrated that proposed approximations\cite{tegmark2016,mediano18} of $\Phi$ correlate with markers of both informational and dynamical complexity\cite{mediano22}.
Meanwhile, FEP and SOC have been combined to jointly explain the physiological regulation of living systems\cite{bettinger23}.\\
\\
As we shall explore further (see \ref{6.2.3}), the information measure we developed in section \ref{5.1} can be associated with the development of localisation (quasi-stable saddle points), metastability, and spectral markers of complexity within random walks\cite{burda10}. 

\section{The MaxCal deviations of stationary distributions}\label{6.2}

\subsection{A path-space based analysis}\label{6.2.1}

Within the single-step systems defined in section \ref{5.1}, \textit{MaxCal} and \textit{stationary} distributions can capture different properties of the system.
In particular, the MaxCal $\mu \times \bm{P}$ captures maximal uncertainty of our system $\mathbb{X}$ at it moves one step forward into the future, while $\mathbb{X} \sim \pi$ (within ergodic systems) describes its long-run evolution.\\
\\
In some systems these align, and for these we can derive some balance equations.
Since the marginal masses $\mu( \bm{x} )$ over a transition network's inputs $\bm{X}^t$ are proportional to the perplexity $\textup{PP}( \bm{x} )$ of their conditional distributions, we retrieve that when $\mu \sim \pi$ the following equation holds:

\begin{equation}\label{pp_balance}
	\textup{PP}( \bm{x} )
	=
	\sum_{\hat{\bm{x}}}
	\textup{PP}( \hat{\bm{x}} )
	\bm{P}(
	\hat{\bm{x}}
	,
	\bm{x}
	).
\end{equation}

Recalling that perplexity is the branching factor of some distribution\cite{jurafsky26}, we might interpret this as suggesting that the effective number of paths into each state are equal to the effective number of paths out: $\forall \bm{x} : \ \# \{ \bm{X}^t \rightarrow \bm{x} \} \approx \# \{ \bm{x} \rightarrow \bm{X}^{t+1} \}$.
In general, we might define a balancing factor which quantifies the effective difference in branching factor between paths into each state and out:

\begin{equation}\label{balance_factor}
	S( \bm{x} )
	=
	\log
	\textup{PP}( \bm{x} )
	-
	\log
	\textup{InPP}( \bm{x} ),
	\qquad
	\textup{InPP}( \bm{x} )
	=
	\sum_{\hat{\bm{x}}}
	\textup{PP}( \hat{\bm{x}} )
	\bm{P}(
	\hat{\bm{x}}
	,
	\bm{x}
	).
\end{equation}

In states for which $\bm{S}(\bm{x}) < 0$ we see that the effective branching space is \textit{shrunk} by a system entering $\bm{x}$, while when $\bm{S}( \bm{x} ) > 0$ the effective path space is \textit{grown.}
We further note that over $\bm{X} \sim \pi$ the expected value $\mathbb{E}_{\pi}\left[ \bm{S}( \bm{X} ) \right]$ is equal to $\mathcal{D}( \pi \lvert \rvert \mu \bm{P} ) - \mathcal{D}( \pi \lvert \rvert \mu )$, which we know is non-positive due to the fact that $\mathcal{D}( \pi \lvert \rvert \mu \bm{P} ) = \mathcal{D}( \pi \bm{P} \lvert \rvert \mu \bm{P} )$.
For this reason we see that chains must spend at least as much time in \textit{shrinking states} as in \textit{dissipative} ones.
Further, the degree to which MaxCal inputs $\bm{X}^t \sim \mu$ move toward the stationary state upon transition $\bm{X}^{t+1} \sim \pi$, characterizes the long run occupancy of our chain between the two types of states.

\subsection{Entropy rate and information}\label{6.2.2}

Another way to understand $\psi(\pi)$ is as a difference between long and short term entropic properties of our memoryless Markovian network, which underscores the study of GRWs and MERWs conducted in the following subsection.
The chain's \textit{marginal entropy} $\mathcal{H}( \pi )$ is the general unpredictability of its movement, while \textit{entropy rate} $h( \pi )$ relates to the average production of entropy per unit step\cite{jaynes80}.
Their sum is thus the overall disorder of the chain's dynamics combined with the average entropy produced per unit time.
Information $\psi( \pi )$ is the distance of this from the theoretical maximum combination that the chain could produce, at any given moment.

\begin{equation}\label{info_ent_rate}
	\psi( \pi )
	=
	\log \kappa
	-
	\mathcal{H}( \pi )
	-
	h( \pi ),
	\qquad
	\log \kappa
	=
	\max \left[
	\mathcal{H}( \bm{X} )
	+
	h( \bm{X} )
	\right].
\end{equation}

The \textit{infomax} principle suggests that efficient systems should maximize mutual information between inputs and outputs\cite{linsker88,linsker90}.
We may take inspiration from this to study the relationship that $\psi$ has with mutual information of our transition network in its stationary distribution, and propose this may be used to characterize different types of networks.
In particular, we note that since $\bm{X}^t \sim \bm{X}^{t+1}$ for $\bm{X}^t \sim \pi$, then $\mathcal{H}^{\pi}( \bm{X}^{t+1} ) = \mathcal{H}^{\pi}( \bm{X}^t )$ and therefore mutual information may be expressed by the difference between marginal entropy and entropy rate:

\begin{equation}\label{mutual_info_stationary}
	i( \pi )
	=
	\mathcal{H}( \pi )
	-
	h( \pi ).
\end{equation}

We, therefore, info some equations which connect our two measures of information:

\begin{eqnarray}
	\psi( \pi )
	+
	i( \pi )
	=
	\log \kappa
	-
	2 h( \pi ),
	\\
	\psi( \pi )
	-
	i( \pi )
	=
	\log \kappa
	-
	2 \mathcal{H}( \pi ).
\end{eqnarray}

We may now deduce, for example, that when $\psi( \pi )$ and $i(\pi)$ are both low (i.e. approximately zero), that the ergodic limit $\pi$ explores maximal available path space while preserving very little information, indicating a high degree of degeneracy.
This aligns with distributions which mix information instantly, discussed further in \ref{6.2.3}.
When at least one of $\psi( \pi )$ and $i( \pi )$ are high, that entropy production is low while the prospective number of paths over any given step forward in time is high.
In the case that $0 \approx \psi( \pi ) \gg i( \pi )$, we know that (approximately) $\mu \sim \pi$ and therefore $\log \kappa$ --- which is large --- is equal to $h( \pi ) + \mathcal{H}( \pi )$.
The implication here would be that much of its value comes from large $\mathcal{H}( \pi )$, which aligns with the circulant chains discussed in section \ref{6.2.3}.
Mutual information, thus, is preserved by highly stable yet expansive movements.\\
\\
In cases where $\psi( \pi )$ is also high, we may either have that $i(\pi)$ is also very high (we shall write $i(\pi) \approx \psi( \pi ) \gg 0$) or that $i( \pi )$ is very low ($0 \approx i(\pi) \ll \psi( \pi )$).
In the former example, we see that the effective number of total paths is approximately twice the effective state space size of the chain over the long run: $\log \kappa \approx 2 \mathcal{H}( \pi )$.
We suggest this indicates a moderate balance between exploitation and exploration.
The comparatively small size of $h( \pi )$ perhaps indicates that state space restriction is achieved by occupancy within the shrinkage states described in section \ref{6.2.1}, posing questions of whether questions such as metastability might be relevant here.\\
\\
The regime $\psi( \pi ) \gg i( \pi ) \approx 0$ appears to indicate a very high degree of latent ``entropic potential''.
The MaxCal marginal $\mu$ must place mass on a significant number of states which ``explode'' the potential of the space, yet the chain explores very little states over the long run.
The key difference of this regime with the former, is the effective number of states characterizing our stationary distribution.
Therefore, while $\psi( \pi ) \gg 0$ informs us of a tendency of the chain to avoid ``explosive'' potential states $\bm{x}$, perhaps $i( \pi )$ informs us whether the chain is pushed toward a single attractor state, or toward a metastable range.

\subsection{Stochastic permutative matrices}\label{6.2.3}

We now turn our attention toward some example regimes and analyze their information production, to illustrate the above points.
First we look at \textit{stochastic permutative matrices,} which are stochastic matricies $\bm{P}$ such that each row $\bm{P}( \bm{x}, \cdot )$ is a permutation of the other, and denote these with $\bm{P}_{\textup{perm}}$.
A specific subtype we will refer to as \textit{circulant} matrices $\bm{P}_{\textup{circ}}$, in which (where applicable) every single row is a right-shift of the one above.

\begin{equation}
	\bm{P}_{\textup{perm}}
	=
	\begin{pmatrix}
		\rho_1 & \rho_2 & \rho_3 & \cdots & \rho_N \\
		\rho_{i_{2, 1}} & \rho_{i_{2, 2}} & \rho_{i_{2, 3}} & \cdots & \rho_{i_{2, N}} \\
		\rho_{i_{3, 1}} & \rho_{i_{3, 2}} & \rho_{i_{3, 3}} & \cdots & \rho_{i_{3, N}} \\
		\vdots & \vdots & \vdots & \ddots & \cdots \\
		\rho_{i_{N, 1}} & \rho_{i_{N, 2}} & \rho_{i_{N, 3}} & \cdots & \rho_{i_{N, N}}
	\end{pmatrix}
	;
	\quad
	\bm{P}_{\textup{circ}}
	=
	\begin{pmatrix}
		\rho_1 & \rho_2 & \rho_3 & \cdots & \rho_N \\
		\rho_N & \rho_1 & \rho_2 & \cdots & \rho_{N-1} \\
		\rho_{N-1} & \rho_N & \rho_1 & \cdots & \rho_{N-2} \\
		\vdots & \vdots & \vdots & \ddots & \cdots \\
		\rho_2 & \rho_3 & \rho_4 & \cdots & \rho_1
	\end{pmatrix}
\end{equation}

Since each row in a permutative matrix $\bm{P}_{\textup{perm}}$ has, up to reordering, an equivalent distribution, they yield equal conditional entropies $h( \rho )$.
The MaxCal distribution, therefore, is uniform $\mu \sim \left(\tfrac{1}{N}\right)_N$.
In addition, we understand that $\kappa = N e^{ h( \rho ) }$ quantifies the total number of paths while the information of any stationary distribution $\pi$ may be expressed as $\psi( \rho ) = \log N - \mathcal{H}( \pi ) = \mathcal{H}( \mu ) - \mathcal{H}( \pi )$.
This reflects the fact that path space cannot be reduced by choosing states $\bm{x}$ with a low ``effective number'' of outgoing paths.
Instead, it is the occupancy of the chain which must be altered.\\
\\
When we look at circulant matrices $\bm{P}_{\textup{circ}}$, these produce uniform stationary distributions, $\pi \sim \mu$, rendering MaxCal deviations to be equal to zero.
As discussed in section \ref{6.2.2} mutual information $i( \pi ) = \log N - \mathcal{H}( \rho )$ grades the dynamics of our system.
It is equal to zero when $\rho$ is uniform, producing a TPM $\bm{P}_{\textup{circ}}$ with entirely uniform rows.
In effect, it entirely eliminates the structure of any incoming signal, imposing maximal uncertainty.
On the other hand, $\mathcal{H}( \rho ) = \epsilon$ for some $0 < \epsilon \ll 1$ produces very high mutual information.
In this setting, $\bm{P}_{\textup{circ}}$ is very close to a cycle, preserving structure of inputs and engaging in movement which is very predictable.\\
\\
In order for information to be large, $\psi( \pi ) \gg 0$, we need overall occupancy of the state space to be biased, i.e. $0 < \mathcal{H}( \pi ) \ll \log N$.
An example which highlights the role of mutual information would be the case of an independent identical distribution, which occurs when $\bm{P}_{\textup{perm}}( \bm{x}, \cdot ) \sim \rho$ for all $\bm{x} \in \Omega_{\bm{X}}$.
Intuitively, we understand that structure is not preserved when samples $\bm{X}^t$ are independently drawn, and we may express this mathematically through the equation $i( \pi ) = \mathcal{H}( \rho ) - \mathcal{H}( \rho ) = 0$.
\textit{MaxCal deviations} $\psi( \pi ) \equiv \psi( \rho )$ thus arise because the intrinsic bias in drawing samples pushes $\mathbb{X}$ to occupy states which reflect this bias, thereby reducing the number of starting points any $\bm{X}^t \rightarrow \bm{X}^{t+1}$ may \textit{start} from.
We might think of the typical set of $\pi \sim \rho$ as a single stable region of the chain $\mathbb{X}$, with deviations being quickly corrected.
Effectively, $\mathbb{X}$ is a system which maintains itself very strictly.\\
\\
For mutual information to also be large, $i( \pi ) \gg 0$, a necessary prerequisite is that $\mathcal{H}( \pi ) \gg 0$.
Thus, $\pi$ must sit in a ``sweet spot'' between being deterministic and uniform.
We can capture this idea by stating that $\textup{Perp}( \pi )$ should be $\mathcal{O}( \sqrt{N} )$, or equivalently that $\mathcal{H}( \pi ) \approx \tfrac{1}{2} \log N$.
For sufficiently small $0 < \epsilon \ll 1$, an example of a matrix $\bm{P}_{\textup{circ}}$ which fulfills this requirement would be one in which the first $\lbrace \sqrt{N} \rbrace$ with $1 - \epsilon$ probability cycle between each other while exiting the region with $\epsilon$ probability, and then each other row matching its distribution with one of the first $\lbrace \sqrt{N} \rbrace$ ones, indicating a regime which balances the cyclical nature of circulant dynamics against the degenerative nature of \textit{iid} ones.
However, the permutative nature of $\bm{P}_{\textup{perm}}$ places significant restrictions on $\psi( \pi )$, as it requires for the $h( \mu ) - h( \rho )$ distance to always equal zero.
As we shall see in the next section, variance of the (effective) degree of each row in $\bm{P}$ appears to underly the development of MaxCal deviations.

\subsection{Maximum-Entropy Random Walks: Information as localization}\label{6.2.4}

A body of literature relevant to understanding the relationship between $\mu$ and $\pi$ is that of Maximum-Entropy Random Walks (MERWs)\cite{burda10}.
A \textit{random walk} is a homogenous, memoryless, ergodic Markov chain over an undirected graph $\mathcal{G}$ overlaying our state space $\Omega_{\bm{X}}$.
In particular, for states $\bm{x}$, $\hat{\bm{x}}$ which are connected there is a non-zero probability of moving from $\bm{x}$ toward $\hat{\bm{x}}$ (or vice versa), while for disconnected nodes there is zero probability.\\
\\
We represent connections between nodes $\bm{x}$, $\hat{\bm{x}}$ with a symmetric adjacency matrix $\bm{A} = ( \bm{A}( \bm{x}, \hat{\bm{x}} ) )_{ \bm{x}, \hat{\bm{x}} }$, and we define the \textit{degree} $d( \bm{x} )$ of state $\bm{x}$ as the number of nodes it is connected to.

\begin{figure}[h]
	\centering
	\begin{tikzpicture}[
		baseline={(current bounding box.center)},
		every node/.style={
			circle,
			fill=black,
			inner sep=0pt,
			minimum size=5pt
		},
		every edge/.style={
			draw,
			thin
		},
		label distance=4pt
		]
		
		\node[label=left:{1}]        (1) at (0, 0)   {};
		\node[label=below left:{5}]  (5) at (2, 0)   {};
		\node[label=above:{4}]       (4) at (3.5, 1.4) {};
		\node[label=right:{3}]       (3) at (4, -0.4) {};
		\node[label=below:{2}]       (2) at (2.8, -1.8) {};
		
		\draw (1) -- (5);
		\draw (5) -- (4);
		\draw (5) -- (3);
		\draw (5) -- (2);
		\draw (4) -- (3);
		\draw (4) -- (2);
		\draw (3) -- (2);
		\draw[->] (5) to [out=100, in=150, looseness=5, min distance=15mm] (5);
		
	\end{tikzpicture}
	\hspace{2cm}
	$\mathbf{A} = \begin{pmatrix}
		0 & 0 & 0 & 0 & 1 \\
		0 & 0 & 1 & 1 & 1 \\
		0 & 1 & 0 & 1 & 1 \\
		0 & 1 & 1 & 0 & 1 \\
		1 & 1 & 1 & 1 & 1
	\end{pmatrix}$
	\caption[A graph and its adjacency matrix]{Here, we have a graph $\mathcal{G}$ over the state space $\Omega = \{ 1, 2, 3, 4, 5 \}$. The adjacency matrix $\bm{A}$ represents connections between nodes. The degree $d(x)$ represents the number of edges a node $x$ belongs to, e.g. $d(5) = 4$.}\label{randomWalk}
\end{figure}
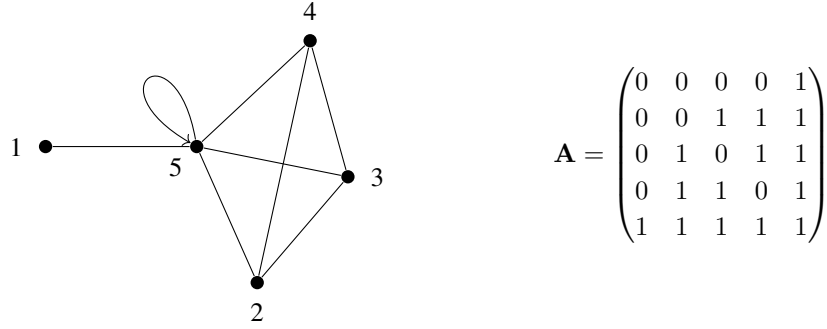

In a \textit{Generic Random Walk (GRW),} entropy is maximized locally and thus the conditional probability distribution from each node $\bm{x}$ is a uniform distribution over its outgoing paths.
For example, the TPM of the GRW associated with the graph from figure \ref{randomWalk} is defined by the following:

\begin{equation*}
	\bm{P}_{\textup{GRW}}
	=
	\begin{pmatrix}
		0 & 0 & 0 & 0 & 1 \\
		0 & 0 & \tfrac{1}{3} & \tfrac{1}{3} & \tfrac{1}{3} \\
		0 & \tfrac{1}{3} & 0 & \tfrac{1}{3} & \tfrac{1}{3} \\
		0 & \tfrac{1}{3} & \tfrac{1}{3} & 0 & \tfrac{1}{3} \\
		\tfrac{1}{5} & \tfrac{1}{5} & \tfrac{1}{5} & \tfrac{1}{5} & \tfrac{1}{5}
	\end{pmatrix}
	,
	\qquad
	\pi_{\textup{GRW}}
	=
	\dfrac{1}{15}
	\begin{pmatrix}
		1 \\
		3 \\
		3 \\
		3 \\
		5
	\end{pmatrix}
\end{equation*}

Here, each $\bm{x}$ will have a conditional entropy value $\log d( \bm{x} )$ and the conditional perplexity will be equal to its degree: $\textup{PP}( \bm{x} ) = d( \bm{x} )$\footnote{
	This provides a very straightforward implementation of the path space interpretations discussed in section \ref{5.2}.
}.
Thus, the MaxCal input marginal $\mu$ is a distribution in which each state $\bm{x}$ is weighted by its degree: $\mu( \bm{x} ) \propto d( \bm{x} )$.
This is precisely the stationary distribution $\pi$ of GRWs, indicating that they produce no local MaxCal deviations.
Correspondingly, over the long run they tend to visit every single state regularly, however there is significant bias in the paths chosen\cite{burda10}.\\
\\
\textit{Maximum-Entropy Random Walks (MERWs)} assign transition probabilities according to the MaxCal principle\cite{jaynes80}.
Rather than maximizing the local uncertainty at each given state $\bm{x}$, they ensure that the probabilities of paths of any given length $L$ are all equal\cite{burda10}.\\
\\
To achieve this, the Perron-Frobenius theorem is employed and transition probabilities are defined in relation to the largest eigenvector $v$ of the adjacency matrix $\bm{A}$.
When we let $\lambda$ represent the eigenvalue for $v$, we can express transition probabilities as a function of both\cite{burda08}:

\begin{equation}
	\bm{P}( \bm{x}, \hat{\bm{x}} )
	=
	\dfrac{
		\bm{A}( \bm{x}, \hat{\bm{x}} )
	}{
		\lambda
	}
	\cdot
	\dfrac{
		v( \hat{\bm{x}} )
	}{
		v( \bm{x} )
	}.
\end{equation}

Meanwhile, the stationary distribution $\pi_{\textup{MERW}}$ has probabilities $\pi( \bm{x} ) \propto v( \bm{x} )^2$.
Further utilizing the example from figure \ref{randomWalk} to illustrate, we may express the transition probabilities and stationary distribution of $\mathcal{G}$'s associated MERW by the following expressions:

\begin{equation*}
	\bm{P}_{\textup{MERW}}
	=
	\dfrac{2 - \sqrt{2}}{4}
	\begin{pmatrix}
		0 & 0 & 0 & 0 & 4 + 2\sqrt{2} \\
		0 & 0 & 2 & 2 & 2\sqrt{2} \\
		0 & 2 & 0 & 2 & 2\sqrt{2} \\
		0 & 2 & 2 & 0 & 2\sqrt{2} \\
		2 - \sqrt{2} & \sqrt{2} & \sqrt{2} & \sqrt{2} & 2
	\end{pmatrix}
	,
	\qquad
	\pi_{\textup{MERW}}
	=
	\dfrac{
		1
	}{
		16 + 10 \sqrt{2}
	}
	\begin{pmatrix}
		1 \\
		3 + 2 \sqrt{2} \\
		3 + 2 \sqrt{2} \\
		3 + 2 \sqrt{2} \\
		6 + 4 \sqrt{2}
	\end{pmatrix}
\end{equation*}

We see, in our toy example, that the odds of being in state $5$ compared to state 1 increase: $\tfrac{\pi_{\textup{MERW}}(5)}{\pi_{\textup{MERW}}(1)} = 6 + 4 \sqrt{2} > 6 = \tfrac{\pi_{\textup{GRW}}(5)}{\pi_{\textup{GRW}}(1)}$, while the odds of being in the region $\{ 2, 3, 4 \}$ compared to state $1$ or state $5$ also do (though not as sharply).
This reflects a broader finding that MERWs cluster into defect-free regions\cite{burda10}.
When we calculate the constrained MaxCal for $\bm{P}_{\textup{MERW}}$, we might propose a mechanism for this.
We see that the effective number of paths outside of $5$ and $\{ 2, 3, 4 \}$ slightly reduces, while the effective number of paths outside of $1$ remains the same.
Thus is due to the reduced uniformity of the rows --- which impacts states $\bm{x}$ with the largest degrees $d( \bm{x} )$ the most:

\begin{equation*}
	\textup{PP}_{\textup{GRW}}
	=
	\begin{pmatrix}
		1 \\
		3 \\
		3 \\
		3 \\
		5
	\end{pmatrix}
	,
	\qquad
	\textup{PP}_{\textup{MERW}}
	\approx
	\begin{pmatrix}
		1 \\
		2.958 \\
		2.958 \\
		2.958 \\
		4.707
	\end{pmatrix}
\end{equation*}

Thus, the effective number of paths outside of $5$ appears to shrink disproportionately compared to the effective number of paths into it, rendering it with a balance term $\bm{S}(5) < 0$, and the remaining states $x$ with balance terms $\bm{S}(x) > 0$.
Accordingly, the weights on our stationary distribution shift toward $5$.
This, perhaps, is related to why the only graphs $\mathcal{G}$ for which $\pi_{\textup{GRW}} \sim \pi_{\textup{MERW}}$ are those for which the degree is constant: $d( \bm{x} ) = d$\cite{burda10}.\\
\\
From an analytical perspective, we note that the entropy rate of an MERW is defined by the first eigenvalue: $h( \pi_{\textup{MERW}} ) = \log \lambda$.
Accordingly, $\lambda = \textup{PP}( \pi )$ characterizes the effective state space size of the chain.
Meanwhile, the entropy may be expressed as $\mathcal{H}( \pi_{\textup{MERW}} ) = - 2 \sum_{\bm{x}} v( \bm{x} )^2 \log v(\bm{x}) = 2 v^{\top} \textup{diag}( - \log v ) v$.
The conditional entropy $h( \bm{x} )$ of any specific state $\bm{x}$ can be expanded as a series of additions and deductions from $h( \pi_{\textup{MERW}} )$:

\begin{equation}
	h( \bm{x} )
	=
	\log \lambda
	+
	\log v( \bm{x} )
	-
	\dfrac{1}{ \lambda v( \bm{x} ) }
	\sum_{\hat{\bm{x}} \leftrightarrow \bm{x}}
	v( \hat{\bm{x}} )
	\log v( \hat{\bm{x}} ).
\end{equation}

These can be expressed as perplexity functions which multiply $\lambda v( \bm{x} )$ by the inverted geometric mean of the first-eigenvalue entries of its connected nodes, raised to the power of $1 / \lambda v( \bm{x} )$.
Employing the balance equation from section \ref{6.2.1} shows us that alignment of these geometric mean terms underlies the condition for $\mu_{\textup{MERW}} \sim \pi_{\textup{MERW}}$, however this is deeper than the scope of this paper.
In general, we can let $\kappa = \sum_{\bm{x}} \textup{PP}( \bm{x} )$ and then express information via the following equation:

\begin{equation}
	\psi( \pi_{\textup{MERW}} )
	=
	\log \kappa
	-
	\log \lambda
	+
	2 v \ \textup{diag}( \log v ) v.
\end{equation}

In general, we have that $\mu_{\textup{MERW}} \nsim \pi_{\textup{MERW}}$.
However, in the case of regular lattices where $d( \bm{x} )$ is constant (i.e. $d( \Omega_{\bm{X}} ) = \{ d \}$) we do, due to the fact that MERWs and GRWs coincide.
Breaking of uniformity induces behaviors such as localization and metastability\cite{burda10}.
Thus, while we have not characterized all situations in which $\mu_{\textup{MERW}} \sim \pi_{\textup{MERW}}$, we can observe (in the general case) some association between it and the emergence of complex dynamics.

\section{Empirically investigating information and complexity}\label{6.3}

Having analytically studied degree heterogeneity and its relationship to MaxCal deviations, we now turn toward empirical investigation.
Our aim is to establish how structural properties of a TPM --- specifically, the mean and variance of its row edges --- shape the information measure $\psi$.
To isolate the role of structure, we make all other attributes maximally agnostic, while taking steps to ensure ergodicity.\\
\\
Full details of the experiment can be found in the link just beneath this paragraph.
We find that the mean-degree $m$ and degree-variance $\sigma^2$ of a TPM, at any scale, characterizes the average value of $\hat{\psi}$ produced.
Further, the maximal ``best estimate'' $\hat{\psi}_{\textup{max}}$ of information, as well as the parameters $m_{\textup{max}}$, $\sigma_{\textup{max}}^2$, associated with it, can be expressed as a function of $N$.
In sections \ref{6.3.1}--\ref{6.3.2} we go through these results in more depth, while in \ref{6.3.3}--\ref{6.3.4} we explore implications and limitations.\\
\\
Full details of the experimental code can be found at \url{https://github.com/AlexanderKearney-eng/IIT-experiments}.

\subsection{Methods}\label{6.3.1}

\begin{algorithm}[ht]
	\caption{\texttt{sweep-space}: Sweep parameter space for information statistics}
	\begin{algorithmic}[1]
		\Require $0 < \varepsilon \ll 1$, $S = 150$, $T = 40$
		\Ensure $\bm{\psi}_{\textup{data}}$, $\bm{i}_{\textup{data}}$, $\bm{\kappa}_{\textup{data}}$
		\For{$\hat{N} = 1, \ldots, 10$}
		\State $N \leftarrow 5 \hat{N}$
		\State $\sigma^2_{\textup{glob}} \leftarrow \tfrac{1}{2}( N - 1 )^2$
		\State $\delta \leftarrow \tfrac{N - 1 + \epsilon}{S}$
		\For{$k = 0, \ldots, S - 1$}
		\State $m \leftarrow 1 + k \delta$
		\State $\sigma_m^2 \leftarrow ( m - 1 )( N - m )$
		\State $\hat{S} \leftarrow \lfloor \tfrac{\sigma_m^2}{ \sigma_N^2 } S \rfloor$
		\For{$l = 0, \ldots, \hat{S} - 1$}
		\State $\sigma^2 \leftarrow \varepsilon + \delta l$
		\State $( \bm{\psi}^{m, \sigma^2}, \bm{i}^{m, \sigma^2}, \bm{\kappa}^{m, \sigma^2} ) \leftarrow \texttt{measure-stats}( N, m, \sigma^2, T )$
		\State $\texttt{data}^{k, l} \leftarrow (\mu, \sigma^2, \bm{\psi}^{\mu, \sigma^2}, \bm{i}^{\mu, \sigma^2}, \bm{\kappa}^{\mu, \sigma^2} )$
		\EndFor
		\EndFor
		\EndFor
		\State \Return $( \bm{\psi}_{\textup{data}}, \bm{i}_{\textup{data}}, \bm{\kappa}_{\textup{data}} ) \leftarrow ( \texttt{data}^{k,l} )_{k, l}$
	\end{algorithmic}
	\label{sweep_space}
\end{algorithm}

Algorithms \ref{info_metrics}, \ref{measure_stats} and \ref{sweep_space} together outline the full data generation process.
At a high level, the aim was to investigate the impact that the degrees of rows in a TPM has on MaxCal deviation $\psi$.
We simultaneously investigated mutual information $i$ to assist in identifying strengths which have the potential to be unique to $\psi$, and we recorded data about the partitioning constant $\kappa$ to investigate its role in any results.
From here, we aimed to make all other attributes as agnostic as possible.\\
\\
To this end, we randomly generated ergodic TPMs $\bm{P}$ while constraining mean degree $m$ and degree variance $\sigma^2$.
Since one cannot deterministically generate vectors $\bm{d} = (d( \bm{x} ))_{\bm{x} \in \Omega_{\bm{X}}}$ without imposing some choice of structure, we treated $\bm{d}$ as a random variable --- maximally random subject to constraints on its mean and variance values.
The mean degree $m$ varied between $1$ and $N$, while variance $\sigma^2$ oscillated between small $\varepsilon$ and its maximal value.
We then generated $40$ samples for each mean-variance pairing, and then collected the mean, median, and skew of each sample for the metrics of interest: $\psi$, $i$, $\kappa$.
This was repeated for state spaces of size $N = 5, 10, \ldots, 50$.\\
\\
Upon plotting the mean and median values of each $\psi$ on heatmaps, the structure revealed a polynomial-like surface.
So, for each value of $N$ we used ordinary least squares methods to fit polynomial surfaces $\hat{\psi}$ of degree 3 to our data, to approximating the impact that $m$ and $\sigma^2$ have on $\psi$'s average.
We then found the maximum points of these surfaces and plotted $\hat{\psi}_{\textup{max}}$, $m_{\textup{max}}$, $\sigma^2_{\textup{max}}$ as functions of $N$, with $\hat{\psi}_{\textup{max}}$ being defined as expected and $( m_{\textup{max}}, \sigma^2_{\textup{max}} ) = \arg \max \hat{\psi}( m, \sigma^2 )$ (for both the mean and median).
Logarithmic, linear, and quadratic appearing relationships were observed.
The parameters of these were learned via linear regression.
Table \ref{max_params} contains the full 

\begin{figure}[ht]
	\centering
	\makebox[\textwidth][c]{
		\begin{minipage}{15.4cm}
			\textbf{A}\hspace{10cm}\textbf{C}\\
			\includegraphics[scale=0.4]{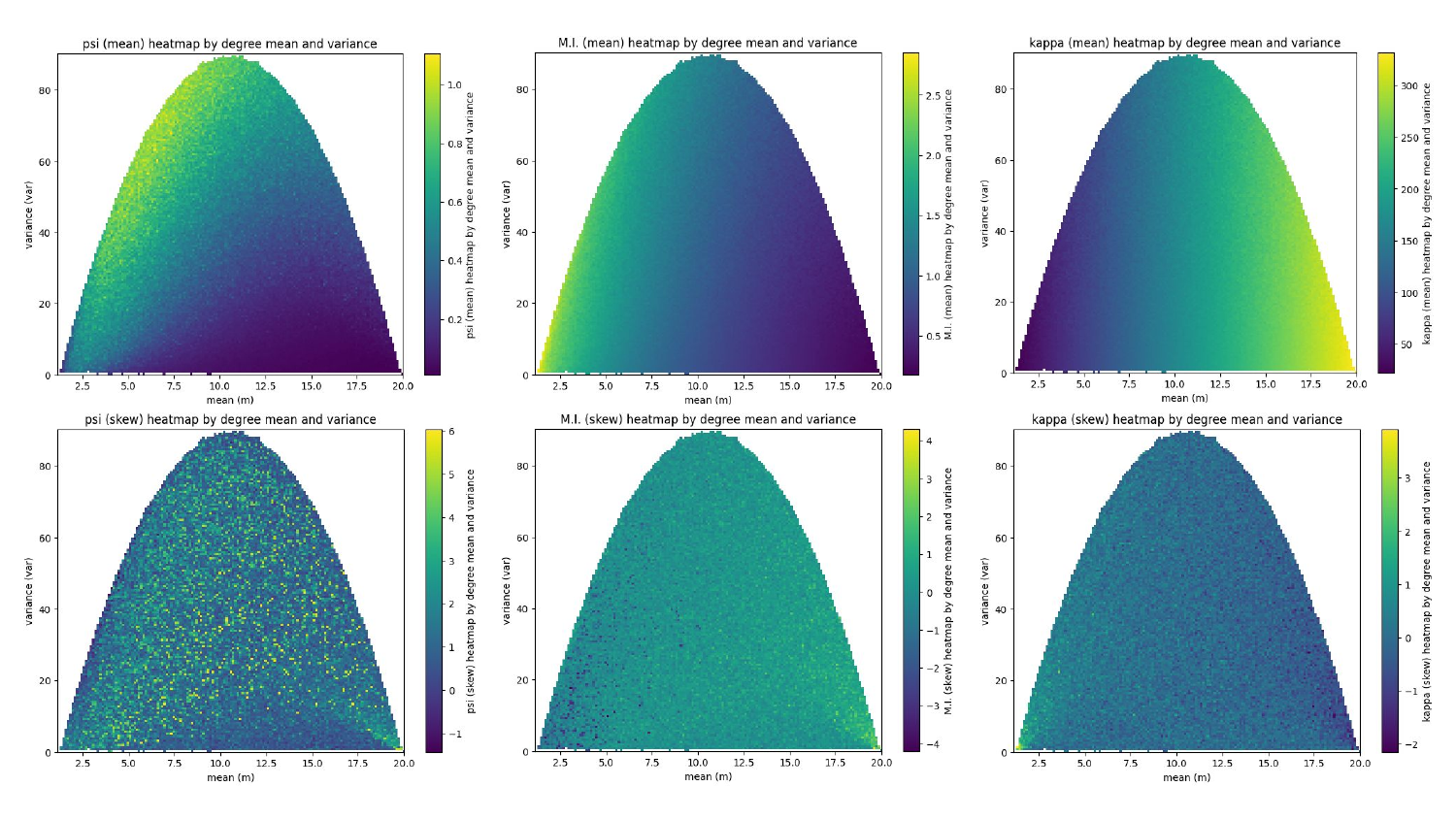}
			\vspace{-0.3cm}
			\includegraphics[scale=0.365]{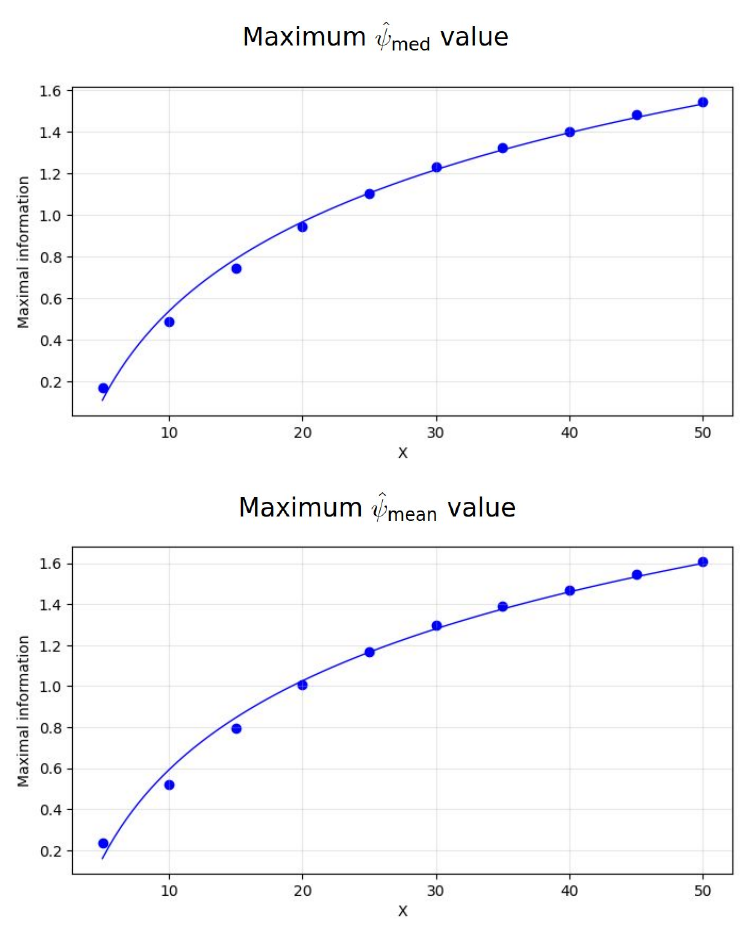}\\
			\textbf{B}\\
			\includegraphics[scale=0.53]{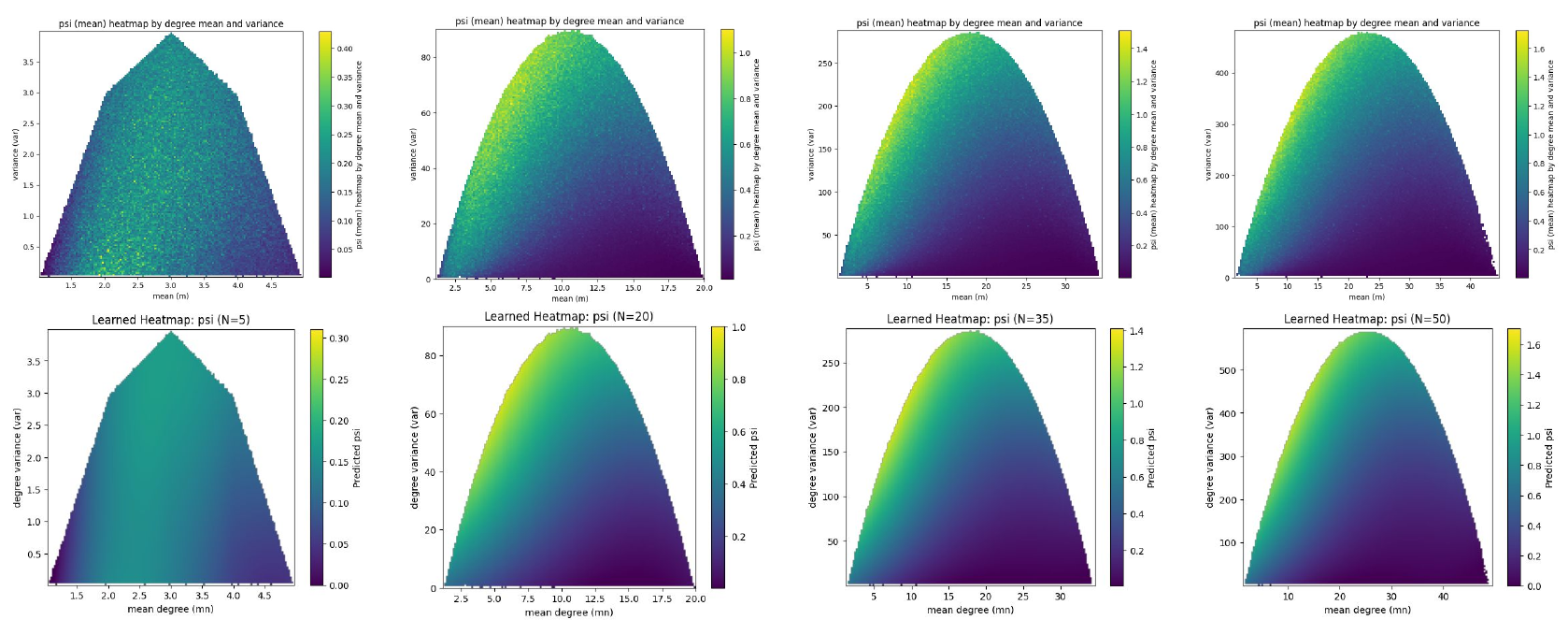}
		\end{minipage}
	}
	\caption[Results from empirical investigation]{Note that the colors are scaled independently across heatmaps. \textbf{A} shows, on the top row, the heatmaps for mean information $\psi^{m, \sigma^2}_{\texttt{mean}}$ with range $(0.01, 1.11)$, mean mutual information $i^{m, \sigma^2}_{\texttt{mean}}$ with range $(0.17, 2.86)$, and mean partitioning constant $\kappa^{m, \sigma^2}_{\texttt{mean}}$ with range $(22, 332)$, for various values of $m$ and$\sigma^2$, for the case $N=20$.
		On the row beneath, we see analogous measures for $\psi^{m, \sigma^2}_{\texttt{sk}} \in ( -1.45 , 6.05 )$; $i^{m, \sigma^2}_{\texttt{sk}} \in ( -4.20 , 4.31 )$; $\kappa^{m, \sigma^2}_{\texttt{sk}} \in ( -2.16 , 3.92 )$.
		We observe that $\psi$ is significantly ``spikier'' and has a more asymmetric skew, indicating greater volatility and bias.
		\space
		\textbf{B} displays the heatmaps $\psi^{m, \sigma^2}_{\texttt{mean}}$ and learned surfaces $\hat{\psi}( m, \sigma^2 )$ for $N \in \{ 5, 20, 35, 50 \}$.
		We have ranges $\psi^5_{\texttt{mean}} \in ( 0.001 , 0.430 )$, $\psi^{20}_{\texttt{mean}} \in ( 0.01 , 1.11 )$, $\psi^{35}_{\texttt{mean}} \in ( 0.01 , 1.51 )$, $\psi^{50}_{\texttt{mean}} \in ( 0.01 , 1.74 )$.
		We observe that as $N$ increases, the low-information region in the bottom-right increases in size, while the high-information strip is increasingly marginalized.
		However, this rate of change slows down.
		With regard to the learned surfaces, they appear to capture the overall shape of our distribution, while appearing to slightly underestimate the high-$\psi$ areas.
		This may arise from $\psi^{m, \sigma^2}_{\tau}$ in such regions having a longer tail.
		\space
		\textbf{C} displays the maximal $\hat{\psi}$ value of our learned surfaces, as a function of $N$, for both the \texttt{mean} and \texttt{median} statistics.
		These plots grow logarithmically with $N$, with functions as described in Table \ref{max_params}.}
	\label{results}
\end{figure}

\subsection{Results and analysis}\label{6.3.2}

\textbf{Bivariate dependence.} Figure \ref{results} displays heatmaps for $\psi$, $i$, $\kappa$ in the case that $N=20$, and demonstrates one instance of a general finding that $\psi$ had the most complex dependence relationship with respect to $m$ and $\sigma^2$.
$\kappa$ and $i$ appear to possess equivalence classes in the direction of some diagonal $r$, with their values instead depending on 
$\kappa$ appeared to increase linearly with $m$, while $i$ appeared to show redundancy in the direction of some gradient $r$, depending instead on some constant $d$ governing an equation $\sigma^2 = rm + d$.
MaxCal deviations $\mu$ adopt an asymmetric relationship.
$\psi$ appears overall to decrease with $m$, however the rate of decrease is much higher when $\sigma^2$ is low.
This produces a \textbf{quasicubic regime,} depicted in figure \ref{results}\\
\\
Equations \ref{info_ent_rate} provide some indication for why this may be the case.
In addition to the path space analysis conducted in section \ref{6.2}, we can deduce that $\hat{\psi}$ is plausibly close to $\log \kappa + i - 2 \mathcal{H}( \pi )$.
Since high degree-variance should indicate a lack of balance (as explored in \ref{6.2.1}), we would expect high-$\sigma^2$ regimes to have low $2 \mathcal{H} ( \pi )$ values.\\
\\
\textbf{Perturbational complexity.} Figure \ref{results} shows the skew and mean values of $\psi$, $i$, $\kappa$ in the case that $N = 20$.
In both cases, the plot for $\psi$ is substantially ``spottier'', and unlike $i$ and $\kappa$ there appears to be no clear pattern or consistency to the skew --- a finding which was replicated for the median, and for other values of $N$.
This appears to suggest that, while the underlying metrics governing the distribution of $\psi$ are reflected in its structure, it can also undergo substantial shifts based on the circumstances of the particular randomness.
This indicates a higher sensitivity to perturbation (of the underlying governing structure, rather than of the state), compared to other measures of information.\\
\\
\textbf{Emergent topological structure.} As $N$ increases, figure \ref{results} shows that random noise appears to dominate less of the variance, and the structure of the surface underlying the noise tends toward some pattern.
Table \ref{table_errors} verifies that global fit of our cubic approximation increases with $N$, while systemic error increases.
With this, we see that our maximal $\hat{\psi}$ value \textbf{moves leftward} as $N$ increases.
However, since analysis outlined in table \ref{max_params} indicated a linear relationship, this could perhaps arise from reduced significance of the constant term.\\
\\
\textbf{Median and mean diverge with $N$.} More interesting is the behavior of $\hat{\psi}_{\textup{max}}$, which appears to scale logarithmically with $N$.
In particular, we see that $e^{\hat{\psi}_{\textup{max}}} \approx B N^A$ for some $A$, $B$ terms described in table \ref{max_params}.
Thus, we see that maximal mean/average information grows according to a power law.
Recalling from section \ref{5.2} that $e^{\psi}$ represents a shrinkage $\Delta_0$ in path space from $\bm{X}^t$ to $\bm{X}^{t+1}$, we shall define $\hat{N}_0 = \Delta_0^{1/A}$ and then express our power law as:

\begin{equation}\label{power_law}
	\left(
	\dfrac{\hat{N}_0}{N}
	\right)^A
	\approx
	B.
\end{equation}

In effect, we claim that when we normalize our ``average deviation size'' by the size of state space, we retrieve a constant value.\\
\\
\textbf{Linear growth.} We can additionally derive from our power law that $\hat{\psi}_{\textup{max}}$ grows with the number of nodes.
If $N = M^n$ for some integer $M$, representing that $\Omega_{\bm{X}} = \bigtimes_{i=1}^n \Omega_i$ is a joint state space over $n$ nodes, then we retrieve $\hat{\psi}_{\textup{max}} \approx a n + b$, with $a = A \log M$ and $b = \log B$.\\
\\
\textbf{Increasing skew.} While small, our results found that the optimal mean degree $m_{\textup{max}}$ grows faster when the mean is used as our best estimate rather than the median.
Further, $\hat{\psi}_{\textup{max}}$ itself also grew faster.
This appears to indicate that mean and median estimates of $\psi$ diverge as $N$ grows.

\begin{table}[ht]
	\centering
	\caption{Parameters at maximal points}
	\begin{tabular}{ l l l c  c }
		\toprule[1pt]
		\textbf{Metric} & \textbf{Aggregation} & \textbf{Formula} & \textbf{RMSE} & $\boldsymbol{R^2}$ \\
		\midrule
		$m_{\max}(N)$ 
		& Mean   & $0.219N + 1.434$ & $4.97 \times 10^{-1}$ & $\bm{0.976}$ \\
		& Median & $0.212N + 1.669$ & $\bm{4.88 \times 10^{-1}}$ & $0.976$ \\
		\addlinespace
		$\sigma^2_{\max}(N)$ 
		& Mean   & $0.191N^2 - 0.992N + 7.165$  & $1.04 \times 10^1$ & $0.995$ \\
		& Median & $0.194N^2 - 1.198N + 10.212$ & $\bm{1.02 \times 10^1}$ & $\bm{0.995}$ \\
		\addlinespace
		$\hat{\psi}_{\max}(N)$ 
		& Mean   & $0.625 \log N + \log 0.429$ & $3.95 \times 10^{-2}$ & $0.992$ \\
		& Median & $0.617 \log N + \log 0.414$ & $\bm{3.10 \times 10^{-2}}$ & $\bm{0.994}$ \\
		\bottomrule[1pt]
	\end{tabular}
	\label{max_params}
\end{table}

\subsection{Implications}\label{6.3.3}

Empirically, our MaxCal deviations $\psi$ appears to track a balance between order and disorder.
It maximizes when the degrees $d( \bm{x} )$ have a reasonably low mean coupled with degree-variance which is maximal --- $m \approx \tfrac{N}{5}$; $\sigma^2 \approx \tfrac{4N(N-5)}{25}$ --- forcing $d \sim p$ to place probability mass almost entirely on $1$ and $N$, with $p(1) \approx \tfrac{4}{5}$ and $p(N) \approx \tfrac{1}{5}$.
The resulting regime, as predicted in section \ref{6.2}, is one which holds real capacity for large dispersal, yet reliably returns to recovery.
This combination allows for $\pi$ to be significantly less entropic than in other regimes, driving high information values.\\
\\
We might draw attention to Burda and Duda's analysis of random walks here.
In regimes where degree across state space is constant, $d( \bm{x} ) \sim \delta(d)$, we see that short-term and indefinite MaxCal objectives align: $\pi_{\textup{GRW}} \sim \pi_{\textup{MERW}}$, producing $\psi \equiv 0$ in both regimes.
We also see on our grid that when $\sigma^2 \approx 0$ we are very likely to produce $\psi = 0$ empirically.
This is not guaranteed, since our TPMs $\bm{P}^{m, \sigma^2}_{\tau}$ may align with acyclic adjacency matrices, however for $A^{\top} = A$ we see alignment.
Likewise, Burda and Duda find that heterogeneity is associated with divergence between MERWs and GRWs\cite{burda10}, which broadly aligns with our requirement of high $\sigma^2$ for $\psi$ to be high.\\
\\
These results raise questions about how information, as defined here, may arise in practice.
In addition to competing spacial objectives explored in chapter \ref{5_IIT_FEP}, one could question whether differing MaxCal objectives over time drive the emergence of information, as defined here, in a sufficiently ``jagged'' state space.

\begin{table}[ht]
	\centering
	\caption[Error metrics by value of $N$]{Polynomial surface fitting error metrics for varying network sizes ($N$).}
	\renewcommand{\arraystretch}{1.2} 
	
	\begin{tabular}{ll cccc}
		\toprule[2pt]
		$\boldsymbol{N}$ & \textbf{Aggregation} & \textbf{MSE} & $\boldsymbol{R^2}$ & \textbf{Max Error} & \textbf{RMSE} \\
		\midrule
		$5$ 
		& Mean   & $\bm{1.05 \times 10^{-3}}$ & $0.6665$ & $\bm{2.05 \times 10^{-1}}$ & $\bm{3.25 \times 10^{-2}}$ \\
		& Median & $5.51 \times 10^{-4}$ & $0.7118$ & $\bm{1.43 \times 10^{-1}}$ & $2.35 \times 10^{-2}$ \\
		\addlinespace
		$10$ 
		& Mean   & $1.84 \times 10^{-3}$ & $0.9091$ & $2.51 \times 10^{-1}$ & $4.29 \times 10^{-2}$ \\
		& Median & $1.26 \times 10^{-3}$ & $0.9247$ & $2.68 \times 10^{-1}$ & $3.55 \times 10^{-2}$ \\
		\addlinespace
		$15$ 
		& Mean   & $2.22 \times 10^{-3}$ & $0.9523$ & $3.11 \times 10^{-1}$ & $4.71 \times 10^{-2}$ \\
		& Median & $1.53 \times 10^{-3}$ & $0.9614$ & $3.90 \times 10^{-1}$ & $3.91 \times 10^{-2}$ \\
		\addlinespace
		$20$ 
		& Mean   & $2.15 \times 10^{-3}$ & $0.9699$ & $3.72 \times 10^{-1}$ & $4.64 \times 10^{-2}$ \\
		& Median & $1.42 \times 10^{-3}$ & $0.9772$ & $4.60 \times 10^{-1}$ & $3.76 \times 10^{-2}$ \\
		\addlinespace
		$25$ 
		& Mean   & $2.01 \times 10^{-3}$ & $0.9787$ & $4.36 \times 10^{-1}$ & $4.48 \times 10^{-2}$ \\
		& Median & $1.35 \times 10^{-3}$ & $0.9838$ & $5.00 \times 10^{-1}$ & $3.67 \times 10^{-2}$ \\
		\addlinespace
		$30$ 
		& Mean   & $1.81 \times 10^{-3}$ & $0.9841$ & $4.53 \times 10^{-1}$ & $4.25 \times 10^{-2}$ \\
		& Median & $1.30 \times 10^{-3}$ & $0.9873$ & $6.85 \times 10^{-1}$ & $3.60 \times 10^{-2}$ \\
		\addlinespace
		$35$ 
		& Mean   & $1.57 \times 10^{-3}$ & $0.9881$ & $3.88 \times 10^{-1}$ & $3.96 \times 10^{-2}$ \\
		& Median & $1.10 \times 10^{-3}$ & $0.9907$ & $5.34 \times 10^{-1}$ & $3.31 \times 10^{-2}$ \\
		\addlinespace
		$40$ 
		& Mean   & $1.44 \times 10^{-3}$ & $0.9901$ & $4.69 \times 10^{-1}$ & $3.79 \times 10^{-2}$ \\
		& Median & $1.06 \times 10^{-3}$ & $0.9920$ & $6.94 \times 10^{-1}$ & $3.26 \times 10^{-2}$ \\
		\addlinespace
		$45$ 
		& Mean   & $1.43 \times 10^{-3}$ & $0.9910$ & $3.18 \times 10^{-1}$ & $3.79 \times 10^{-2}$ \\
		& Median & $\bm{1.08 \times 10^{-3}}$ & $\bm{0.9926}$ & $4.23 \times 10^{-1}$ & $\bm{3.28 \times 10^{-2}}$ \\
		\addlinespace
		$50$ 
		& Mean   & $1.50 \times 10^{-3}$ & $\bm{0.9912}$ & $5.72 \times 10^{-1}$ & $3.87 \times 10^{-2}$ \\
		& Median & $1.18 \times 10^{-3}$ & $0.9925$ & $6.89 \times 10^{-1}$ & $3.44 \times 10^{-2}$ \\
		\bottomrule[1.5pt]
	\end{tabular}
	\label{table_errors}
\end{table}

\subsection{Limitations}\label{6.3.4}

Though some insight has been drawn regarding the behavior of \textit{average} MaxCal deviations in $N \times N$ settings, notable limitations still exist.
We have not thoroughly assessed the distribution of $\psi_{\tau}^{m, \sigma^2}$ values for each $(m, \sigma^2)$-pairing, and so the precise mechanics behind our result remain speculative.
We note that preliminary analysis indicates a ``spiky'' distribution of $\kappa$ drives extreme values, while low marginal entropy $\mathcal{H}( \pi )$ may exaggerate the effects.
However, further study is required to investigate these claims.\\
\\
We note further that as $N$ increased, the global fit of our approximation ($R^2$) increased, however so did MSE, as outlined in Table \ref{table_errors}.
Residual heatmaps, contained in the full analysis, displayed substantially less random noise (with respect to the underlying shape), however also contained regions where labels were consistently underestimated, as well as regions with consistent overestimates.
It could be that a more complex, piecewise surface is required for defining $\hat{\psi}$.\\
\\
Overall, the distribution of $\psi_{\tau}^{( m, \sigma^2 )}$ merits further investigation before any precise hypotheses are drawn.
We note further that these results are restricted to ergodic, homogeneous Markovian regimes, limiting the scope of conclusions which can be drawn.

\section{Concluding remarks}\label{6.4}

Overall, when our deviations metric $\psi$ is investigated across empirical and theoretical contexts, we observe some association with markers of dynamical complexity, as well as specific state space topologies which indicate a capacity to ``collapse and recover''.
It appears to be vulnerable to perturbation within $\bm{P}$, the causal structure itself, indicating that high $\phi$ is likely to occur when perturbing nodes in $\mathcal{G}_{\textup{static}}$ materially disturbs outcomes within the regime.
A picture emerges of movement $\pi \rightarrow \pi$ which appears stable, yet is fragile in the sense of depending upon every moving part, to exist.
There is, of course, to possibility of defining ``moving parts'' by their capacity to perturb (an approach which would mirror the logic of IIT 4.0\cite{albantakis2023}), which allows us to consider complexity from a reductive (rather than constructive) perspective.\\
\\
An interesting result from the experiments conducted in section \ref{6.3} is that $\psi$ is unlikely to achieve its maximal values in undirected state space.
For each $m$, maximal variance $\sigma^2 \approx \sigma_m^2$ forces a regimes in which $d \in \{ 1, N \}$ are exceedingly unlikely.
In the undirected case, the only structure which meets this requirement is a ``star'' in which all states lead toward and from one value.
We can prove analytically here that $\psi$ is $O(- \log N)$.
However, we can also observe it produces a mean of $m_0 = 2 - \tfrac{1}{N}$ and note that this will increasingly shift toward the bottom left corner of our heatmap as $N$ grows.
Thus, we should generally expect an asymmetric connectivity matrix in high-$\psi$ regimes.\\
\\
One should note that to achieve $\Phi > 0$, from a MaxCal perspective, one does not merely need $\psi( \pi ) > 0$.
Instead, a repertoire of transitions $\bm{Y}^t \rightarrow \bm{V}^{t+1}$ should exist, each with non-negligible, irreducible $\psi_{\bm{Y} \rightarrow \bm{V}} > 0$.
For this to occur in a stationary, homogeneous Markov chain, the behaviors (metastability; localization; irregularity; non-predictability) and structures (heterogeneously sparse state space; ``elastic determinism'') associated with $\psi > 0$ would presumably be replicated across each of these trajectories.
While further study is required, a picture on the intuitive front certainly emerges of consumption, synergy, and exploitation-exploration.
There are also some grounds to suspect criticality or SOC.
In particular, since $\psi_{\bm{Y} \rightarrow \bm{V}}$ in each case measures deviation of each trajectory from its ``maximally relaxed'' path (given the causal structure $\bm{P}_{\bm{Y} \rightarrow \bm{V}}$), and their estimators $\hat{\psi}$ are approximated by $B N^A$ for large $N$, one might expect some scale-free structures to emerge of the form deviations across $\bm{X}$ and its subsystems $\bm{Y}$ take.\\
\\
Further research could focus on the ``shape'' of MaxCal deviations across structures other than homogeneous Markov chains, to investigate whether the broad findings are replicated, or where they differ.
Additionally, one could could consider modeling the circumstances in which deviations combine across timescales.\\
\\
For example, an MERW might be likely to deviate locally from its ``single-step MaxCal'' ensemble, while adhering to MaxCal via $\pi$ at the global level.
Thus, modeling via a homogeneous Markov chain may be appropriate.
``Information'' here thus might reflect something structural about the window of time a system's ``canonical state space'' exists over\footnote{
	This could be consistent with the way MERWs can be used to derive the steady states of quantum wavefunctions\cite{burda10}.
}.
Conversely, in an inhomogeneous system in which $\bm{P}_t$ changes, one might rely on other tools.
As we did in section \ref{5.5}, a mean-field approximation $\bar{\bm{P}}$ could be used over some window $\mathbb{T} = \{ 0, 1, \ldots, T \}$, and then the MaxCal deviation of its stationary approximation $\bar{\pi}$ could characterize one facet of the system, while the fluctuations of its empirical distribution $\rho$ could be assessed with respect to $\bar{\pi}$\footnote{
	Further modeling of MaxCal across some sequence of chains $( \mathbb{X}_1, \mathbb{T}_1 ), ( \mathbb{X}_2, \mathbb{T}_2 ), \ldots, ( \mathbb{X}_L, \mathbb{T}_L )$, could also be accomplished.
}.
A pertinent question would be under which circumstances do these deviations occur with and without each other.\\
\\
Overall, the conceptual space of this method is enrich due to the breadth of MaxEnt \& MaxCal research across physics \cite{davis15,presse13} and inference\cite{berger96,giffin09,haarnoja18}.
Further researching IIT, FEP, or computational neuroscience more broadly from this perspective, could advance the interdisciplinary nature of these fields, while arguably holding scope to catalyze mathematical and computational progress.

%% file: 7-conc/7_conc.tex
\chapter{Toward a unified model of cognition}\label{7_conc}

\textsc{Prior to outlining} the case for a unified model, we shall review our progress.
In chapter \ref{2_IIT}, we formalized our understanding of IIT (3.0) through the use of cause and effect functions to simplify its original perturbative approach.
Chapter \ref{3_bays} saw us review Bayesian brain theories, including predictive coding and active inference.\\
\\
The theoretical case for unification presented its face in chapter \ref{4_info_dev}, where DBNs helped us to understand IIT's techniques as assessing constrained entropy maximization over a transition network.
Following this, in chapter \ref{4.3}, we explored the mathematical dualism of active inference with constrained entropy maximization (CMEP).
The significance here is not conclusive proof of congruence, but rather a shared framework for understanding two previously disparate theories.\\
\\
A series of theoretical ventures then formed the basis of chapter \ref{5_IIT_FEP}.
Initially, we defined information $\psi$ over a single step forward in time directly as deviation from a MaxCal ensemble $\mu \times \bm{P}$.
\textit{Perplexity} helped us interpret this from the lens of path space compression.
Sections \ref{5.3}--\ref{5.5} then focused on reconciling $\psi$ with FEP directly.\\
\\
First, we thought of transition log-probabilities as Glauber-style energies, and constructed a free energy functional $\mathcal{F}$ from this.
When assumed to be dependent, $\bm{P} = \bm{P}_{\gamma}$, on some blanket path $\gamma \equiv \bm{b}^t \rightarrow \bm{b}^{t+1}$, we can identify $\mathcal{F}$ as the divergence of a VFE functional $\mathcal{F}_V$ which should be minimized over repeated trajectories.\\
\\
The stronger frameworks, arguably, came from modeling $\psi$ over longer term trajectories.
After applying mean-field assumptions to $\bm{P}$, we applied CLT for Markov chains to derive a view of $\psi_{\textup{clt}}$ as model complexity.
In our LDP view over an Ising model, $\psi_{\textup{clt}}$ was instead an accuracy penalty traded off against a cost of model updating, while $\mathcal{F}_V$ took the form of L2-regularized loss.
While just two plausible unifications among many, these highlight a degree of mathematical congruence across each of the theories.\\
\\
To investigate the role of MaxCal deviations in complexity, we returned to our single-step view and investigated stationary distributions, finding that $\psi$ can plausibly be associated with traits such as localization, collapse/recover dynamics, critical regimes of sparsity, and metastability.
While further study is required, we derived a vision of $\Phi_{\textup{MaxCal}}$ as a metric which captures emergence, consumption, and symmetry.
Scale-free invariance of deviation ``volume'' $\psi$ is likely to occur too, due to the power law we derived.\\
\\
In addition to all of this, we probed some connections between our core theories and machine learning, including chaos in deep neural networks, and MaxEnt in algorithms such as Soft Actor-Critic.

\begin{figure}[ht]
	\centering
	\includegraphics[height=5cm]{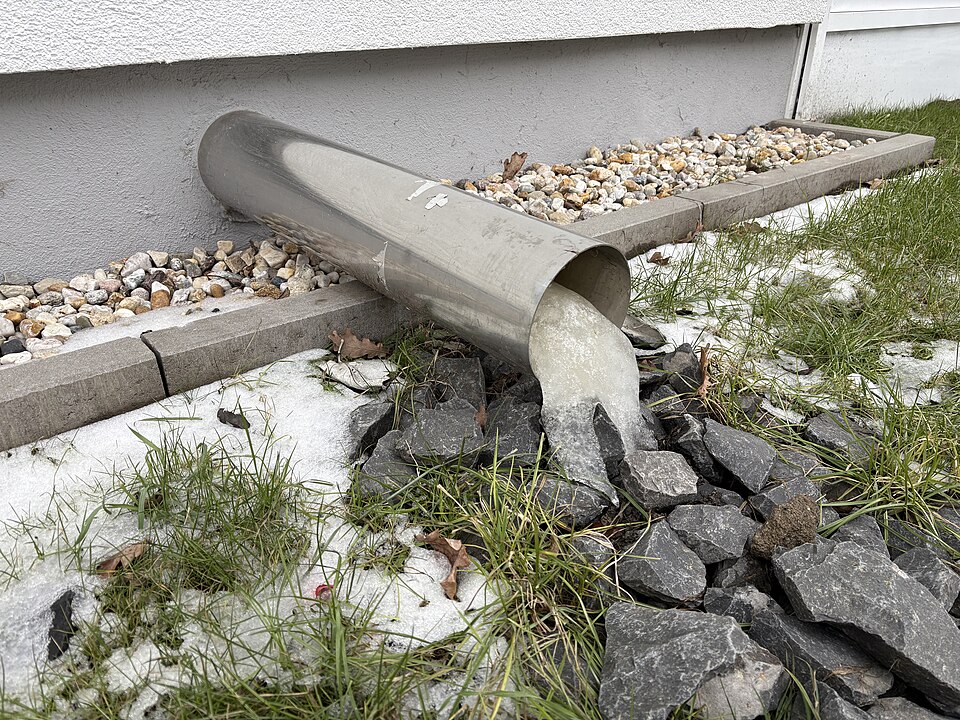}
	\hspace{0.2cm}
	\includegraphics[height=5cm]{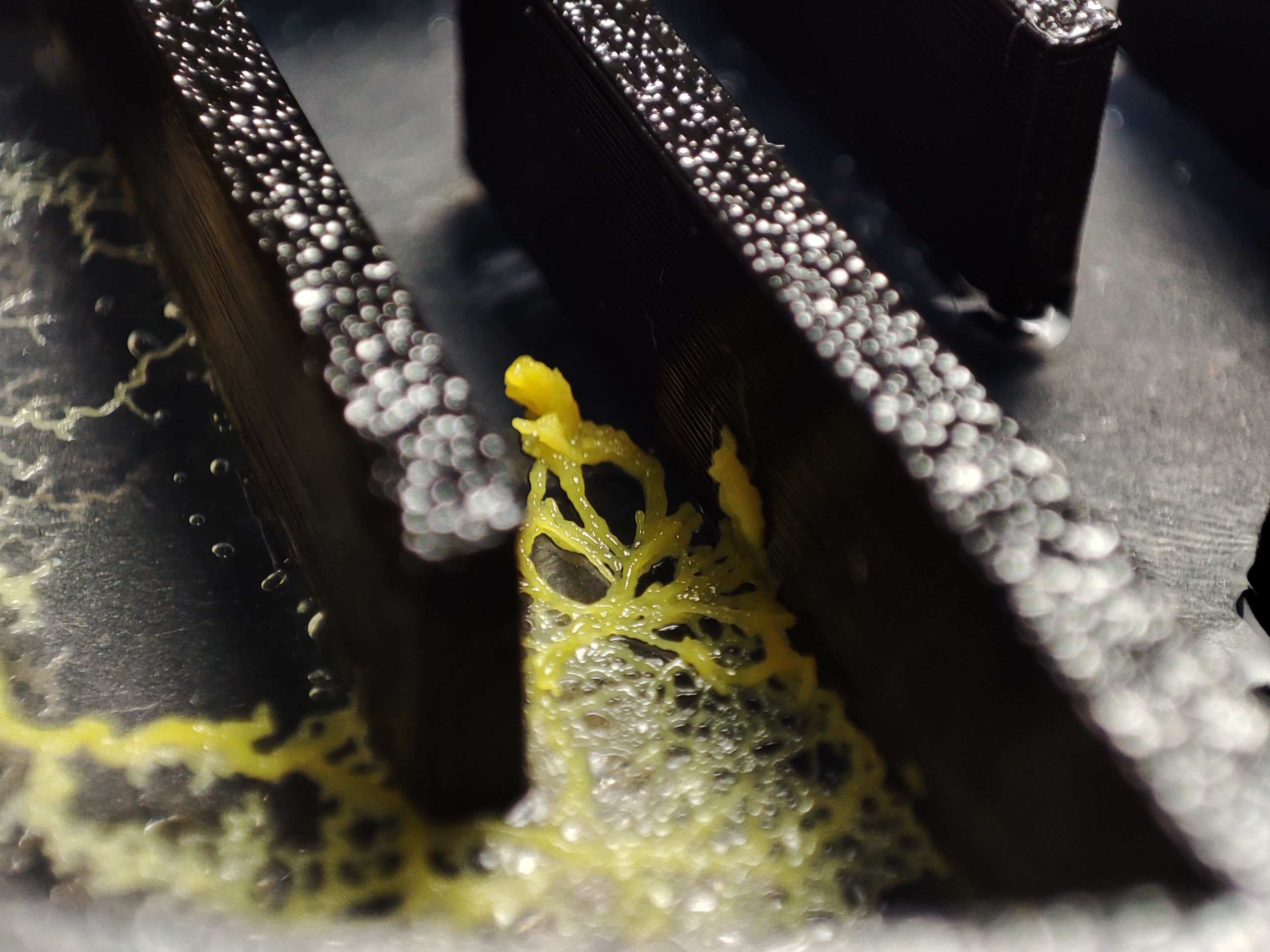}
	\caption[Comparing active and inactive systems]{
		Pipes and mazes constrain maximal path entropy by limiting movement in physical space.
		Water will naturally relax into the most disordered regime possible, given forces which are externally imposed.
		Physarum polycephalum, conversely, generates its own forces and appears to move via an exploration-exploitation balance reflected in its fractal-like branching structure.
		\textit{(Right) This image was taken by Tim Tim (VD fr) and the CCA 4.0 license can be found \href{https://commons.wikimedia.org/wiki/File:Maze_-_Australian_physarum_polycephalum.jpg}{here}.}
	}
	\label{water_mold_comp}
\end{figure}

With all this noted, we can now forward an outline of a MaxCal-oriented unification theory.
First, we note that while entropy is combinatorial in nature\cite{jaynes57,giffin09}, MaxCal is dynamic\cite{jaynes80,dixit18}.
A very intuitive example, included in figure \ref{water_mold_comp}, is water flowing through a pipe.
Left to its own devices, water would spread chaotically in all directions.
The pipe, however, imposes physical boundaries that enclose this disorder within a confined space, allowing water to be channeled for functional purposes.\\
\\
One could argue that controlling a dynamic system is a matter of determining appropriate boundaries.
Learning, from this angle, can be viewed as creating the optimal landscape to channel chaos effectively.
Since MaxCal trajectories represent a system in a ``relaxed'' case, doing this reduces the effort (free energy) required to meet some objective.
The distinction made in active inference\cite{pezzulo24}, and some new paradigms in machine learning\cite{dawid24}, is that learning is performed endogenously.
Deep neural networks are sculpted by engineers who impose an external objective.
Agents possess internal drives and choose their own trajectories.\\
\\
For an agent to minimize its deviation the MaxCal objective, it must be able to ``see'' its residuals in some capacity.
This suggests a natural link to perturbational complexity and Fluctuation-Dissipation Theorem (FDT) violation based theories, both of which result in propagation of these.
On the IIT side, if information itself is a measure of MaxCal deviation, then perception can be understood as a consequence of this regime.\\
\\
It is important to note that IIT does not attempt to answer functionalist topics in neuroscience, instead focusing its efforts on the more ontological, \textit{``what does it mean to feel something?''}\cite{nagel74}
Unification, thus, should not attempt to conflate IIT with FEP or FDT-violation theories.
Instead, one might see each of these frameworks as observing something useful and distinct about the brain.
By respecting the distinct contributions of each framework, exploiting their complementary strengths, and systematically mapping points of contact between each, perhaps we may construct a theoretical neuroscience which is ``greater than the sum of its parts'' and move ever so slightly closer toward a complete, unified brain theory. $\square$

%% file: appendix/app.tex
\chapter*{Appendix}\label{app}

\section*{Full algorithm for sweeping parameter space}

\begin{algorithm}[h]
	\caption{\texttt{info-metrics}: Retrieve information metrics from a TPM}
	\begin{algorithmic}[1]
		\Require state space $\Omega$ of size $N > 1$, degree distribution $p$
		\Ensure $(\psi, i, \kappa)$ triple of MaxCal deviation, mutual information, and partitioning constant.
		\State Generate degree values $d( \bm{x} )$ iid $\sim p$
		\State Generate ergodic $\bm{P}$ with row-$\bm{x}$ degree $d( \bm{x} )$
		\State Compute stationary distribution $\pi$
		\State Compute MaxCal input marginal $\mu$
		\State $\psi \leftarrow \mathcal{D}( \pi \lvert \rvert \mu )$
		\State $i \leftarrow \mathcal{I}^{\pi}( \bm{X}^{t+1} ; \bm{X}^t )$
		\State $\kappa \leftarrow \sum_{\bm{X}} \textup{PP}( \bm{x} )$
		\State \Return $( \psi, i, \kappa )$
	\end{algorithmic}
	\label{info_metrics}
\end{algorithm}

\begin{algorithm}[h]
	\caption{\texttt{measure-stats}: Measure statistics at parameter cell}
	\begin{algorithmic}[1]
		\Require state space $\Omega$ of size $N > 1$, mean degree $m$, degree-variance $\sigma^2$, trial count $T$
		\Ensure summary statistics \texttt{mean}, \texttt{median}, \texttt{skew} for each $( \psi, i, \kappa )$ triple at $( N, m, \sigma^2 )$
		\State Solve MaxEnt $p( d ) \propto \exp( \lambda_1 d + \lambda_2 d^2 )$ with $\mathbb{E}[d] = m$, $\mathbb{E}[d^2] = m^2 + \sigma^2$
		\For{$\tau = 1, \ldots, T$}
		\State $( \psi_{\tau}, i_{\tau}, \kappa_{\tau} ) \leftarrow \texttt{info-metrics}( N, p )$
		\EndFor
		\State $\bm{\psi}^{ \mu, \sigma^2 } \leftarrow ( \texttt{median}( \psi_{\tau} ), \texttt{mean}( \psi_{\tau} ), \texttt{skew}( \psi_{\tau} ) )$
		\State $\bm{i}^{\mu, \sigma^2} \leftarrow ( \texttt{median}( i_{\tau} ), \texttt{mean}( i_{\tau} ), \texttt{skew}( i_{\tau} ) )$
		\State $\bm{\kappa}^{ \mu, \sigma^2 } \leftarrow ( \texttt{median}( \kappa_{\tau} ), \texttt{mean}( \kappa_{\tau} ), \texttt{skew}( \kappa_{\tau} ) )$
		\State \Return $( \bm{\psi}^{ m, \sigma^2 }, \bm{i}^{ m, \sigma^2 }, \bm{\kappa}^{ m, \sigma^2 } )$
	\end{algorithmic}
	\label{measure_stats}
\end{algorithm}